\documentclass[aps,
pra,
twocolumn,
notitlepage,
superscriptaddress,
longbibliography]{revtex4-1}

\usepackage{bm}
\usepackage{amsthm}
\usepackage{amsmath,amssymb}
\usepackage[dvipdfmx]{graphicx}
\usepackage{braket}
\usepackage{enumerate}
\usepackage{ascmac}
\usepackage{mathrsfs}

\usepackage[
colorlinks=true,
linkcolor=blue,
filecolor=blue,
urlcolor=blue
]{hyperref}
\usepackage{mathtools,bbding}
\usepackage{CJKutf8}
\usepackage{soul}
\usepackage{chngcntr}

\theoremstyle{definition}
\newtheorem{theorem}{Theorem}

\newtheorem{definition}{Definition}
\newtheorem{lemma}{Lemma}
\newtheorem{corollary}{Corollary}
\newtheorem{example}{Example}
\newtheorem{remark}{Remark}
\newtheorem{claim}{Claim}
\newtheorem{proposition}{Proposition}

\newcommand{\pd}[3]{
 \if 1#1 \frac{\partial #2}{\partial #3}
 \else \frac{\partial^{#1} #2}{\partial #3^{#1}}\fi}
 \newcommand{\od}[3]{
 \if 1#1 \frac{{\mathrm d} #2}{{\mathrm d} #3}
 \else \frac{{\mathrm d}^{#1} #2}{{\mathrm d}#3^{#1}}\fi}

\newcommand{\ee}[0]{{\mathrm e}}

\newcommand\Rap{R_{\mathrm{ap}}}
\newcommand\Rex{R_{\mathrm{ex}}}

\newcommand\calD{{\cal D}}
\newcommand\calE{{\cal E}}

\newcommand\calH{{\cal H}}

\newcommand\calK{{\cal K}}
\newcommand\calL{{\cal L}}

\newcommand\calQ{{\cal Q}}

\newcommand\calU{{\cal U}}
\newcommand\calV{{\cal V}}

\newcommand{\sym}[0]{{\rm Sym}}

\newcommand{\floor}[1]{\lfloor {#1} \rfloor}

\newcommand{\bal}{\begin{equation}\begin{aligned}}
\newcommand{\eal}{\end{aligned}\end{equation}}

\allowdisplaybreaks[4]

\begin{document}
%\begin{CJK}{UTF8}{min}

%\preprint{}

%TC:ignore
\title{
The i.i.d. State Convertibility in the Resource Theory of Asymmetry for Finite Groups
}



\author{Tomohiro Shitara}
\thanks{Both authors contributed equally to this work}
\email[]{tomohiro.shitara@ntt.com}
\affiliation{NTT Computer and Data Science Laboratories, NTT Inc.,
3-9-11 Midori-cho, Musashino, Tokyo 180-8585, Japan}
\affiliation{NTT Research Center for Theoretical Quantum Information, NTT Inc., 3-1 Morinosato Wakamiya, Atsugi, Kanagawa, 243-0198, Japan}
%%%%
\author{Yosuke Mitsuhashi}
\thanks{Both authors contributed equally to this work}
\email[]{yosuke.mitsuhashi@riken.jp}
\thanks{Current affiliation: RIKEN}
\affiliation{Department of Basic Science, University of Tokyo, 3-8-1 Komaba, Meguro-ku, Tokyo, 153-8902, Japan}
%%%%
\author{Hiroyasu Tajima}
\thanks{Current affiliation: Kyushu University}
%\email[]{hiroyasu.tajima@uec.ac.jp}
\email[]{hiroyasu.tajima@inf.kyushu-u.ac.jp}
%\thanks{Both authors contributed equally to this work.}
\affiliation{Department of Communication Engineering and Informatics,
University of Electro-Communications, 1-5-1 Chofugaoka, Chofu, Tokyo, 182-8585, Japan}
\affiliation{JST, PRESTO, 4-1-8 Honcho, Kawaguchi, Saitama, 332-0012, Japan}

%\thanks{}


% \date{\today}

\begin{abstract}

We identify exact and approximate conversion rates between i.i.d.~pure states under covariant operations in the resource theory of asymmetry for symmetries described by finite groups. 
We establish the formula for the exact conversion rate by completely specifying the relevant set of resource measures. 
The exact conversion is generally asymptotically irreversible due to the existence of multiple independent resource measures, and we find the necessary and sufficient condition for asymptotic reversibility. 
On the other hand, we show that the approximate conversion rates diverge or vanish, which implies that the asymmetry can be infinitely amplified if we allow a vanishingly small error. 
We reveal the underlying mechanism of such a counterintuitive phenomenon by utilizing the properties of maximally asymmetric states.

%In recent years, there has been active research toward understanding the connection between symmetry and physics from the viewpoint of quantum information theory. This approach stems from the resource theory of asymmetry (RTA), a general framework treating quantum dynamics with symmetry, and scopes various fields ranging from the fundamentals of physics, such as thermodynamics and black hole physics, to the limitations of information processing, such as quantum computation, quantum measurement, and error-correcting codes. Despite its importance, in RTA, the resource measures characterizing the asymptotic conversion rate between i.i.d. states are not known except for $U(1)$ and $\mathbb Z_2$. In this work, we derive both the approximate and exact conversion rates between i.i.d. pure states in RTA where the symmetry is described by a finite group. The approximate conversion rate either diverges or equals zero, which implies that the asymmetry can be amplified infinitely if we allow a vanishingly small error. We reveal the underlying mechanism of such a counterintuitive phenomenon, by showing the existence of the uniform state that acts as a perfect catalysis. On the other hand, the exact conversion rate is determined by a set of resource measures. The exact conversion is in general irreversible due to multiple independent resource measures, but we also find the condition for reversibility. These results are expected to significantly broaden the scope of the application of RTA.
\end{abstract}

%TC:endignore

%\keywords{}

\maketitle

% References should be done using the~\cite, \ref, and \label commands
%------------------------------------------------------------------------------------%
%------------------------------------------------------------------------------------%

\textit{\textbf{Introduction}}---
%Symmetry, RTA
Symmetry is one of the most powerful guiding principles in modern physics. The importance of symmetry has been understood since the age of analytical mechanics, and today it plays a central role in almost every field in physics, from condensed matter physics~\cite{McGreevy2023} to high energy physics~\cite{Georgi2000}. In recent years, active research has been conducted in order to understand the connection between symmetry and physics from the point of view of quantum information theory~\cite{Marvian2016,Tajima2018,Woods2018,Tajima2020, Marvian2019,Lostaglio2019,Korzekwa2013,Ahmadi2013,Marvian2012,Tajima2019,Kuramochi2023,Emori2023,Tajima2022,Tajima2021,Zhou2021,Yang2022,Liu2023}. The foundation of this approach is an emerging field of quantum information theory called the resource theory of asymmetry (RTA)~\cite{Gour2008,Marvian2013,Marvian2012a,Zhang2017,Takagi2019,Marvian2020,Marvian2022,Yamaguchi2023,Yamaguchi2023a,Kudo2023}.

%Resource theory
The RTA is a branch of the resource theory~\cite{Gour2008,Marvian2013,Marvian2012a,Zhang2017,Takagi2019,Marvian2020,Marvian2022,Yamaguchi2023,Yamaguchi2023a,Kudo2023,Chitambar2019,Bennett1996,Bennett1996a,Vedral1997,Brandao2013,Brandao2015,Horodecki2013,Aberg2006,Baumgratz2014}, an active area in the quantum information theory. Resource theories define quantum ``resources'' in a variety of situations under some constraints on easily available operations, and examine, in a general framework, what can be done with the help of resources. 
There are several types of resource theories, depending on which property of the system is considered a resource~\cite{Chitambar2019}, such as entanglement theory~\cite{Bennett1996,Bennett1996a,Vedral1997} and quantum thermodynamics~\cite{Brandao2013,Brandao2015,Horodecki2013}, which treat entanglement and athermality as resources, respectively.
In particular, the RTA deals with the ``degree of symmetry breaking'' of a state as a resource, allowing us to treat the restrictions imposed by the symmetry in a unified framework~\cite{Gour2008,Marvian2013,Marvian2012a,Zhang2017,Takagi2019,Marvian2020,Marvian2022, Yamaguchi2023,Yamaguchi2023a,Kudo2023}.
The RTA has been applied to a substantial number of symmetry-related topics such as speed limits~\cite{Marvian2016}, implementation of quantum computation gates~\cite{Tajima2018,Tajima2020,Tajima2022,Tajima2021}, clocks~\cite{Woods2018}, coherence broadcasting~\cite{Marvian2019,Lostaglio2019}, measurement theory~\cite{Korzekwa2013,Ahmadi2013,Marvian2012,Tajima2019,Kuramochi2023,Emori2023,Tajima2022},  quantum error correction~\cite{Tajima2022,Tajima2021,Zhou2021,Yang2022,Liu2023}, coherence cost of thermodynamic process~\cite{Faist2015,Tajima2022,Tajima2025}, quantum Mpemba effects~\cite{Summer2025,Yamashika2025}, and black hole physics~\cite{Tajima2022,Tajima2021}. 

%iid conversion rateを特徴づける量
Despite its importance and potential, the RTA still has many open problems in its foundations.
One of the most important problems is to identify a resource measure or a set of resource measures that characterizes the convertibility between the independent and identically distributed (i.i.d.) states.
For example, the entanglement entropy~\cite{Bennett1996} in the case of entanglement and the Helmholtz free energy~\cite{Brandao2013} in the case of quantum thermodynamics (for isothermal processes) determine the optimal conversion rates between the i.i.d.~states. In the RTA, on the other hand, this problem has been solved only for $U(1)$ symmetry~\cite{Gour2008, Marvian2020,Marvian2022} and $\mathbb Z_2$ symmetry~\cite{Gour2008}, the simplest of the continuous and discrete symmetries, respectively.
This fact means that the most fundamental quantities in the RTA have not yet been revealed, which prevents further analysis of the RTA.

%Summary of this letter
In this work, we solve this problem for symmetries described by finite groups. Concretely, we consider transforming $N$ copies of a pure state to $M$ copies of another pure state, and identify the maximal ratio $M/N$ in two cases: exact conversion and approximate conversion. We show that when we consider an exact transformation without any errors, the conversion rate is given by the minimum of the ratio of resource measures defined for each group element.
On the other hand, when we allow for vanishingly small errors in the transformation, we show that the conversion rate takes only the extreme values of $0$ or $\infty$, showing a striking difference from the former case. %These results are different from those for continuous symmetries, and we explain the underlying mechanism for the difference between finite-group symmetries and continuous symmetries. In a nutshell, the difference arises from the existence of maximally asymmetric states possessing infinite amounts of asymmetry.
This approximate result is also different from the case of continuous symmetries, which can be intuitively understood by the nonexistence of maximally asymmetric states for continuous symmetries. 
%We show that the Pontryagin duality~\cite{}, a well-known duality relation in the group theory, gives a clear explanation of this problem.

%-----------------------------------------------------------------------------------%
\textit{\textbf{Formulations}}---
Symmetry is mathematically specified by a group $G$, and it appears in a physical system through a projective unitary representation $U(g)$, which is a unitary operator on the Hilbert space satisfying $U(g)U(g')=\omega(g,g')U(gg')$ with some phase factor $\omega(g, g')\in\mathbb{C}$ such that $|\omega(g, g')|=1$ for all $g, g'\in G$. 
The representation is called a (non-projective) unitary representation when the phase factor is trivial, i.e., $\omega(g, g')=1$ for all $g, g'\in G$. 
For simplicity, we restrict ourselves to the non-projective case in the main text, and we guide readers to the Supplemental Material (SM)~\footnote{See Supplemental Material. URL to be added.} for the projective case.

To define the RTA~\cite{Gour2008, Marvian2012a, Marvian2013}, we introduce free states and free operations, which correspond to easily available states and easily implementable operations, respectively. 
First, we say that a state $\rho$ is free if the state is a \textit{symmetric state}, which is invariant under the conjugation actions of all symmetry operations $U(g)$, i.e., $\rho$ satisfies $\mathcal{U}_g(\rho)=\rho$ for all $g\in G$, where $\calU_g(\cdot)=U(g) \cdot U^\dagger (g)$. 
Next, we say that a CPTP map $\calE$ is a free operation if it is a \textit{$G$-covariant operation}, which commutes with conjugation actions of all symmetry operations, i.e., $\calE$ satisfies $\calU_g'\circ\calE=\calE\circ\calU_g$ for all $g\in G$, where $\calU_g(\cdot):=U(g)\ \cdot\  U(g)^\dag$ and $\calU'_g(\cdot):=U'(g) \ \cdot \  U'(g)^\dag$ are the conjugation actions of projective unitary representations $U$ and $U'$ on the input and output systems, respectively.

\textit{\textbf{State conversion}}---
%single-shot conversion, information theoretical setting
One of the most important issues in resource theories is to determine whether one state $\rho$ can be converted to another state $\sigma$ using only free operations. 
We consider two kinds of conversions, exact conversion and approximate conversion. 
We say that a state $\rho$ is exactly ($\epsilon$-approximately resp.) convertible to a state $\sigma$ if there exists a $G$-covariant operation $\calE$ satisfying $\calE(\rho)=\sigma$ ($T(\calE(\rho), \sigma)\leq\epsilon$ resp.) with the trace distance $T(\rho_1, \rho_2):=\|\rho_1-\rho_2\|_1/2$, which we denote by $\rho\xrightarrow{G\text{-cov.}}\sigma$ ($\rho\xrightarrow{G\text{-cov.}}_\epsilon\sigma$ resp.). 
%We say that $\rho$ is $\epsilon$-approximately convertible to $\sigma$ if there exists a $G$-covariant operation $\calE$ satisfying $T(\calE(\rho), \sigma)\leq\epsilon$ with the trace distance $T(\rho_1, \rho_2):=\|\rho_1-\rho_2\|_1/2$, which we denote by $\rho\xrightarrow{G\text{-cov.}}_\epsilon\sigma$.
%We note that the exact conversion corresponds to the case of $\epsilon=0$ in the approximate conversion. 
%We note that the condition $\rho\xrightarrow{G\text{-cov.}}_\epsilon\sigma$ is equivalent to the existence of a (not necessarily covariant) CPTP map $\calE'$ such that $T(\calE'(\calU_g(\rho)),\calU'_g(\sigma))\le \epsilon$ for all $g\in G$, which allows us to interpret the operational meaning of the covariant operation from the information-theoretic viewpoint~\cite{Marvian2013}.

%iid conversion rate
In this Letter, we focus on the conversion rate $M/N$ between i.i.d.~states $\rho^{\otimes N}$ and $\sigma^{\otimes M}$. 
In this conversion, we consider the tensor product of representations $U^{\otimes N}(g):=U(g)^{\otimes N}$ as the input and output representations. 
We say that the conversion rate $r$ is exactly ($\epsilon$-approximately resp.) achievable if $\rho^{\otimes N}$ can be convertible to $\sigma^{\otimes \floor{rN}}$ with no error (within error $\epsilon$ resp.)
for any $N$ greater than a sufficiently large $N_0$.
%for sufficiently large $N$. 
%, i.e., $\rho^{\otimes N}\xrightarrow{G\text{-cov.}}_\epsilon\sigma^{\otimes \floor{rN}}$ 
%In the case of $\epsilon=0$, we say that the rate $r$ is exactly achievable. 
We define the optimal exact conversion rate $\Rex(\rho\rightarrow\sigma)$ as the supremum of the exactly achievable conversion rates. 
We define the optimal approximate conversion rate $\Rap(\rho\rightarrow\sigma)$ as the supremum of the rates $r$ that are $\epsilon$-approximately achievable for all $\epsilon>0$. 
Although only the approximate one is usually discussed in the context of conversion rate, we also consider the exact conversion rate, following the previous studies on the RTA for the $\mathbb Z_2$ symmetry~\cite{Gour2008}.

We briefly review the result of Ref.~\cite{Marvian2013}, which provides several useful tools in the resource theory of asymmetry. 
%(or as the state of the art in the field of the RTA?)
The first one is the symmetry subgroup defined by $\sym_G(\rho) \coloneqq \{g\in G | \mathcal{U}_g(\rho)=\rho\}$. 
The symmetry subgroup roughly constrains state convertibility, since it becomes monotonically larger under covariant operations, i.e., $\sym_G(\rho)\subset\sym_G(\sigma)$ if $\rho\xrightarrow{G\text{-cov.}}\sigma$. 
We note, however, that the converse does not hold in general. 
The second one is the characteristic function $\chi_\rho: G\to\mathbb{C}$, which is defined by $\chi_\rho(g):=\mathrm{tr}(\rho U(g))$ for all $g\in G$. 
This function tells us the condition on the convertibility of states. 
Concretely, a pure state $\psi$ can be converted to another pure state $\phi$ by some covariant operation if and only if there exists some positive semi-definite function $f: G\to\mathbb{C}$ such that $\chi_\psi=f \chi_\phi$, where $f$ is called positive semi-definite if $\sum_{g, h\in G}\alpha(g)^* \alpha(h) f(g^{-1} h) \ge 0$ for all $\alpha: G\to\mathbb{C}$. 
%It is also shown that there exists a state $\rho$ with characteristic function $\chi$ if and only if $\chi(e)=1$ and $\chi$ is positive semi-definite when $G$ is a finite group. 

\textit{\textbf{Key concepts}}---
We introduce two concepts that play important roles in this Letter. 
First, we define a function $L_\psi: G\to[0, \infty]$ for a pure state $\psi$ by 
\begin{align}
    L_\psi(g):=-\ln(|\mathrm{tr}(\psi U(g))|) \ \forall g\in G, 
\end{align}
which is equivalently expressed as $L_\psi(g)=-\ln(|\chi_\psi(g)|)$ using the characteristic function $\chi_\psi(g)$. 
By using the properties of the characteristic function, we find that the function $L_\psi$ has the following four desirable properties as a resource measure: (i) positivity, i.e, $L_\psi(g)\geq 0$ for all $g\in G$; (ii) faithfulness, i.e., $L_\psi(g)=0$ for all $g\in G$ if and only if $\psi$ is a symmetric (free) state; (iii) additivity, i.e., $L_{\psi_1\otimes \psi_2}(g)=L_{\psi_1}(g)+L_{\psi_2}(g)$ for all $g\in G$; and (iv) monotonicity, i.e., $L_\psi(g)\ge L_\phi(g)$ for all $g\in G$ whenever $\psi\xrightarrow{G\text{-cov.}}\phi$ (see Lemma~\ref{SMlem:measure_L_properties} in SM~\cite{Note1} for details).

Second, we define \textit{maximally asymmetric states}, which have the maximum amount of asymmetry among all the states with some given symmetry subgroup. 
Concretely, we say that a state $\xi$ is a maximally asymmetric state with a symmetry subgroup $H$ if $\xi$ is a pure state satisfying $|\chi_\xi(g)|=1$, i.e., $L_\xi(g)=0$ if $g\in H$ and $|\chi_\xi(g)|=0$, i.e., $L_\xi(g)=\infty$ otherwise, which includes the catalytic states in Refs.~\cite{Marvian2013, Luijk2023} as a special case where $H=\{e\}$. 
By using the properties of the characteristic functions mentioned above, we can easily confirm that maximally asymmetric states can be exactly converted to any state with the same or a larger symmetry subgroup, but on the other hand, maximally asymmetric states cannot be generated from any other states with the same or a smaller symmetry subgroup. 
We note that maximally asymmetric states do not exist in the case of continuous symmetries, since the characteristic function is continuous in $g\in G$.

We comment on the explicit construction and physical realization of maximally asymmetric states. 
With the left regular representation $F$ of $G$ on the space spanned by an orthonormal basis $\{\ket{\eta_g}\}_{g\in G}$, which satisfies $F(g)\ket{\eta_h}=\ket{\eta_{gh}}$, we can confirm that a state $\gamma$ defined by $\ket{\gamma}:=|H|^{-\frac{1}{2}}\sum_{h\in H} \ket{\eta_h}$ is a maximally asymmetric state with symmetry subgroup $\sym_G(\gamma)=H$ (see Lemma~\ref{SMlem:max_asymm_construction} in SM~\cite{Note1} for details). 
We note that maximally asymmetric states may appear in physical systems without using the left regular representations. 
%As a simple example, we consider the permutation symmetry $S_n$ on $n$ qubits with representation $U$ of $\tau\in S_n$ that brings $i$th qubit to $\tau(i)$th qubit, i.e., $U(\tau)\ket{x_1 x_2 ... x_n}=\ket{x_{\tau^{-1}(1)} x_{\tau^{-1}(2)} ... x_{\tau^{-1}(n)}}$
For example, for the symmetric group $S_2$ with the qubit permutation representation $U$ on a two-qubit system, $\ket{01}$ is a maximally asymmetric state, which can be confirmed by noting that $U((21))\ket{01}=\ket{10}$ is orthogonal to $\ket{01}$. 
For the $S_3$ symmetry on a three-qubit system, six-qubit state $\ket{001}\otimes \ket{010}$ is a maximally asymmetric state, because the states $U(\tau)^{\otimes 2}(\ket{001}\otimes \ket{010})$ with $\tau\in S_3$ are orthogonal to each other.

\textit{\textbf{Main Result 1: Exact conversion rate}}---
First, we present the formula that gives the exact conversion rate between two pure states.

\begin{theorem} \label{thm:exact_conversion}
    Let $G$ be a finite group, and $\psi$ and $\phi$ be pure states. Then, the exact conversion rate from  $\psi$ to $\phi$ is given by
    \begin{align}
        \Rex(\psi\rightarrow\phi) = \min_{g\in G} \frac{L_{\psi}(g)}{L_{\phi}(g)}, \label{ExactRate}
    \end{align}
    where we define $c/0:=\infty$ {when $0\leq c\leq \infty$ and $\infty/\infty:=\infty$ for the value of $L_{\psi}(g)/L_{\phi}(g)$ in extreme cases}.
    Moreover, for any $r\in (0, \Rex(\psi\to\phi))$, the conversion rate $r$ is exactly achievable if the number $N$ of copies of the input state $\psi$ satisfies 
    \begin{align}
        N\ge \frac{\ln\left(\displaystyle\frac{|G|}{|\sym_G(\psi)|}\right)}{\left(1-\displaystyle\frac{r}{\Rex(\psi\rightarrow\phi)}\right)\Delta(\psi)}, \label{eq:input_num_exact}
    \end{align}
    where we exclude the trivial case where $\psi$ is a symmetric state, and $\Delta(\psi):=\min_{g\in G-\sym_G(\psi)} L_\psi(g)$, i.e., $\Delta(\psi)$ is the smallest nonzero value of $\{L_\psi(g)\}_{g\in G}$. 
\end{theorem}

%Rが0,正, 無限になる場合について言及
%symmetry subgroupの条件を満たさなければRが0になっちゃうこととか

Our main contribution is to show that $\Rex(\psi\rightarrow\phi) \geq \min_{g\in G} L_{\psi}(g)/L_{\phi}(g)$, because the inverse inequality directly follows from the additivity and monotonicity of $L$. 
This theorem implies that the conversion is asymptotically reversible, i.e., $\Rex(\psi\rightarrow\phi)=\Rex(\phi\rightarrow\psi)^{-1}$, if and only if $L_\psi\propto L_\phi$. 
It is worth noting that only the information on the amplitudes of the characteristic functions suffices to determine the i.i.d.~exact conversion rate, contrasting to the single-copy convertibility~\cite{Marvian2013}, where information on the phases of the characteristic functions is also essential. 
We also note that when $G=\mathbb Z_2$, Eq.~\eqref{ExactRate} reproduces the previous result~\cite{Gour2008} by Gour and Spekkens.

We note that Theorem~\ref{thm:exact_conversion} captures the following three trivial cases. 
First, when $\sym_G(\psi)\not\subset\sym_G(\phi)$, there exists some $g\in G$ such that $L_\psi(g)=0$ and $L_\phi(g)>0$, which implies that $\Rex(\psi\to\phi)=0$. 
Second, when $\psi$ and $\phi$ have an identical symmetry subgroup $H$, $\psi$ is a maximally asymmetric state, and $\phi$ is an arbitrary state, we have $L_\psi(g)=L_\phi(g)=0$ for all $g\in H$ and $L_\psi(g)=\infty$ for all $g\in G-H$, which implies that $\Rex(\psi\to\phi)=\infty$. 
Third, in contrast, 
%when $\phi$ is a maximally asymmetric state with symmetry subgroup $H$ that is strictly smaller than $G$ and $\psi$ is not a maximally asymmetric state with symmetry subgroup $H$, 
when $\psi$ and $\phi$ have an identical symmetry subgroup $H$, $\phi$ is a maximally asymmetric state, and $\psi$ is not a maximally asymmetric state, 
%is not,
%is not a maximally asymmetric state whose symmetry subgroup includes that of $\phi$,
there exists some $g\in G-H$ such that $L_\psi(g)<\infty$ and $L_\phi(g)=\infty$, which implies that $\Rex(\psi\to\phi)=0$.

%This result implies that the exact conversion rate is determined by the ratio of $L_\psi(g)$. 
%Since there are multiple and possibly independent resource measures $L_\psi(g)$ for different $g$'s, we need to take the minimum over $G$. 
%We note that this result implies that $\mathrm{Sym}_G(\psi)\subset\mathrm{Sym}_G(\phi)$ holds whenever $\psi\xrightarrow{G\text{-cov.}}_0\phi$ is possible. 

%From Theorem~\ref{thm:exact_conversion}, we can also discuss the reversibility and irreversibility of the i.i.d. conversion. We see from Eq.~\eqref{ExactRate} that the conversion is asymptotically reversible, i.e., $\Rex(\psi\rightarrow\phi)=\Rex(\phi\rightarrow\psi)^{-1}$, if and only if $L_\psi(g)\propto L_\phi(g)$. 
%We can also confirm that maximally asymmetric states cannot be exactly generated at a non-zero rate from a state $\psi$ with $L_\psi(g) < \infty$ for some $g\in G$.

To illustrate this result, we take state conversions under the qubit permutation symmetries $S_2$ and $S_3$ on $2$ and $3$ qubits as examples. 
First, we consider the $S_2$-covariant conversion rate between two-qubit states in the form of $\ket{\psi(\theta)}=(\ket{10}+\ee^{i\theta}\ket{01})/\sqrt{2}$ with $\theta\in\mathbb{R}$. 
%\ym{($\ket{\psi(\theta)}:=\cos(\theta)\ket{01}+\sin(\theta)\ket{10}$ may be a better example to illustrate the special property of the uniform state.)}
Since $L_{\psi(\theta)}(\tau)$ is given by $0$ and $-\ln(|\cos(\theta)|)$ for $\tau=(12)$, and $(21)$, respectively, Theorem~\ref{thm:exact_conversion} states that 
\begin{align}
    &\Rex(\psi(\theta_1)\rightarrow\psi(\theta_2)) \nonumber\\
    &=\begin{cases}
        \displaystyle\frac{\ln(|\cos(\theta_1)|)}{\ln(|\cos(\theta_2)|)} & \textrm{if }\cos(\theta_1)\neq 0 \textrm{ and } 0<|\cos(\theta_2)|<1\\
        \infty & \textrm{if }\cos(\theta_1)=0 \textrm{ or } |\cos(\theta_2)|=1\\
        0 & \textrm{otherwise}
    \end{cases} \label{eq:exact_ex1}
\end{align}
for all $\theta_1, \theta_2\in\mathbb{R}$. 
The first case in Eq.~\eqref{eq:exact_ex1} means that the conversion is generally asymptotically reversible. 
For example, if we set $\theta_1=\pi/3$ and $\theta_2=\pi/4$, we have $\Rex(\psi(\theta_1)\rightarrow\psi(\theta_2))=2$ and $\Rex(\psi(\theta_2)\rightarrow\psi(\theta_1))=1/2$, which means that this conversion is asymptotically reversible. 
We can confirm that $L_{\psi(\theta_1)}\propto L_{\psi(\theta_2)}$ holds in this case. 
The second case in Eq.~\eqref{eq:exact_ex1} means that once we have a maximally asymmetric state $(\ket{01}+i\ket{10})/\sqrt{2}$, we can transform it to infinitely many copies of any state $\ket{\psi(\theta)}$. 
On the other hand, the third case in Eq.~\eqref{eq:exact_ex1} means that we cannot generate even a single copy of a maximally asymmetric state from a state that is not maximally asymmetric.

Next, we consider the $S_3$-covariant conversion rate between three-qubit states in the form of $\ket{\psi(\theta)}=(\ket{100}+\ee^{i\theta}\ket{010})/\sqrt{2}$ with $\theta\in\mathbb{R}$. 
Since $L_{\psi(\theta)}(\tau)$ is given by $0$, $-\ln(|\cos(\theta)|)$, and $\ln(2)$ for $\tau=(123)$, $(213)$, and otherwise, respectively, Theorem~\ref{thm:exact_conversion} states that 
\begin{align}
    &\Rex(\psi(\theta_1)\rightarrow\psi(\theta_2)) \nonumber\\
    &=\begin{cases}
        \displaystyle\frac{\ln(|\cos(\theta_1)|)}{\ln(|\cos(\theta_2)|)} & \textrm{if }|\cos(\theta_1)|>|\cos(\theta_2)|>0,\\
        1 & \textrm{if }\cos(\theta_1)=0 \\
        0 & \textrm{otherwise}
    \end{cases} 
\end{align}
for all $\theta_1, \theta_2\in\mathbb{R}$, which implies that this conversion is generally asymptotically irreversible except in the special cases of $|\cos(\theta_1)|=|\cos(\theta_2)|$. 
For example, if we set $\theta_1=\pi/3$ and $\theta_2=\pi/4$, we have $\Rex(\psi(\theta_1)\rightarrow\psi(\theta_2))=1$ and $\Rex(\psi(\theta_2)\rightarrow\psi(\theta_1))=1/2$, which means that this conversion is asymptotically irreversible, contrasting to the case above.

Although we guide readers to Sec.~\ref{SMsec:exact_conversion} of SM~\cite{Note1} for the general proof, we present the proof idea by taking a simple case where $\sym_G(\psi)=\sym_G(\phi)=\{e\}$, and $L_\psi(g)\neq\infty$ and $L_\phi(g)\neq \infty$ for all $g\in G$. 
In this case, Eq.~\eqref{ExactRate} can be rewritten as $\Rex(\psi\to\phi)=\min_{g\in G-\{e\}} L_\psi(g)/L_\phi(g)$, where the right-hand side does not involve extreme cases such as $c/0$ or $\infty/\infty$. 
We focus on the proof of $\Rex(\psi\to\phi)\geq \min_{g\in G-\{e\}} L_\psi(g)/L_\phi(g)$, since the inverse inequality immediately follows from the additivity and monotonicity of $L$. 
We take arbitrary $r<R:=\min_{g\in G-\{e\}} L_\psi(g)/L_\phi(g)$, and show that $\phi^{\otimes N}$ can be converted to $\psi^{\otimes \lfloor rN\rfloor}$ for sufficiently large $N$. 
For this proof, it is sufficient to show that the function $f(g):=\chi_{\psi^{\otimes N}}(g)/\chi_{\phi^{\otimes \floor{rN}}}(g)=\chi_\psi(g)^N/\chi_\phi(g)^{\floor{rN}}$ is positive semi-definite. 
For any function $\alpha: G\to\mathbb{C}$, we note that 
\begin{align}
    &\sum_{g, h\in G} \alpha^*(g)\alpha(h)f(g^{-1}h) \nonumber\\
    &=\sum_{g\in G} |\alpha(g)|^2 +\sum_{g, h\in G : g\ne h} \alpha^*(g)\alpha(h)f(g^{-1}h). 
\end{align}
Since $f(g)$ satisfies $f(e)=1$ and $|f(g)|\leq \ee^{-N(L_\psi(g)-rL_\phi(g))}\leq \ee^{-N(1-r/R) \Delta(\psi)}$ for all $g\in G-\{e\}$, the second term is exponentially small in $N$, while the first term is a constant, which implies that $f$ is positive semi-definite for sufficiently large $N$. 
Equation~\eqref{eq:input_num_exact} can be obtained by considering the condition on $N$ more precisely. 

%For a conversion rate $r$ to be achievable, the function $f(g)=\chi_\psi(g)^N/\chi_\phi(g)^{\floor{rN}}$ should be positive semi-definite, i.e., for any function $\alpha(g)$, the following inequality is satisfied: 
%\begin{align}
%    \sum_{g,h} \alpha^*(g)f(g^{-1}h)\alpha(h)\ge 0. 
%\end{align}
%We divide the sum in the left-hand side into two parts as 
%\begin{align}
%    &\sum_{g\in G} |\alpha(g)|^2 +\sum_{g,h\in G : g\ne h} \alpha^*(g)f(g^{-1}h)\alpha(h) \nonumber \\
%    \simeq& \|\alpha\|_2^2 + \sum_{g\in G : g\ne h} \kakko{\frac{\chi_\psi(g^{-1}h)}{\chi_\phi(g^{-1}h)^r}}^N\alpha^*(g)\alpha(h).
%\end{align}
% O(q^N)
% 文章で説明
%Then, the contribution from the second term is exponentially small in $N$ if $r\le L_\psi(g)/L_\phi(g)$, which assures the positivity for sufficiently large $N$.

%-----------------------------------------------------------------------------------%
%-----------------------------------------------------------------------------------%
\textit{\textbf{Main Result 2: Approximate conversion rate}}---
Next, we give the formula for the approximate conversion rate.
% レート有限N

\begin{theorem} \label{thm:approx_conversion}
    Let $G$ be a finite group, $\psi$ be a pure state, and $\sigma$ be a general (not necessarily pure) state. 
    Then, the approximate conversion rate from $\psi$ to $\sigma$ is given by 
    \begin{align}
        \Rap(\psi\to\sigma)=
        \begin{cases}
            \infty & \textrm{if }\sym_G(\psi)\subset\sym_G(\sigma)\\
            0 & {\rm otherwise}.
        \end{cases}
    \end{align} 
    Moreover, when $\sym_G(\psi)\subset\sym_G(\sigma)$, for any $\epsilon\in (0, 1)$, any conversion rate $r\in (0, \infty)$ is $\epsilon$-approximately achievable if the number $N$ of copies of the input state $\psi$ satisfies 
    \begin{align}
        N\geq\frac{-\ln\left(2\displaystyle\frac{|\sym_G(\psi)|}{|G|}\left[\frac{|\sym_G(\sigma)|}{|G|}(1-\epsilon^2)\right]^{\frac{1}{2}}\epsilon\right)}{\Delta(\psi)}, \label{eq:input_num_approx}
    \end{align}
    where we exclude the trivial case where $\psi$ is a symmetric state, and $\Delta(\psi):=\min_{g\in G-\sym_G(\psi)} L_\psi(g)$, i.e., $\Delta(\psi)$ is the smallest nonzero value of $\{L_\psi(g)\}_{g\in G}$. 
\end{theorem}

This theorem implies that by preparing sufficiently many copies of a state, we can achieve an arbitrarily high conversion rate with an arbitrarily small error as long as the symmetry subgroup of the input state is included in that of the output state.
%More concretely, there exists a sequence of positive real numbers $\{\epsilon_N\}$ satisfying $\lim_{N\rightarrow\infty}\epsilon_N=0$ such that 
%\begin{align}
    %\psi^{\otimes N}\xrightarrow{G{\textrm -}{\rm cov.}}_{\epsilon_N}\sigma^{\otimes M} \label{ApproximateConversion}
%\end{align}
%for an arbitrary number $M>0$, where the error $\epsilon_N$ is independent of $M$.
This result implies that asymmetry can be infinitely amplified by covariant operations if we allow a vanishingly small error. 
As a relation between exact and approximate conversion rates, we note that for any pure states $\psi$ and $\phi$, we have $\Rap(\psi\to\phi)=\infty$ if $\Rex(\psi\to\phi)>0$, which is the contraposition of the trivial statement that $\Rex(\psi\to\phi)=0$ if $\Rap(\psi\to\phi)=0$. 
However, as we show in the following example, the converse is not true, i.e., $\Rap(\psi\to\phi)$ can take $\infty$ even if $\Rex(\psi\to\phi)=0$.

To illustrate this result, we consider the conversions between two-qubit states $\ket{\psi(\theta)}:=(\ket{01}+\ee^{i\theta}\ket{10})/\sqrt{2}$ under the $S_2$ symmetry, which is the same example that we considered in Theorem~\ref{thm:exact_conversion}. 
Since the symmetry subgroup of $\psi(\theta)$ is given by $\sym_G(\psi(\theta))=\{(12), (21)\}=S_2$ when $|\cos(\theta)|=1$ and $\sym_G(\psi(\theta))=\{(12)\}=\{e\}$ otherwise, the approximate conversion rate is given by 
\begin{align}
    &\Rap(\psi(\theta_1)\to\psi(\theta_2)) \nonumber\\
    &=\begin{cases}
        0 & \textrm{if } |\cos(\theta_1)|=1 \textrm{ and } |\cos(\theta_2)|\neq 1\\
        \infty & \textrm{otherwise}.
    \end{cases}
\end{align}
For example, if we set $\theta_1=\pi/4$ and $\theta_2=\pi/2$, we have $\Rap(\psi(\theta_1)\to\psi(\theta_2))=\infty$, although $\psi(\theta_2)$ is a maximally asymmetric state. 
On the other hand, Theorem~\ref{thm:exact_conversion} implies that $\Rex(\psi(\theta_1)\to\psi(\theta_2))=0$, which gives a counterexample of the statement that $\Rap(\psi\to\phi)=0$ if $\Rex(\psi\to\phi)=0$.

Although we guide readers to Sec.~\ref{SMsec:approx_conversion} in SM~\cite{Note1} for a detailed proof, we give a rough sketch of the proof. 
First, we prove that $\Rap(\psi\to\sigma)=\infty$ if $\sym_G(\psi)\subset\sym_G(\sigma)$. 
The key point is that we can convert $\psi^{\otimes N}$ to a state close to a maximally asymmetric state with symmetry subgroup $\sym_G(\sigma)$ within an exponentially small error $\epsilon$ in $N$, although exact conversion to the maximally asymmetric state itself is impossible in general.
Once we have a maximally asymmetric state with symmetry subgroup $\sym_G(\sigma)$, we can convert it to any state with the same symmetry subgroup, which includes $\sigma^{\otimes M}$ for all $M\in\mathbb{N}$. 
The latter conversion from a maximally asymmetric state is guaranteed by  Theorem~\ref{thm:exact_conversion} for pure $\sigma$, but is also valid even when the target state $\sigma$ is mixed.
By combining these two conversions, for any $M\in\mathbb{N}$ and $\epsilon>0$, we can construct a covariant operation that converts $\psi^{\otimes N}$ to $\sigma^{\otimes M}$ within error $\epsilon$ for sufficiently large $N$. 
By analyzing the lower bound of $N$ more precisely, we can obtain the condition shown in Eq.~\eqref{eq:input_num_approx}. 

%We take arbitrary $\epsilon>0$ and show that there exists some $N\in\mathbb{N}$ such that for any $M\in\mathbb{N}$, $\psi^{\otimes N}\xrightarrow{G{\textrm -}{\rm cov.}}_\epsilon \sigma^{\otimes M}$ by constructing a map in two steps. 
%In the first step, we construct a maximally asymmetric state $\gamma$ with symmetry subgroup $\sym_G(\psi)$ on the representation space of the left regular representation. 
%Then, we can take a state $\zeta$ satisfying $T(\zeta, \gamma)=\epsilon$ and $L_\zeta(g)=O(\ln(\epsilon^{-1}))$. 
%By using Theorem~\ref{thm:exact_conversion}, we get $\Rex(\psi\to\zeta)=\Omega(1/\ln(\epsilon^{-1}))$, and we find the minimal number $N$ of copies of $\psi$ to achieve the exact conversion rate $r\in (0, \Rex(\psi\to\zeta))$. 
%By taking $r$ such that $Nr\geq 1$, we decide $N$, which implies that we can take a $G$-covariant map $\calE_1$ such that $\calE_1(\psi^{\otimes N})=\zeta$. 
%In the second step, we take a $G$-covariant map $\calE_2$ such that $\calE_2(\gamma)=\psi^{\otimes M}$ for all $M$, where the existence is guaranteed by the property of maximally asymmetric states, as we mentioned above. 
%By combining these two maps, we construct a map $\calE:=\calE_2\circ\calE_1$. 
%Then, by using the information-processing inequality, we get $T(\calE(\psi^{\otimes N}), \sigma^{\otimes M})=T(\calE_2(\calE_1(\psi^{\otimes N})), \calE_2(\gamma))\leq T(\calE_1(\psi^{\otimes N}), \gamma)=T(\zeta, \gamma)=\epsilon$. 

Next, we prove that $\Rap(\psi\to\sigma)=0$ if $\sym_G(\psi)\not\subset\sym_G(\sigma)$. 
For the proof of this, we show a stronger statement that even if we have arbitrarily many copies of $\psi$, we cannot convert them into a single copy of $\sigma$ within some fixed error $\epsilon$ by any covariant maps. 
We take $\epsilon:=(1-|\chi_\sigma(g)|)/4>0$ with some $g\in\sym_G(\psi)-\sym_G(\sigma)$. 
By the definition of the characteristic function, we have $T(\calE(\psi^{\otimes N}), \sigma)\geq (|\chi_{\calE(\psi^{\otimes N})}(g)|-|\chi_\sigma(g)|)/2$. 
For any $G$-covariant map $\calE$, we have
$g\in \sym_G(\psi)=\sym_G(\psi^{\otimes N})\subset \sym_G(\calE(\psi^{\otimes N}))$, which implies that $|\chi_{\calE(\psi^{\otimes N})}(g)|=1$. 
Therefore, we get $T(\calE(\psi^{\otimes N}), \sigma)\geq (1-|\chi_\sigma(g)|)/2=2\epsilon>\epsilon$, which means that we cannot convert $\psi^{\otimes N}$ to $\sigma$ within error $\epsilon$ by any covariant maps.

%-----------------------------------------------------------------------------------%
\textit{\textbf{Application}}---Our work has a direct application to the theory of the comparison of quantum statistical models~\cite{Matsumoto2010,Buscemi2012}.
Let $E=\{\rho_\theta\}_{\theta\in\Theta}$ be a family of quantum states parameterized by $\theta\in\Theta$, which is called a quantum statistical model.
We define the expected gain $G_\theta(\Gamma, E):=\mathrm{tr}(g_\theta \Gamma(\rho_\theta))$, where a Hermitian operator $g_\theta$ and a CPTP map $\Gamma$ represent the gain and the decision, respectively.
To compare the information contents of quantum statistical models $E=\{\rho_\theta\}_{\theta\in\Theta}$ and $E'=\{\rho'_\theta\}_{\theta\in\Theta}$,  a partial order have been introduced. 
For $\epsilon>0$, a quantum statistical model $E$ is said to be q-$\epsilon$-deficient relative to $E'$%~\footnote{For a more general setting~\cite{Matsumoto2010}, the parameter $\epsilon$ can be dependent on $\theta$.}
, if for any $\Gamma'$, there exists $\Gamma$ such that for any $\theta\in\Theta$, $G_\theta(\Gamma,E) \ge G_\theta(\Gamma',E')-\epsilon/2$ holds for all gains $g_\theta$ satisfying $0\le g_\theta \le I$, which is denoted by denoted by $E\ge^\mathrm{q}_\epsilon E'$.

Our results provide the necessary and sufficient conditions for q-$\epsilon$-deficiency between pure-state models $E_N=\{ \calU_g(\psi)^{\otimes N} \}$ and $E'_{M}=\{ \calU_g(\phi)^{\otimes M} \}$ generated by a finite group $G$. 
First, Theorem~\ref{thm:exact_conversion} implies that $E_N\ge^\mathrm{q}_0 E'_{\floor{rN}}$ for sufficiently large $N$ if and only if $r<\Rex(\psi\to\phi)$. 
Second, Theorem~\ref{thm:approx_conversion} implies that for any $\epsilon, r>0$, $E_N\ge^\mathrm{q}_\epsilon E'_{\floor{rN}}$ holds for sufficiently large $N$ if and only if $\sym_G(\psi)\subset\sym_G(\phi)$.
This can be shown by using a randomization criteria~\cite{Matsumoto2010} that $E\ge^\mathrm{q}_\epsilon E'$ if and only if there exists a CPTP map $\Lambda$ such that $T(\Lambda(\rho_\theta),\rho'_\theta)\le \epsilon$ for all $\theta\in\Theta$, which is equivalent to the convertibility via $G$-covariant operations~\cite{Marvian2013}.

%-----------------------------------------------------------------------------------%
\textit{\textbf{Conclusion and future direction}}---
In this letter, we have specified the exact and approximate conversion rates between i.i.d.~pure states when the symmetry is described by a finite group. 
It should be stressed that our results are valid even for noncommutative symmetries. 
In Theorem~\ref{thm:exact_conversion}, we have shown that the exact conversion rate is determined by the minimum of the ratios of multiple independent resource measures $\{L_\psi(g)\}_{g\in G}$, which makes the conversion generally asymptotically irreversible. 
We have also found that the conversion is asymptotically reversible if and only if the set of the resource measures of the input state is proportional to those of the output states. 
On the other hand, in Theorem~\ref{thm:approx_conversion}, we have proven that the approximate conversion rate either diverges or vanishes depending on the inclusion relation of the symmetry subgroups of the input and output states. 
This fact implies that if we allow a vanishingly small error, the asymmetry can be arbitrarily amplified.
We have clarified the reason for this by using the notion of maximally asymmetric states. 
These results are very special for finite-group symmetries, significantly differing from those for Lie-group symmetries~\cite{Gour2008,Marvian2020,Marvian2022,Yamaguchi2024}.

We comment on three interesting future directions. 
First, it is important to compare another mechanism of amplifying the asymmetry by using correlated catalysts, which has been investigated for $U(1)$ symmetry~\cite{Takagi2022,Shiraishi2024}. 
Clarifying the relationship between these two mechanisms will be useful for understanding the essence of asymmetry and coherence.
Second, extending our results to the case of infinite-dimensional Hilbert spaces remains outstanding. 
Such an extension is necessary to analyze rotationally symmetric bosonic codes~\cite{Grimsmo2020,Endo2025} from the information-theoretical viewpoint.
Finally, it is also demanding to generalize our results to symmetries described by discrete but infinite groups such as $\mathbb{Z}^n$, which is physically motivated because it appears as translational symmetries in an $n$-dimensional lattice system.
%One physically motivated example is a discrete translational symmetry described by $\mathbb Z^n$.
Such a generalization would be helpful to analyze the crystal phase in condensed matter physics, and the translationally symmetric bosonic codes such as Gottesman-Kitaev-Preskill code~\cite{Gottesman2001,Grimsmo2021,Lachance-Quirion2024} and the squeezed cat code~\cite{Schlegel2022,Hillmann2023,Endo2024}. 
%We conclude this paper by leaving these as future works.\\

%%候補
% 物理的な応用として：等間隔の準位、離散時間発展、
% 物性での対称性の破れの指標になっている。

%As for future works, there are several interesting directions. First, it remains elusive to give rigorous proof of extension of our result for finite groups to discrete groups such as the discrete translational symmetry on lattices.
%We expect that the argument in this paper for finite groups also holds for finitely generated groups, due to the similarity between the structure of finite groups and that of finitely generated groups. However, for a rigorous proof, a careful treatment of the infinity is necessary. Another yet important direction is to apply the RTA for finite-group symmetry to various fields of physics. One possible application is the discrete rotational symmetry in the rotationally symmetric bosonic code.

%TC:ignore
\textit{\textbf{Acknowledgments}}---We are grateful to Ryuji Takagi, Henrik Wilming, Koji Yamaguchi, Iman Marvian, Suguru Endo, Yasunari Suzuki, and Yui Kuramochi for valuable discussions.
T.S. acknowledges supports from JST Moonshot R\&D Grant No. JPMJMS2061, JST CREST Grant No. JPMJCR23I4, and MEXT Q-LEAP Grant No. JPMXS0120319794, Japan.
Y.M. is supported by JSPS KAKENHI Grant No. JP23KJ0421.
H.T. is supported by MEXT KAKENHI Grant-in-Aid for Transformative Research Areas B ``Quantum Energy Innovation” Grant Numbers JP24H00830 and JP24H00831, JSPS Grants-in-Aid for Scientific Research JP19K14610, No. JP22H05250 and JST PRESTO No. JPMJPR2014, JST MOONSHOT No. JPMJMS2061.

\appendix 

\makeatletter
\def\bibsection{}  
\makeatother

\bibliography{references.bib}

\begin{thebibliography}{57}%
\makeatletter
\providecommand \@ifxundefined [1]{%
 \@ifx{#1\undefined}
}%
\providecommand \@ifnum [1]{%
 \ifnum #1\expandafter \@firstoftwo
 \else \expandafter \@secondoftwo
 \fi
}%
\providecommand \@ifx [1]{%
 \ifx #1\expandafter \@firstoftwo
 \else \expandafter \@secondoftwo
 \fi
}%
\providecommand \natexlab [1]{#1}%
\providecommand \enquote  [1]{``#1''}%
\providecommand \bibnamefont  [1]{#1}%
\providecommand \bibfnamefont [1]{#1}%
\providecommand \citenamefont [1]{#1}%
\providecommand \href@noop [0]{\@secondoftwo}%
\providecommand \href [0]{\begingroup \@sanitize@url \@href}%
\providecommand \@href[1]{\@@startlink{#1}\@@href}%
\providecommand \@@href[1]{\endgroup#1\@@endlink}%
\providecommand \@sanitize@url [0]{\catcode `\\12\catcode `\$12\catcode `\&12\catcode `\#12\catcode `\^12\catcode `\_12\catcode `\%12\relax}%
\providecommand \@@startlink[1]{}%
\providecommand \@@endlink[0]{}%
\providecommand \url  [0]{\begingroup\@sanitize@url \@url }%
\providecommand \@url [1]{\endgroup\@href {#1}{\urlprefix }}%
\providecommand \urlprefix  [0]{URL }%
\providecommand \Eprint [0]{\href }%
\providecommand \doibase [0]{http://dx.doi.org/}%
\providecommand \selectlanguage [0]{\@gobble}%
\providecommand \bibinfo  [0]{\@secondoftwo}%
\providecommand \bibfield  [0]{\@secondoftwo}%
\providecommand \translation [1]{[#1]}%
\providecommand \BibitemOpen [0]{}%
\providecommand \bibitemStop [0]{}%
\providecommand \bibitemNoStop [0]{.\EOS\space}%
\providecommand \EOS [0]{\spacefactor3000\relax}%
\providecommand \BibitemShut  [1]{\csname bibitem#1\endcsname}%
\let\auto@bib@innerbib\@empty
%</preamble>
\bibitem [{\citenamefont {McGeevy}(2023)}]{McGreevy2023}%
  \BibitemOpen
  \bibfield  {author} {\bibinfo {author} {\bibfnamefont {John}\ \bibnamefont {McGeevy}},\ }\bibfield  {title} {\enquote {\bibinfo {title} {{Generalized Symmetries in Condensed Matter}},}\ }\href {\doibase 10.1146/annurev-conmatphys} {\bibfield  {journal} {\bibinfo  {journal} {Annual Review of Condensed Matter Physics}\ }\textbf {\bibinfo {volume} {17}},\ \bibinfo {pages} {53} (\bibinfo {year} {2023})}\BibitemShut {NoStop}%
\bibitem [{\citenamefont {Georgi}(2000)}]{Georgi2000}%
  \BibitemOpen
  \bibfield  {author} {\bibinfo {author} {\bibfnamefont {Howard}\ \bibnamefont {Georgi}},\ }\href {\doibase 10.1201/9780429499210} {\emph {\bibinfo {title} {{Lie Algebras in Particle Physics: from Isospin to Unified Theories}}}}\ (\bibinfo  {publisher} {Taylor {\&} Francis},\ \bibinfo {address} {Boca Raton},\ \bibinfo {year} {2000})\BibitemShut {NoStop}%
\bibitem [{\citenamefont {Marvian}\ \emph {et~al.}(2016)\citenamefont {Marvian}, \citenamefont {Spekkens},\ and\ \citenamefont {Zanardi}}]{Marvian2016}%
  \BibitemOpen
  \bibfield  {author} {\bibinfo {author} {\bibfnamefont {Iman}\ \bibnamefont {Marvian}}, \bibinfo {author} {\bibfnamefont {Robert~W.}\ \bibnamefont {Spekkens}}, \ and\ \bibinfo {author} {\bibfnamefont {Paolo}\ \bibnamefont {Zanardi}},\ }\bibfield  {title} {\enquote {\bibinfo {title} {{Quantum speed limits, coherence, and asymmetry}},}\ }\href {\doibase 10.1103/PhysRevA.93.052331} {\bibfield  {journal} {\bibinfo  {journal} {Physical Review A}\ }\textbf {\bibinfo {volume} {93}},\ \bibinfo {pages} {052331} (\bibinfo {year} {2016})}\BibitemShut {NoStop}%
\bibitem [{\citenamefont {Tajima}\ \emph {et~al.}(2018)\citenamefont {Tajima}, \citenamefont {Shiraishi},\ and\ \citenamefont {Saito}}]{Tajima2018}%
  \BibitemOpen
  \bibfield  {author} {\bibinfo {author} {\bibfnamefont {Hiroyasu}\ \bibnamefont {Tajima}}, \bibinfo {author} {\bibfnamefont {Naoto}\ \bibnamefont {Shiraishi}}, \ and\ \bibinfo {author} {\bibfnamefont {Keiji}\ \bibnamefont {Saito}},\ }\bibfield  {title} {\enquote {\bibinfo {title} {{Uncertainty Relations in Implementation of Unitary Operations}},}\ }\href {\doibase 10.1103/PhysRevLett.121.110403} {\bibfield  {journal} {\bibinfo  {journal} {Physical Review Letters}\ }\textbf {\bibinfo {volume} {121}},\ \bibinfo {pages} {110403} (\bibinfo {year} {2018})}\BibitemShut {NoStop}%
\bibitem [{\citenamefont {Woods}\ \emph {et~al.}(2018)\citenamefont {Woods}, \citenamefont {Silva},\ and\ \citenamefont {Oppenheim}}]{Woods2018}%
  \BibitemOpen
  \bibfield  {author} {\bibinfo {author} {\bibfnamefont {Mischa~P.}\ \bibnamefont {Woods}}, \bibinfo {author} {\bibfnamefont {Ralph}\ \bibnamefont {Silva}}, \ and\ \bibinfo {author} {\bibfnamefont {Jonathan}\ \bibnamefont {Oppenheim}},\ }\bibfield  {title} {\enquote {\bibinfo {title} {{Autonomous Quantum Machines and Finite-Sized Clocks}},}\ }\href {\doibase 10.1007/s00023-018-0736-9} {\bibfield  {journal} {\bibinfo  {journal} {Annales Henri Poincar{\'{e}}}\ }\textbf {\bibinfo {volume} {20}},\ \bibinfo {pages} {125} (\bibinfo {year} {2018})}\BibitemShut {NoStop}%
\bibitem [{\citenamefont {Tajima}\ \emph {et~al.}(2020)\citenamefont {Tajima}, \citenamefont {Shiraishi},\ and\ \citenamefont {Saito}}]{Tajima2020}%
  \BibitemOpen
  \bibfield  {author} {\bibinfo {author} {\bibfnamefont {Hiroyasu}\ \bibnamefont {Tajima}}, \bibinfo {author} {\bibfnamefont {Naoto}\ \bibnamefont {Shiraishi}}, \ and\ \bibinfo {author} {\bibfnamefont {Keiji}\ \bibnamefont {Saito}},\ }\bibfield  {title} {\enquote {\bibinfo {title} {{Coherence cost for violating conservation laws}},}\ }\href {\doibase 10.1103/PhysRevResearch.2.043374} {\bibfield  {journal} {\bibinfo  {journal} {Physical Review Research}\ }\textbf {\bibinfo {volume} {2}},\ \bibinfo {pages} {043374} (\bibinfo {year} {2020})}\BibitemShut {NoStop}%
\bibitem [{\citenamefont {Marvian}\ and\ \citenamefont {Spekkens}(2019)}]{Marvian2019}%
  \BibitemOpen
  \bibfield  {author} {\bibinfo {author} {\bibfnamefont {Iman}\ \bibnamefont {Marvian}}\ and\ \bibinfo {author} {\bibfnamefont {Robert~W.}\ \bibnamefont {Spekkens}},\ }\bibfield  {title} {\enquote {\bibinfo {title} {{No-Broadcasting Theorem for Quantum Asymmetry and Coherence and a Trade-off Relation for Approximate Broadcasting}},}\ }\href {\doibase 10.1103/PhysRevLett.123.020404} {\bibfield  {journal} {\bibinfo  {journal} {Physical Review Letters}\ }\textbf {\bibinfo {volume} {123}},\ \bibinfo {pages} {020404} (\bibinfo {year} {2019})}\BibitemShut {NoStop}%
\bibitem [{\citenamefont {Lostaglio}\ and\ \citenamefont {M{\"{u}}ller}(2019)}]{Lostaglio2019}%
  \BibitemOpen
  \bibfield  {author} {\bibinfo {author} {\bibfnamefont {Matteo}\ \bibnamefont {Lostaglio}}\ and\ \bibinfo {author} {\bibfnamefont {Markus~P.}\ \bibnamefont {M{\"{u}}ller}},\ }\bibfield  {title} {\enquote {\bibinfo {title} {{Coherence and Asymmetry Cannot be Broadcast}},}\ }\href {\doibase 10.1103/PhysRevLett.123.020403} {\bibfield  {journal} {\bibinfo  {journal} {Physical Review Letters}\ }\textbf {\bibinfo {volume} {123}},\ \bibinfo {pages} {020403} (\bibinfo {year} {2019})}\BibitemShut {NoStop}%
\bibitem [{\citenamefont {Korzekwa}(2013)}]{Korzekwa2013}%
  \BibitemOpen
  \bibfield  {author} {\bibinfo {author} {\bibfnamefont {Kamil}\ \bibnamefont {Korzekwa}},\ }\emph {\bibinfo {title} {{Resource theory of asymmetry}}},\ \href@noop {} {Ph.D. thesis},\ \bibinfo  {school} {Imperial College London} (\bibinfo {year} {2013})\BibitemShut {NoStop}%
\bibitem [{\citenamefont {Ahmadi}\ \emph {et~al.}(2013)\citenamefont {Ahmadi}, \citenamefont {Jennings},\ and\ \citenamefont {Rudolph}}]{Ahmadi2013}%
  \BibitemOpen
  \bibfield  {author} {\bibinfo {author} {\bibfnamefont {Mehdi}\ \bibnamefont {Ahmadi}}, \bibinfo {author} {\bibfnamefont {David}\ \bibnamefont {Jennings}}, \ and\ \bibinfo {author} {\bibfnamefont {Terry}\ \bibnamefont {Rudolph}},\ }\bibfield  {title} {\enquote {\bibinfo {title} {{The Wigner-Araki-Yanase theorem and the quantum resource theory of asymmetry}},}\ }\href {\doibase 10.1088/1367-2630/15/1/013057} {\bibfield  {journal} {\bibinfo  {journal} {New Journal of Physics}\ }\textbf {\bibinfo {volume} {15}},\ \bibinfo {pages} {013057} (\bibinfo {year} {2013})}\BibitemShut {NoStop}%
\bibitem [{\citenamefont {Marvian}\ and\ \citenamefont {Spekkens}(2012)}]{Marvian2012}%
  \BibitemOpen
  \bibfield  {author} {\bibinfo {author} {\bibfnamefont {Iman}\ \bibnamefont {Marvian}}\ and\ \bibinfo {author} {\bibfnamefont {Robert~W.}\ \bibnamefont {Spekkens}},\ }\bibfield  {title} {\enquote {\bibinfo {title} {{An information-theoretic account of the Wigner-Araki-Yanase theorem}},}\ }\href {http://arxiv.org/abs/1212.3378} {\bibfield  {journal} {\bibinfo  {journal} {arXiv}\ ,\ \bibinfo {pages} {1212.3378}} (\bibinfo {year} {2012})}\BibitemShut {NoStop}%
\bibitem [{\citenamefont {Tajima}\ and\ \citenamefont {Nagaoka}(2019)}]{Tajima2019}%
  \BibitemOpen
  \bibfield  {author} {\bibinfo {author} {\bibfnamefont {Hiroyasu}\ \bibnamefont {Tajima}}\ and\ \bibinfo {author} {\bibfnamefont {Hiroshi}\ \bibnamefont {Nagaoka}},\ }\bibfield  {title} {\enquote {\bibinfo {title} {{Coherence-variance uncertainty relation and coherence cost for quantum measurement under conservation laws}},}\ }\href {http://arxiv.org/abs/1909.02904} {\bibfield  {journal} {\bibinfo  {journal} {arXiv}\ ,\ \bibinfo {pages} {1909.02904}} (\bibinfo {year} {2019})}\BibitemShut {NoStop}%
\bibitem [{\citenamefont {Kuramochi}\ and\ \citenamefont {Tajima}(2023)}]{Kuramochi2023}%
  \BibitemOpen
  \bibfield  {author} {\bibinfo {author} {\bibfnamefont {Yui}\ \bibnamefont {Kuramochi}}\ and\ \bibinfo {author} {\bibfnamefont {Hiroyasu}\ \bibnamefont {Tajima}},\ }\bibfield  {title} {\enquote {\bibinfo {title} {{Wigner-Araki-Yanase Theorem for Continuous and Unbounded Conserved Observables}},}\ }\href {\doibase 10.1103/PhysRevLett.131.210201} {\bibfield  {journal} {\bibinfo  {journal} {Physical Review Letters}\ }\textbf {\bibinfo {volume} {131}},\ \bibinfo {pages} {210201} (\bibinfo {year} {2023})}\BibitemShut {NoStop}%
\bibitem [{\citenamefont {Emori}\ and\ \citenamefont {Tajima}(2023)}]{Emori2023}%
  \BibitemOpen
  \bibfield  {author} {\bibinfo {author} {\bibfnamefont {Haruki}\ \bibnamefont {Emori}}\ and\ \bibinfo {author} {\bibfnamefont {Hiroyasu}\ \bibnamefont {Tajima}},\ }\bibfield  {title} {\enquote {\bibinfo {title} {{Error and Disturbance as Irreversibility with Applications: Unified Definition, Wigner--Araki--Yanase Theorem and Out-of-Time-Order Correlator}},}\ }\href {http://arxiv.org/abs/2309.14172} {\bibfield  {journal} {\bibinfo  {journal} {arXiv}\ ,\ \bibinfo {pages} {2309.14172}} (\bibinfo {year} {2023})}\BibitemShut {NoStop}%
\bibitem [{\citenamefont {Tajima}\ \emph {et~al.}(2022)\citenamefont {Tajima}, \citenamefont {Takagi},\ and\ \citenamefont {Kuramochi}}]{Tajima2022}%
  \BibitemOpen
  \bibfield  {author} {\bibinfo {author} {\bibfnamefont {Hiroyasu}\ \bibnamefont {Tajima}}, \bibinfo {author} {\bibfnamefont {Ryuji}\ \bibnamefont {Takagi}}, \ and\ \bibinfo {author} {\bibfnamefont {Yui}\ \bibnamefont {Kuramochi}},\ }\bibfield  {title} {\enquote {\bibinfo {title} {{Universal trade-off structure between symmetry, irreversibility, and quantum coherence in quantum processes}},}\ }\href {http://arxiv.org/abs/2206.11086} {\bibfield  {journal} {\bibinfo  {journal} {arXiv}\ ,\ \bibinfo {pages} {2206.11086}} (\bibinfo {year} {2022})}\BibitemShut {NoStop}%
\bibitem [{\citenamefont {Tajima}\ and\ \citenamefont {Saito}(2021)}]{Tajima2021}%
  \BibitemOpen
  \bibfield  {author} {\bibinfo {author} {\bibfnamefont {Hiroyasu}\ \bibnamefont {Tajima}}\ and\ \bibinfo {author} {\bibfnamefont {Keiji}\ \bibnamefont {Saito}},\ }\bibfield  {title} {\enquote {\bibinfo {title} {{Universal limitation of quantum information recovery: symmetry versus coherence}},}\ }\href {http://arxiv.org/abs/2103.01876} {\bibfield  {journal} {\bibinfo  {journal} {arXiv}\ ,\ \bibinfo {pages} {2103.01876}} (\bibinfo {year} {2021})}\BibitemShut {NoStop}%
\bibitem [{\citenamefont {Zhou}\ \emph {et~al.}(2021)\citenamefont {Zhou}, \citenamefont {Liu},\ and\ \citenamefont {Jiang}}]{Zhou2021}%
  \BibitemOpen
  \bibfield  {author} {\bibinfo {author} {\bibfnamefont {Sisi}\ \bibnamefont {Zhou}}, \bibinfo {author} {\bibfnamefont {Zi~Wen}\ \bibnamefont {Liu}}, \ and\ \bibinfo {author} {\bibfnamefont {Liang}\ \bibnamefont {Jiang}},\ }\bibfield  {title} {\enquote {\bibinfo {title} {{New perspectives on covariant quantum error correction}},}\ }\href {\doibase 10.22331/Q-2021-08-09-521} {\bibfield  {journal} {\bibinfo  {journal} {Quantum}\ }\textbf {\bibinfo {volume} {5}},\ \bibinfo {pages} {521} (\bibinfo {year} {2021})}\BibitemShut {NoStop}%
\bibitem [{\citenamefont {Yang}\ \emph {et~al.}(2022)\citenamefont {Yang}, \citenamefont {Mo}, \citenamefont {Renes}, \citenamefont {Chiribella},\ and\ \citenamefont {Woods}}]{Yang2022}%
  \BibitemOpen
  \bibfield  {author} {\bibinfo {author} {\bibfnamefont {Yuxiang}\ \bibnamefont {Yang}}, \bibinfo {author} {\bibfnamefont {Yin}\ \bibnamefont {Mo}}, \bibinfo {author} {\bibfnamefont {Joseph~M.}\ \bibnamefont {Renes}}, \bibinfo {author} {\bibfnamefont {Giulio}\ \bibnamefont {Chiribella}}, \ and\ \bibinfo {author} {\bibfnamefont {Mischa~P.}\ \bibnamefont {Woods}},\ }\bibfield  {title} {\enquote {\bibinfo {title} {{Optimal universal quantum error correction via bounded reference frames}},}\ }\href {\doibase 10.1103/PhysRevResearch.4.023107} {\bibfield  {journal} {\bibinfo  {journal} {Physical Review Research}\ }\textbf {\bibinfo {volume} {4}},\ \bibinfo {pages} {023107} (\bibinfo {year} {2022})}\BibitemShut {NoStop}%
\bibitem [{\citenamefont {Liu}\ and\ \citenamefont {Zhou}(2023)}]{Liu2023}%
  \BibitemOpen
  \bibfield  {author} {\bibinfo {author} {\bibfnamefont {Zi~Wen}\ \bibnamefont {Liu}}\ and\ \bibinfo {author} {\bibfnamefont {Sisi}\ \bibnamefont {Zhou}},\ }\bibfield  {title} {\enquote {\bibinfo {title} {{Approximate symmetries and quantum error correction}},}\ }\href {\doibase 10.1038/s41534-023-00788-4} {\bibfield  {journal} {\bibinfo  {journal} {npj Quantum Information}\ }\textbf {\bibinfo {volume} {9}},\ \bibinfo {pages} {119} (\bibinfo {year} {2023})}\BibitemShut {NoStop}%
\bibitem [{\citenamefont {Gour}\ and\ \citenamefont {Spekkens}(2008)}]{Gour2008}%
  \BibitemOpen
  \bibfield  {author} {\bibinfo {author} {\bibfnamefont {Gilad}\ \bibnamefont {Gour}}\ and\ \bibinfo {author} {\bibfnamefont {Robert~W.}\ \bibnamefont {Spekkens}},\ }\bibfield  {title} {\enquote {\bibinfo {title} {{The resource theory of quantum reference frames: Manipulations and monotones}},}\ }\href {\doibase 10.1088/1367-2630/10/3/033023} {\bibfield  {journal} {\bibinfo  {journal} {New Journal of Physics}\ }\textbf {\bibinfo {volume} {10}},\ \bibinfo {pages} {033023} (\bibinfo {year} {2008})}\BibitemShut {NoStop}%
\bibitem [{\citenamefont {Marvian}\ and\ \citenamefont {Spekkens}(2013)}]{Marvian2013}%
  \BibitemOpen
  \bibfield  {author} {\bibinfo {author} {\bibfnamefont {Iman}\ \bibnamefont {Marvian}}\ and\ \bibinfo {author} {\bibfnamefont {Robert~W.}\ \bibnamefont {Spekkens}},\ }\bibfield  {title} {\enquote {\bibinfo {title} {{The theory of manipulations of pure state asymmetry: I. Basic tools, equivalence classes and single copy transformations}},}\ }\href {\doibase 10.1088/1367-2630/15/3/033001} {\bibfield  {journal} {\bibinfo  {journal} {New Journal of Physics}\ }\textbf {\bibinfo {volume} {15}},\ \bibinfo {pages} {033001} (\bibinfo {year} {2013})}\BibitemShut {NoStop}%
\bibitem [{\citenamefont {Marvian}(2012)}]{Marvian2012a}%
  \BibitemOpen
  \bibfield  {author} {\bibinfo {author} {\bibfnamefont {Iman}\ \bibnamefont {Marvian}},\ }\emph {\bibinfo {title} {{Symmetry, Asymmetry and Quantum Information}}},\ \href@noop {} {Ph.D. thesis},\ \bibinfo  {school} {University of Waterloo} (\bibinfo {year} {2012})\BibitemShut {NoStop}%
\bibitem [{\citenamefont {Zhang}\ \emph {et~al.}(2017)\citenamefont {Zhang}, \citenamefont {Yadin}, \citenamefont {Hou}, \citenamefont {Cao}, \citenamefont {Liu}, \citenamefont {Huang}, \citenamefont {Maity}, \citenamefont {Vedral}, \citenamefont {Li}, \citenamefont {Guo},\ and\ \citenamefont {Girolami}}]{Zhang2017}%
  \BibitemOpen
  \bibfield  {author} {\bibinfo {author} {\bibfnamefont {Chao}\ \bibnamefont {Zhang}}, \bibinfo {author} {\bibfnamefont {Benjamin}\ \bibnamefont {Yadin}}, \bibinfo {author} {\bibfnamefont {Zhi~Bo}\ \bibnamefont {Hou}}, \bibinfo {author} {\bibfnamefont {Huan}\ \bibnamefont {Cao}}, \bibinfo {author} {\bibfnamefont {Bi~Heng}\ \bibnamefont {Liu}}, \bibinfo {author} {\bibfnamefont {Yun~Feng}\ \bibnamefont {Huang}}, \bibinfo {author} {\bibfnamefont {Reevu}\ \bibnamefont {Maity}}, \bibinfo {author} {\bibfnamefont {Vlatko}\ \bibnamefont {Vedral}}, \bibinfo {author} {\bibfnamefont {Chuan~Feng}\ \bibnamefont {Li}}, \bibinfo {author} {\bibfnamefont {Guang~Can}\ \bibnamefont {Guo}}, \ and\ \bibinfo {author} {\bibfnamefont {Davide}\ \bibnamefont {Girolami}},\ }\bibfield  {title} {\enquote {\bibinfo {title} {{Detecting metrologically useful asymmetry and entanglement by a few local measurements}},}\ }\href {\doibase 10.1103/PhysRevA.96.042327} {\bibfield  {journal} {\bibinfo  {journal} {Physical Review A}\ }\textbf
  {\bibinfo {volume} {96}},\ \bibinfo {pages} {042327} (\bibinfo {year} {2017})}\BibitemShut {NoStop}%
\bibitem [{\citenamefont {Takagi}(2019)}]{Takagi2019}%
  \BibitemOpen
  \bibfield  {author} {\bibinfo {author} {\bibfnamefont {Ryuji}\ \bibnamefont {Takagi}},\ }\bibfield  {title} {\enquote {\bibinfo {title} {{Skew informations from an operational view via resource theory of asymmetry}},}\ }\href {\doibase 10.1038/s41598-019-50279-w} {\bibfield  {journal} {\bibinfo  {journal} {Scientific Reports}\ }\textbf {\bibinfo {volume} {9}},\ \bibinfo {pages} {14562} (\bibinfo {year} {2019})}\BibitemShut {NoStop}%
\bibitem [{\citenamefont {Marvian}(2020)}]{Marvian2020}%
  \BibitemOpen
  \bibfield  {author} {\bibinfo {author} {\bibfnamefont {Iman}\ \bibnamefont {Marvian}},\ }\bibfield  {title} {\enquote {\bibinfo {title} {{Coherence distillation machines are impossible in quantum thermodynamics}},}\ }\href {\doibase 10.1038/s41467-019-13846-3} {\bibfield  {journal} {\bibinfo  {journal} {Nature Communications}\ }\textbf {\bibinfo {volume} {11}},\ \bibinfo {pages} {25} (\bibinfo {year} {2020})}\BibitemShut {NoStop}%
\bibitem [{\citenamefont {Marvian}(2022)}]{Marvian2022}%
  \BibitemOpen
  \bibfield  {author} {\bibinfo {author} {\bibfnamefont {Iman}\ \bibnamefont {Marvian}},\ }\bibfield  {title} {\enquote {\bibinfo {title} {{Operational Interpretation of Quantum Fisher Information in Quantum Thermodynamics}},}\ }\href {\doibase 10.1103/PhysRevLett.129.190502} {\bibfield  {journal} {\bibinfo  {journal} {Physical Review Letters}\ }\textbf {\bibinfo {volume} {129}},\ \bibinfo {pages} {190502} (\bibinfo {year} {2022})}\BibitemShut {NoStop}%
\bibitem [{\citenamefont {Yamaguchi}\ and\ \citenamefont {Tajima}(2023{\natexlab{a}})}]{Yamaguchi2023}%
  \BibitemOpen
  \bibfield  {author} {\bibinfo {author} {\bibfnamefont {Koji}\ \bibnamefont {Yamaguchi}}\ and\ \bibinfo {author} {\bibfnamefont {Hiroyasu}\ \bibnamefont {Tajima}},\ }\bibfield  {title} {\enquote {\bibinfo {title} {{Smooth Metric Adjusted Skew Information Rates}},}\ }\href {http://arxiv.org/abs/2211.12522} {\bibfield  {journal} {\bibinfo  {journal} {Quantum}\ }\textbf {\bibinfo {volume} {7}},\ \bibinfo {pages} {1012} (\bibinfo {year} {2023}{\natexlab{a}})}\BibitemShut {NoStop}%
\bibitem [{\citenamefont {Yamaguchi}\ and\ \citenamefont {Tajima}(2023{\natexlab{b}})}]{Yamaguchi2023a}%
  \BibitemOpen
  \bibfield  {author} {\bibinfo {author} {\bibfnamefont {Koji}\ \bibnamefont {Yamaguchi}}\ and\ \bibinfo {author} {\bibfnamefont {Hiroyasu}\ \bibnamefont {Tajima}},\ }\bibfield  {title} {\enquote {\bibinfo {title} {{Beyond i.i.d. in the Resource Theory of Asymmetry: An Information-Spectrum Approach for Quantum Fisher Information}},}\ }\href {\doibase https://doi.org/10.1103/PhysRevLett.131.200203} {\bibfield  {journal} {\bibinfo  {journal} {Physical Review Letters}\ }\textbf {\bibinfo {volume} {131}},\ \bibinfo {pages} {200203} (\bibinfo {year} {2023}{\natexlab{b}})}\BibitemShut {NoStop}%
\bibitem [{\citenamefont {Kudo}\ and\ \citenamefont {Tajima}(2023)}]{Kudo2023}%
  \BibitemOpen
  \bibfield  {author} {\bibinfo {author} {\bibfnamefont {Daigo}\ \bibnamefont {Kudo}}\ and\ \bibinfo {author} {\bibfnamefont {Hiroyasu}\ \bibnamefont {Tajima}},\ }\bibfield  {title} {\enquote {\bibinfo {title} {{Fisher information matrix as a resource measure in the resource theory of asymmetry with general connected-Lie-group symmetry}},}\ }\href {\doibase 10.1103/PhysRevA.107.062418} {\bibfield  {journal} {\bibinfo  {journal} {Physical Review A}\ }\textbf {\bibinfo {volume} {107}},\ \bibinfo {pages} {062418} (\bibinfo {year} {2023})}\BibitemShut {NoStop}%
\bibitem [{\citenamefont {Chitambar}\ and\ \citenamefont {Gour}(2019)}]{Chitambar2019}%
  \BibitemOpen
  \bibfield  {author} {\bibinfo {author} {\bibfnamefont {Eric}\ \bibnamefont {Chitambar}}\ and\ \bibinfo {author} {\bibfnamefont {Gilad}\ \bibnamefont {Gour}},\ }\bibfield  {title} {\enquote {\bibinfo {title} {{Quantum resource theories}},}\ }\href {\doibase 10.1103/RevModPhys.91.025001} {\bibfield  {journal} {\bibinfo  {journal} {Reviews of Modern Physics}\ }\textbf {\bibinfo {volume} {91}},\ \bibinfo {pages} {025001} (\bibinfo {year} {2019})}\BibitemShut {NoStop}%
\bibitem [{\citenamefont {Bennett}\ \emph {et~al.}(1996{\natexlab{a}})\citenamefont {Bennett}, \citenamefont {Bernstein}, \citenamefont {Popescu},\ and\ \citenamefont {Schumacher}}]{Bennett1996}%
  \BibitemOpen
  \bibfield  {author} {\bibinfo {author} {\bibfnamefont {Charles~H}\ \bibnamefont {Bennett}}, \bibinfo {author} {\bibfnamefont {Herbert~J}\ \bibnamefont {Bernstein}}, \bibinfo {author} {\bibfnamefont {Sandu}\ \bibnamefont {Popescu}}, \ and\ \bibinfo {author} {\bibfnamefont {Benjamin}\ \bibnamefont {Schumacher}},\ }\bibfield  {title} {\enquote {\bibinfo {title} {{Concentrating partial entanglement by local operations}},}\ }\href {\doibase https://doi.org/10.1103/PhysRevA.53.2046} {\bibfield  {journal} {\bibinfo  {journal} {Physical Review A}\ }\textbf {\bibinfo {volume} {53}},\ \bibinfo {pages} {2046} (\bibinfo {year} {1996}{\natexlab{a}})}\BibitemShut {NoStop}%
\bibitem [{\citenamefont {Bennett}\ \emph {et~al.}(1996{\natexlab{b}})\citenamefont {Bennett}, \citenamefont {Divincenzo}, \citenamefont {Smolin},\ and\ \citenamefont {Wootters}}]{Bennett1996a}%
  \BibitemOpen
  \bibfield  {author} {\bibinfo {author} {\bibfnamefont {Charles~H}\ \bibnamefont {Bennett}}, \bibinfo {author} {\bibfnamefont {David~P}\ \bibnamefont {Divincenzo}}, \bibinfo {author} {\bibfnamefont {John~A}\ \bibnamefont {Smolin}}, \ and\ \bibinfo {author} {\bibfnamefont {William~K}\ \bibnamefont {Wootters}},\ }\bibfield  {title} {\enquote {\bibinfo {title} {{Mixed-state entanglement and quantum error correction}},}\ }\href {\doibase https://doi.org/10.1103/PhysRevA.54.3824} {\bibfield  {journal} {\bibinfo  {journal} {Physical Review A}\ }\textbf {\bibinfo {volume} {54}},\ \bibinfo {pages} {3824} (\bibinfo {year} {1996}{\natexlab{b}})}\BibitemShut {NoStop}%
\bibitem [{\citenamefont {Vedral}\ \emph {et~al.}(1997)\citenamefont {Vedral}, \citenamefont {Plenio}, \citenamefont {Rippin},\ and\ \citenamefont {Knight}}]{Vedral1997}%
  \BibitemOpen
  \bibfield  {author} {\bibinfo {author} {\bibfnamefont {Vlatko}\ \bibnamefont {Vedral}}, \bibinfo {author} {\bibfnamefont {Martin~B.}\ \bibnamefont {Plenio}}, \bibinfo {author} {\bibfnamefont {M~A}\ \bibnamefont {Rippin}}, \ and\ \bibinfo {author} {\bibfnamefont {Peter~L.}\ \bibnamefont {Knight}},\ }\bibfield  {title} {\enquote {\bibinfo {title} {{Quantifying Entanglement}},}\ }\href {\doibase https://doi.org/10.1103/PhysRevLett.78.2275} {\bibfield  {journal} {\bibinfo  {journal} {Physical Review Letters}\ }\textbf {\bibinfo {volume} {78}},\ \bibinfo {pages} {2275} (\bibinfo {year} {1997})}\BibitemShut {NoStop}%
\bibitem [{\citenamefont {Brand{\~{a}}o}\ \emph {et~al.}(2013)\citenamefont {Brand{\~{a}}o}, \citenamefont {Horodecki}, \citenamefont {Oppenheim}, \citenamefont {Renes},\ and\ \citenamefont {Spekkens}}]{Brandao2013}%
  \BibitemOpen
  \bibfield  {author} {\bibinfo {author} {\bibfnamefont {Fernando~G.S.L.}\ \bibnamefont {Brand{\~{a}}o}}, \bibinfo {author} {\bibfnamefont {Michał}\ \bibnamefont {Horodecki}}, \bibinfo {author} {\bibfnamefont {Jonathan}\ \bibnamefont {Oppenheim}}, \bibinfo {author} {\bibfnamefont {Joseph~M.}\ \bibnamefont {Renes}}, \ and\ \bibinfo {author} {\bibfnamefont {Robert~W.}\ \bibnamefont {Spekkens}},\ }\bibfield  {title} {\enquote {\bibinfo {title} {{Resource theory of quantum states out of thermal equilibrium}},}\ }\href {\doibase 10.1103/PhysRevLett.111.250404} {\bibfield  {journal} {\bibinfo  {journal} {Physical Review Letters}\ }\textbf {\bibinfo {volume} {111}},\ \bibinfo {pages} {250404} (\bibinfo {year} {2013})}\BibitemShut {NoStop}%
\bibitem [{\citenamefont {Brand{\~{a}}o}\ \emph {et~al.}(2015)\citenamefont {Brand{\~{a}}o}, \citenamefont {Horodecki}, \citenamefont {Ng}, \citenamefont {Oppenheim},\ and\ \citenamefont {Wehner}}]{Brandao2015}%
  \BibitemOpen
  \bibfield  {author} {\bibinfo {author} {\bibfnamefont {Fernando}\ \bibnamefont {Brand{\~{a}}o}}, \bibinfo {author} {\bibfnamefont {Michał}\ \bibnamefont {Horodecki}}, \bibinfo {author} {\bibfnamefont {Nelly}\ \bibnamefont {Ng}}, \bibinfo {author} {\bibfnamefont {Jonathan}\ \bibnamefont {Oppenheim}}, \ and\ \bibinfo {author} {\bibfnamefont {Stephanie}\ \bibnamefont {Wehner}},\ }\bibfield  {title} {\enquote {\bibinfo {title} {{The second laws of quantum thermodynamics}},}\ }\href {\doibase 10.1073/pnas.1411728112} {\bibfield  {journal} {\bibinfo  {journal} {Proceedings of the National Academy of Sciences of the United States of America}\ }\textbf {\bibinfo {volume} {112}},\ \bibinfo {pages} {3275} (\bibinfo {year} {2015})}\BibitemShut {NoStop}%
\bibitem [{\citenamefont {Horodecki}\ and\ \citenamefont {Oppenheim}(2013)}]{Horodecki2013}%
  \BibitemOpen
  \bibfield  {author} {\bibinfo {author} {\bibfnamefont {Michał}\ \bibnamefont {Horodecki}}\ and\ \bibinfo {author} {\bibfnamefont {Jonathan}\ \bibnamefont {Oppenheim}},\ }\bibfield  {title} {\enquote {\bibinfo {title} {{Fundamental limitations for quantum and nanoscale thermodynamics}},}\ }\href {\doibase 10.1038/ncomms3059} {\bibfield  {journal} {\bibinfo  {journal} {Nature Communications}\ }\textbf {\bibinfo {volume} {4}},\ \bibinfo {pages} {2059} (\bibinfo {year} {2013})}\BibitemShut {NoStop}%
\bibitem [{\citenamefont {{\AA}berg}(2006)}]{Aberg2006}%
  \BibitemOpen
  \bibfield  {author} {\bibinfo {author} {\bibfnamefont {Johan}\ \bibnamefont {{\AA}berg}},\ }\bibfield  {title} {\enquote {\bibinfo {title} {{Quantifying Superposition}},}\ }\href {\doibase https://doi.org/10.48550/arXiv.quant-ph/0612146} {\bibfield  {journal} {\bibinfo  {journal} {arXiv}\ ,\ \bibinfo {pages} {quant--ph/0612146}} (\bibinfo {year} {2006})}\BibitemShut {NoStop}%
\bibitem [{\citenamefont {Baumgratz}\ \emph {et~al.}(2014)\citenamefont {Baumgratz}, \citenamefont {Cramer},\ and\ \citenamefont {Plenio}}]{Baumgratz2014}%
  \BibitemOpen
  \bibfield  {author} {\bibinfo {author} {\bibfnamefont {Tillmann}\ \bibnamefont {Baumgratz}}, \bibinfo {author} {\bibfnamefont {Marcus}\ \bibnamefont {Cramer}}, \ and\ \bibinfo {author} {\bibfnamefont {Martin~B.}\ \bibnamefont {Plenio}},\ }\bibfield  {title} {\enquote {\bibinfo {title} {{Quantifying coherence}},}\ }\href {\doibase https://doi.org/10.1103/PhysRevLett.113.140401} {\bibfield  {journal} {\bibinfo  {journal} {Physical Review Letters}\ }\textbf {\bibinfo {volume} {113}},\ \bibinfo {pages} {140401} (\bibinfo {year} {2014})}\BibitemShut {NoStop}%
\bibitem [{\citenamefont {Faist}\ \emph {et~al.}(2015)\citenamefont {Faist}, \citenamefont {Oppenheim},\ and\ \citenamefont {Renner}}]{Faist2015}%
  \BibitemOpen
  \bibfield  {author} {\bibinfo {author} {\bibfnamefont {Philippe}\ \bibnamefont {Faist}}, \bibinfo {author} {\bibfnamefont {Jonathan}\ \bibnamefont {Oppenheim}}, \ and\ \bibinfo {author} {\bibfnamefont {Renato}\ \bibnamefont {Renner}},\ }\bibfield  {title} {\enquote {\bibinfo {title} {{Gibbs-preserving maps outperform thermal operations in the quantum regime}},}\ }\href {\doibase https://doi.org/10.1088/1367-2630/17/4/043003} {\bibfield  {journal} {\bibinfo  {journal} {New Journal of Physics}\ }\textbf {\bibinfo {volume} {17}},\ \bibinfo {pages} {043003} (\bibinfo {year} {2015})}\BibitemShut {NoStop}%
\bibitem [{\citenamefont {Tajima}\ and\ \citenamefont {Takagi}(2025)}]{Tajima2025}%
  \BibitemOpen
  \bibfield  {author} {\bibinfo {author} {\bibfnamefont {Hiroyasu}\ \bibnamefont {Tajima}}\ and\ \bibinfo {author} {\bibfnamefont {Ryuji}\ \bibnamefont {Takagi}},\ }\bibfield  {title} {\enquote {\bibinfo {title} {{Gibbs-preserving operations requiring infinite amount of quantum coherence}},}\ }\href {\doibase 10.1103/PhysRevLett.134.170201} {\bibfield  {journal} {\bibinfo  {journal} {Physical Review Letters}\ }\textbf {\bibinfo {volume} {134}},\ \bibinfo {pages} {170201} (\bibinfo {year} {2025})}\BibitemShut {NoStop}%
\bibitem [{\citenamefont {Summer}\ \emph {et~al.}(2025)\citenamefont {Summer}, \citenamefont {Moroder}, \citenamefont {Bettmann}, \citenamefont {Turkeshi}, \citenamefont {Marvian},\ and\ \citenamefont {Goold}}]{Summer2025}%
  \BibitemOpen
  \bibfield  {author} {\bibinfo {author} {\bibfnamefont {Alessandro}\ \bibnamefont {Summer}}, \bibinfo {author} {\bibfnamefont {Mattia}\ \bibnamefont {Moroder}}, \bibinfo {author} {\bibfnamefont {Laetitia~P.}\ \bibnamefont {Bettmann}}, \bibinfo {author} {\bibfnamefont {Xhek}\ \bibnamefont {Turkeshi}}, \bibinfo {author} {\bibfnamefont {Iman}\ \bibnamefont {Marvian}}, \ and\ \bibinfo {author} {\bibfnamefont {John}\ \bibnamefont {Goold}},\ }\bibfield  {title} {\enquote {\bibinfo {title} {{A resource theoretical unification of Mpemba effects: classical and quantum}},}\ }\href {http://arxiv.org/abs/2507.16976} {\bibfield  {journal} {\bibinfo  {journal} {arXiv}\ ,\ \bibinfo {pages} {2507.16976}} (\bibinfo {year} {2025})}\BibitemShut {NoStop}%
\bibitem [{\citenamefont {Yamashika}\ \emph {et~al.}(2025)\citenamefont {Yamashika}, \citenamefont {Endo},\ and\ \citenamefont {Tajima}}]{Yamashika2025}%
  \BibitemOpen
  \bibfield  {author} {\bibinfo {author} {\bibfnamefont {Shion}\ \bibnamefont {Yamashika}}, \bibinfo {author} {\bibfnamefont {Shimpei}\ \bibnamefont {Endo}}, \ and\ \bibinfo {author} {\bibfnamefont {Hiroyasu}\ \bibnamefont {Tajima}},\ }\bibfield  {title} {\enquote {\bibinfo {title} {{Quantum Fisher Information as a Measure of Symmetry Breaking in Quantum Many-Body Systems}},}\ }\href {http://arxiv.org/abs/2509.07468} {\bibfield  {journal} {\bibinfo  {journal} {arXiv}\ ,\ \bibinfo {pages} {2509.07468}} (\bibinfo {year} {2025})}\BibitemShut {NoStop}%
\bibitem [{Note1()}]{Note1}%
  \BibitemOpen
  \bibinfo {note} {See Supplemental Material. URL to be added.}\BibitemShut {Stop}%
\bibitem [{\citenamefont {Luijk}\ \emph {et~al.}(2023)\citenamefont {Luijk}, \citenamefont {Werner},\ and\ \citenamefont {Wilming}}]{Luijk2023}%
  \BibitemOpen
  \bibfield  {author} {\bibinfo {author} {\bibfnamefont {Lauritz~van}\ \bibnamefont {Luijk}}, \bibinfo {author} {\bibfnamefont {Reinhard~F.}\ \bibnamefont {Werner}}, \ and\ \bibinfo {author} {\bibfnamefont {Henrik}\ \bibnamefont {Wilming}},\ }\bibfield  {title} {\enquote {\bibinfo {title} {{Covariant catalysis requires correlations and good quantum reference frames degrade little}},}\ }\href {\doibase 10.22331/q-2023-11-06-1166} {\bibfield  {journal} {\bibinfo  {journal} {Quantum}\ }\textbf {\bibinfo {volume} {7}},\ \bibinfo {pages} {1166} (\bibinfo {year} {2023})}\BibitemShut {NoStop}%
\bibitem [{\citenamefont {Matsumoto}(2015)}]{Matsumoto2010}%
  \BibitemOpen
  \bibfield  {author} {\bibinfo {author} {\bibfnamefont {Keiji}\ \bibnamefont {Matsumoto}},\ }\bibfield  {title} {\enquote {\bibinfo {title} {{A quantum version of randomization criterion}},}\ }\href {http://arxiv.org/abs/1012.2650} {\bibfield  {journal} {\bibinfo  {journal} {arXiv}\ ,\ \bibinfo {pages} {1012.2650}} (\bibinfo {year} {2015})}\BibitemShut {NoStop}%
\bibitem [{\citenamefont {Buscemi}(2012)}]{Buscemi2012}%
  \BibitemOpen
  \bibfield  {author} {\bibinfo {author} {\bibfnamefont {Francesco}\ \bibnamefont {Buscemi}},\ }\bibfield  {title} {\enquote {\bibinfo {title} {{Comparison of Quantum Statistical Models: Equivalent Conditions for Sufficiency}},}\ }\href {\doibase 10.1007/s00220-012-1421-3} {\bibfield  {journal} {\bibinfo  {journal} {Communications in Mathematical Physics}\ }\textbf {\bibinfo {volume} {310}},\ \bibinfo {pages} {625} (\bibinfo {year} {2012})}\BibitemShut {NoStop}%
\bibitem [{\citenamefont {Yamaguchi}\ \emph {et~al.}(2024)\citenamefont {Yamaguchi}, \citenamefont {Mitsuhashi}, \citenamefont {Shitara},\ and\ \citenamefont {Tajima}}]{Yamaguchi2024}%
  \BibitemOpen
  \bibfield  {author} {\bibinfo {author} {\bibfnamefont {Koji}\ \bibnamefont {Yamaguchi}}, \bibinfo {author} {\bibfnamefont {Yosuke}\ \bibnamefont {Mitsuhashi}}, \bibinfo {author} {\bibfnamefont {Tomohiro}\ \bibnamefont {Shitara}}, \ and\ \bibinfo {author} {\bibfnamefont {Hiroyasu}\ \bibnamefont {Tajima}},\ }\bibfield  {title} {\enquote {\bibinfo {title} {{Quantum geometric tensor determines the i.i.d. conversion rate in the resource theory of asymmetry for any compact Lie group}},}\ }\href {http://arxiv.org/abs/2411.04766} {\bibfield  {journal} {\bibinfo  {journal} {arXiv}\ ,\ \bibinfo {pages} {2411.04766}} (\bibinfo {year} {2024})}\BibitemShut {NoStop}%
\bibitem [{\citenamefont {Takagi}\ and\ \citenamefont {Shiraishi}(2022)}]{Takagi2022}%
  \BibitemOpen
  \bibfield  {author} {\bibinfo {author} {\bibfnamefont {Ryuji}\ \bibnamefont {Takagi}}\ and\ \bibinfo {author} {\bibfnamefont {Naoto}\ \bibnamefont {Shiraishi}},\ }\bibfield  {title} {\enquote {\bibinfo {title} {{Correlation in Catalysts Enables Arbitrary Manipulation of Quantum Coherence}},}\ }\href {\doibase 10.1103/PhysRevLett.128.240501} {\bibfield  {journal} {\bibinfo  {journal} {Physical Review Letters}\ }\textbf {\bibinfo {volume} {128}},\ \bibinfo {pages} {240501} (\bibinfo {year} {2022})}\BibitemShut {NoStop}%
\bibitem [{\citenamefont {Shiraishi}\ and\ \citenamefont {Takagi}(2024)}]{Shiraishi2024}%
  \BibitemOpen
  \bibfield  {author} {\bibinfo {author} {\bibfnamefont {Naoto}\ \bibnamefont {Shiraishi}}\ and\ \bibinfo {author} {\bibfnamefont {Ryuji}\ \bibnamefont {Takagi}},\ }\bibfield  {title} {\enquote {\bibinfo {title} {{Arbitrary Amplification of Quantum Coherence in Asymptotic and Catalytic Transformation}},}\ }\href {\doibase 10.1103/PhysRevLett.132.180202} {\bibfield  {journal} {\bibinfo  {journal} {Physical Review Letters}\ }\textbf {\bibinfo {volume} {132}},\ \bibinfo {pages} {180202} (\bibinfo {year} {2024})}\BibitemShut {NoStop}%
\bibitem [{\citenamefont {Grimsmo}\ \emph {et~al.}(2020)\citenamefont {Grimsmo}, \citenamefont {Combes},\ and\ \citenamefont {Baragiola}}]{Grimsmo2020}%
  \BibitemOpen
  \bibfield  {author} {\bibinfo {author} {\bibfnamefont {Arne~L.}\ \bibnamefont {Grimsmo}}, \bibinfo {author} {\bibfnamefont {Joshua}\ \bibnamefont {Combes}}, \ and\ \bibinfo {author} {\bibfnamefont {Ben~Q.}\ \bibnamefont {Baragiola}},\ }\bibfield  {title} {\enquote {\bibinfo {title} {{Quantum Computing with Rotation-Symmetric Bosonic Codes}},}\ }\href {\doibase 10.1103/PhysRevX.10.011058} {\bibfield  {journal} {\bibinfo  {journal} {Physical Review X}\ }\textbf {\bibinfo {volume} {10}},\ \bibinfo {pages} {011058} (\bibinfo {year} {2020})}\BibitemShut {NoStop}%
\bibitem [{\citenamefont {Endo}\ \emph {et~al.}(2025)\citenamefont {Endo}, \citenamefont {Suzuki}, \citenamefont {Tsubouchi}, \citenamefont {Asaoka}, \citenamefont {Yamamoto}, \citenamefont {Matsuzaki},\ and\ \citenamefont {Tokunaga}}]{Endo2025}%
  \BibitemOpen
  \bibfield  {author} {\bibinfo {author} {\bibfnamefont {Suguru}\ \bibnamefont {Endo}}, \bibinfo {author} {\bibfnamefont {Yasunari}\ \bibnamefont {Suzuki}}, \bibinfo {author} {\bibfnamefont {Kento}\ \bibnamefont {Tsubouchi}}, \bibinfo {author} {\bibfnamefont {Rui}\ \bibnamefont {Asaoka}}, \bibinfo {author} {\bibfnamefont {Kaoru}\ \bibnamefont {Yamamoto}}, \bibinfo {author} {\bibfnamefont {Yuichiro}\ \bibnamefont {Matsuzaki}}, \ and\ \bibinfo {author} {\bibfnamefont {Yuuki}\ \bibnamefont {Tokunaga}},\ }\bibfield  {title} {\enquote {\bibinfo {title} {{Quantum error mitigation for rotation-symmetric bosonic codes with symmetry expansion}},}\ }\href {\doibase 10.1103/PhysRevA.111.062402} {\bibfield  {journal} {\bibinfo  {journal} {Physical Review A}\ }\textbf {\bibinfo {volume} {111}},\ \bibinfo {pages} {062402} (\bibinfo {year} {2025})}\BibitemShut {NoStop}%
\bibitem [{\citenamefont {Gottesman}\ \emph {et~al.}(2001)\citenamefont {Gottesman}, \citenamefont {Kitaev},\ and\ \citenamefont {Preskill}}]{Gottesman2001}%
  \BibitemOpen
  \bibfield  {author} {\bibinfo {author} {\bibfnamefont {Daniel}\ \bibnamefont {Gottesman}}, \bibinfo {author} {\bibfnamefont {Alexei}\ \bibnamefont {Kitaev}}, \ and\ \bibinfo {author} {\bibfnamefont {John}\ \bibnamefont {Preskill}},\ }\bibfield  {title} {\enquote {\bibinfo {title} {{Encoding a qubit in an oscillator}},}\ }\href {\doibase 10.1103/PhysRevA.64.012310} {\bibfield  {journal} {\bibinfo  {journal} {Physical Review A}\ }\textbf {\bibinfo {volume} {64}},\ \bibinfo {pages} {123101} (\bibinfo {year} {2001})}\BibitemShut {NoStop}%
\bibitem [{\citenamefont {Grimsmo}\ and\ \citenamefont {Puri}(2021)}]{Grimsmo2021}%
  \BibitemOpen
  \bibfield  {author} {\bibinfo {author} {\bibfnamefont {Arne~L.}\ \bibnamefont {Grimsmo}}\ and\ \bibinfo {author} {\bibfnamefont {Shruti}\ \bibnamefont {Puri}},\ }\bibfield  {title} {\enquote {\bibinfo {title} {{Quantum Error Correction with the Gottesman-Kitaev-Preskill Code}},}\ }\href {\doibase 10.1103/PRXQuantum.2.020101} {\bibfield  {journal} {\bibinfo  {journal} {PRX Quantum}\ }\textbf {\bibinfo {volume} {2}},\ \bibinfo {pages} {020101} (\bibinfo {year} {2021})}\BibitemShut {NoStop}%
\bibitem [{\citenamefont {Lachance-Quirion}\ \emph {et~al.}(2024)\citenamefont {Lachance-Quirion}, \citenamefont {Lemonde}, \citenamefont {Simoneau}, \citenamefont {St-Jean}, \citenamefont {Lemieux}, \citenamefont {Turcotte}, \citenamefont {Wright}, \citenamefont {Lacroix}, \citenamefont {Fr{\'{e}}chette-Viens}, \citenamefont {Shillito}, \citenamefont {Hopfmueller}, \citenamefont {Tremblay}, \citenamefont {Frattini}, \citenamefont {Camirand~Lemyre},\ and\ \citenamefont {St-Jean}}]{Lachance-Quirion2024}%
  \BibitemOpen
  \bibfield  {author} {\bibinfo {author} {\bibfnamefont {Dany}\ \bibnamefont {Lachance-Quirion}}, \bibinfo {author} {\bibfnamefont {Marc~Antoine}\ \bibnamefont {Lemonde}}, \bibinfo {author} {\bibfnamefont {Jean~Olivier}\ \bibnamefont {Simoneau}}, \bibinfo {author} {\bibfnamefont {Lucas}\ \bibnamefont {St-Jean}}, \bibinfo {author} {\bibfnamefont {Pascal}\ \bibnamefont {Lemieux}}, \bibinfo {author} {\bibfnamefont {Sara}\ \bibnamefont {Turcotte}}, \bibinfo {author} {\bibfnamefont {Wyatt}\ \bibnamefont {Wright}}, \bibinfo {author} {\bibfnamefont {Amélie}\ \bibnamefont {Lacroix}}, \bibinfo {author} {\bibfnamefont {Joëlle}\ \bibnamefont {Fr{\'{e}}chette-Viens}}, \bibinfo {author} {\bibfnamefont {Ross}\ \bibnamefont {Shillito}}, \bibinfo {author} {\bibfnamefont {Florian}\ \bibnamefont {Hopfmueller}}, \bibinfo {author} {\bibfnamefont {Maxime}\ \bibnamefont {Tremblay}}, \bibinfo {author} {\bibfnamefont {Nicholas~E.}\ \bibnamefont {Frattini}}, \bibinfo {author} {\bibfnamefont {Julien}\ \bibnamefont
  {Camirand~Lemyre}}, \ and\ \bibinfo {author} {\bibfnamefont {Philippe}\ \bibnamefont {St-Jean}},\ }\bibfield  {title} {\enquote {\bibinfo {title} {{Autonomous Quantum Error Correction of Gottesman-Kitaev-Preskill States}},}\ }\href {\doibase 10.1103/PhysRevLett.132.150607} {\bibfield  {journal} {\bibinfo  {journal} {Physical Review Letters}\ }\textbf {\bibinfo {volume} {132}},\ \bibinfo {pages} {150607} (\bibinfo {year} {2024})}\BibitemShut {NoStop}%
\bibitem [{\citenamefont {Schlegel}\ \emph {et~al.}(2022)\citenamefont {Schlegel}, \citenamefont {Minganti},\ and\ \citenamefont {Savona}}]{Schlegel2022}%
  \BibitemOpen
  \bibfield  {author} {\bibinfo {author} {\bibfnamefont {David~S.}\ \bibnamefont {Schlegel}}, \bibinfo {author} {\bibfnamefont {Fabrizio}\ \bibnamefont {Minganti}}, \ and\ \bibinfo {author} {\bibfnamefont {Vincenzo}\ \bibnamefont {Savona}},\ }\bibfield  {title} {\enquote {\bibinfo {title} {{Quantum error correction using squeezed Schr{\"{o}}dinger cat states}},}\ }\href {\doibase 10.1103/PhysRevA.106.022431} {\bibfield  {journal} {\bibinfo  {journal} {Physical Review A}\ }\textbf {\bibinfo {volume} {106}},\ \bibinfo {pages} {022431} (\bibinfo {year} {2022})}\BibitemShut {NoStop}%
\bibitem [{\citenamefont {Hillmann}\ and\ \citenamefont {Quijandr{\'{i}}a}(2023)}]{Hillmann2023}%
  \BibitemOpen
  \bibfield  {author} {\bibinfo {author} {\bibfnamefont {Timo}\ \bibnamefont {Hillmann}}\ and\ \bibinfo {author} {\bibfnamefont {Fernando}\ \bibnamefont {Quijandr{\'{i}}a}},\ }\bibfield  {title} {\enquote {\bibinfo {title} {{Quantum error correction with dissipatively stabilized squeezed-cat qubits}},}\ }\href {\doibase 10.1103/PhysRevA.107.032423} {\bibfield  {journal} {\bibinfo  {journal} {Physical Review A}\ }\textbf {\bibinfo {volume} {107}},\ \bibinfo {pages} {032423} (\bibinfo {year} {2023})}\BibitemShut {NoStop}%
\bibitem [{\citenamefont {Endo}\ \emph {et~al.}(2024)\citenamefont {Endo}, \citenamefont {Anai}, \citenamefont {Matsuzaki}, \citenamefont {Tokunaga},\ and\ \citenamefont {Suzuki}}]{Endo2024}%
  \BibitemOpen
  \bibfield  {author} {\bibinfo {author} {\bibfnamefont {Suguru}\ \bibnamefont {Endo}}, \bibinfo {author} {\bibfnamefont {Keitaro}\ \bibnamefont {Anai}}, \bibinfo {author} {\bibfnamefont {Yuichiro}\ \bibnamefont {Matsuzaki}}, \bibinfo {author} {\bibfnamefont {Yuuki}\ \bibnamefont {Tokunaga}}, \ and\ \bibinfo {author} {\bibfnamefont {Yasunari}\ \bibnamefont {Suzuki}},\ }\bibfield  {title} {\enquote {\bibinfo {title} {{Projective squeezing for translation symmetric bosonic codes}},}\ }\href {http://arxiv.org/abs/2403.14218} {\bibfield  {journal} {\bibinfo  {journal} {arXiv}\ ,\ \bibinfo {pages} {2403.14218}} (\bibinfo {year} {2024})}\BibitemShut {NoStop}%
\end{thebibliography}%

\if0
%--------------------------------------------------------------------%
%--------------------------------------------------------------------%
%% END MATTER %%

\newpage
\onecolumngrid
\begin{center}
    {\bf\large End Matter}
\end{center}
\vspace{1cm}
\twocolumngrid

%--------------------------------------------------------------------%
%--------------------------------------------------------------------%

%--------------------------------------------------------------------%
%--------------------------------------------------------------------%
\fi
%% SUPPLEMENTAL MATERIAL %%

\clearpage

\onecolumngrid
\renewcommand{\thetheorem}{S\arabic{theorem}}
\renewcommand{\thelemma}{S\arabic{lemma}}
\renewcommand{\thecorollary}{S\arabic{corollary}}
\renewcommand{\thesection}{S\arabic{section}}
\setcounter{equation}{0}
\setcounter{figure}{0}
\setcounter{theorem}{0}
\setcounter{lemma}{0}
\setcounter{page}{1}
\setcounter{section}{0}
\counterwithout{equation}{section}
\renewcommand{\theequation}{S.\arabic{equation}}
\renewcommand{\thefigure}{S.\arabic{figure}}

%%%% With affiliation %%%%
\begin{center}
{\large \bf Supplemental Material: \protect\\ 
The i.i.d.~State Convertibility in the Resource Theory of Asymmetry for Finite Groups}\\
\vspace*{0.3cm}
Tomohiro Shitara$^{1,2}$, Yosuke Mitsuhashi$^{3}$ and Hiroyasu Tajima$^{4,5}$\\
\vspace*{0.1cm}

$^{1}$NTT Computer and Data Science Laboratories, NTT Inc., 3-9-11 Midori-cho, Musashino, Tokyo 180-8585, Japan

$^{2}$NTT Research Center for Theoretical Quantum Information, NTT Inc., 3-1 Morinosato Wakamiya, Atsugi, Kanagawa, 243-0198, Japan

$^{3}$Department of Basic Science, University of Tokyo, 3-8-1 Komaba, Meguro-ku, Tokyo, 153-8902, Japan

$^{4}$Department of Communication Engineering and Informatics, University of Electro-Communications, 1-5-1 Chofugaoka, Chofu, Tokyo, 182-8585, Japan

$^{5}$JST, PRESTO, 4-1-8 Honcho, Kawaguchi, Saitama, 332-0012, Japan
\end{center}

\if0
%%%%without affiliation
\begin{center}
{\large \bf Supplemental Material: \protect\\ 
The i.i.d. State Convertibility in the Resource Theory of Asymmetry for Finite Groups}\\
\vspace*{0.3cm}
Tomohiro Shitara, Yosuke Mitsuhashi and Hiroyasu Tajima\\
\end{center}
\fi

%-----------------------------------------------------------------------------------%

\tableofcontents

\bigskip

This Supplemental Material aims to provide the complete proofs of Theorems~\ref{thm:exact_conversion} and \ref{thm:approx_conversion} in the main text. 
Concretely, we determine the conversion rate via covariant operations in both exact and approximate cases. 
Moreover, we give a finite-size analysis, i.e., present a sufficient condition on the number of copies of the input state for achieving a given conversion rate (within a given error in the approximate case).

First, we introduce the mathematical notions and the notations for them that we use in this Supplemental Material. 
We denote the set of all natural numbers by $\mathbb{N}$, which excludes $0$. 
We denote the identity element of a group by $e$ and Napier's constant by $\ee$. 
We denote the set difference of $A$ and $B$ by $A-B$. 
When $H$ is a subgroup of a group $G$, we denote the set of left cosets of $H$ in $G$ by $G/H$. 
We denote the identity operator on a linear space $\calH$ by $I_\calH$. 
We denote the set of linear operators on $\calH$ by $\calL(\calH)$. 
We say that $U$ is a unitary representation of a group $G$ if $U$ is a map from $G$ to a unitary group and satisfies $U(gh)=U(g)U(h)$ for all $g, h\in G$. 
More generally, we say that $U$ is a projective unitary representation of a group $G$ if $U$ satisfies $U(g)U(h)=\omega(g, h)U(gh)$ for all $g, h\in G$ with some map $\omega$ from $G\times G$ to $\mathbb{C}$ satisfying $|\omega(g, h)|=1$. 
When we distinguish unitary representations from projective unitary representations, we express them as non-projective unitary representations.

Next, we rigorously present the two main concepts in our work, exact and approximate conversion rates via covariant operations. 
We say that a linear map $\calE$ is $(G, U, V)$-covariant if $\calE\circ\calU_g=\calV_g\circ\calE$ for all $g\in G$, where $\calU_g(\cdot):=U(g)\cdot U(g)^\dag$ and $\calV_g(\cdot):=V(g)\cdot V(g)^\dag$ for $g\in G$. 
For quantum states $\rho$ and $\sigma$, we denote $\rho\xrightarrow{(G, U, V)\text{-cov.}}\sigma$ if there exists a $(G, U, V)$-covariant map $\calE$ such that $\calE(\rho)=\sigma$. 
We say that rate $r\geq 0$ is exactly achievable in the conversion from $\rho$ to $\sigma$ if 
\begin{align}
    \exists N_0\in\mathbb{N} \textrm{ s.t. } \forall N\geq N_0,\ \rho^{\otimes N}\xrightarrow{\left(G, U^{\otimes N}, V^{\otimes \floor{rN}}\right)\text{-cov.}}\sigma^{\otimes \floor{rN}}, 
\end{align}
where $U^{\otimes N}$ is a map defined by $U^{\otimes N}(g):=U(g)^{\otimes N}$ for all $g\in G$. 
We define the exact conversion rate by the supremum of the exactly achievable conversion rates, which we denote by $R_{\textrm{ex}, G, U, V} (\rho\to\sigma)$, i.e., 
\begin{align}
    R_{\textrm{ex}, G, U, V}(\rho\rightarrow\sigma)
    :=\sup\left\{r\geq 0\ |\ \exists N_0\in\mathbb{N}\mbox{ s.t. }\forall N\geq N_0,\ \rho^{\otimes N}\xrightarrow{\left(G, U^{\otimes N}, V^{\otimes \floor{rN}}\right)\text{-cov.}}\sigma^{\otimes \floor{rN}}\right\}, 
\end{align}
which corresponds to $\Rex(\rho\to\sigma)$ in the main text.

Correspondingly, for $\epsilon\geq 0$, we denote $\rho\xrightarrow{(G, U, V)\text{-cov.}}_\epsilon\sigma$ if there exists a $(G, U, V)$-covariant map $\calE$ such that $T(\calE(\rho), \sigma)\leq\epsilon$, where $T$ is the trace distance, i.e., $T(\rho, \sigma):=\|\rho-\sigma\|_1/2$. 
We say that rate $r\geq 0$ is $\epsilon$-approximately achievable in the conversion from $\rho$ to $\sigma$ if 
\begin{align}
    \exists N_0\in\mathbb{N} \textrm{ s.t. } \forall N\geq N_0,\ \rho^{\otimes N}\xrightarrow{\left(G, U^{\otimes N}, V^{\otimes \floor{rN}}\right)\text{-cov.}}_\epsilon\sigma^{\otimes \floor{rN}}. 
\end{align}
We define the approximate conversion rate by the supremum of the values $r$ such that $r$ is $\epsilon$-approximately achievable for all $\epsilon>0$, which we denote by $R_{\textrm{ap}, G, U, V} (\rho\to\sigma)$, i.e., 
\begin{align}
    R_{\textrm{ap}, G, U, V}(\rho\rightarrow\sigma)
    :=\sup\left\{r\geq 0\ |\ \forall\epsilon>0,\ \exists N_0\in\mathbb{N}\mbox{ s.t. }\forall N\geq N_0,\ \rho^{\otimes N}\xrightarrow{\left(G, U^{\otimes N}, V^{\otimes \floor{rN}}\right)\text{-cov.}}_\epsilon\sigma^{\otimes \floor{rN}}\right\}, 
\end{align}
which corresponds to $\Rap(\rho\to\sigma)$ in the main text.

Finally, we explicitly describe our results. 
As for exact conversion, when $\psi$ and $\phi$ are two pure states, the exact conversion rate $R_{\textrm{ex}, G, U, V}(\psi\rightarrow\phi)$ is given by the minimum of the ratios $L_{\psi, G, U}(g)/L_{\phi, G, V}(g)$ over $g\in G$, where $L$ is defined for a general state $\rho$, a group $G$, and its projective unitary representation $U$ by 
\begin{align}
    L_{\rho, G, U}(g):=-\ln(|\mathrm{tr}(\rho U(g))|)\ \forall g\in G, \label{SMeq:asymmetry_measure}
\end{align}
which corresponds to $L_\rho(g)$ in the main text. 
We show this result in the case of non-projective unitary representations in Theorem~\ref{SMthm:exact_conversion} and generalize it to the case of projective unitary representations in Theorem~\ref{SMthm:exact_projective}. 
In both cases, we also present a sufficient condition on the number $N$ of the copies of the input state $\psi$ for $\rho^{\otimes N}\xrightarrow{\left(G, U^{\otimes N}, V^{\otimes \floor{rN}}\right)\text{-cov.}}\sigma^{\otimes \floor{rN}}$ for all $r\in (0, R_{\textrm{ex}, G, U, V}(\psi\rightarrow\phi))$.

As for approximate conversion, when $\psi$ is a pure state and $\sigma$ is a general state, the approximate conversion rate $R_{\textrm{ap}, G, U, V}(\psi\rightarrow\sigma)$ takes $\infty$ if $\sym_{G, U}(\psi)\subset\sym_{G, V}(\sigma)$, and it takes $0$ otherwise, where $\sym_{G, U}(\rho)$ is the symmetry subgroup of a general state $\rho$ defined for a group $G$ and its projective unitary representation $U$ by 
\begin{align}
    \sym_{G, U}(\rho):=\{g\in G\ |\ U(g) \rho U(g)^\dag=\rho\}, \label{SMeq:symmetry_subgroup}
\end{align}
which corresponds to $\sym_G(\rho)$ in the main text. 
We show this result in the case of non-projective unitary representations in Theorem~\ref{SMthm:approx_conversion} and generalize it to the case of projective unitary representations in Theorem~\ref{SMthm:approx_projective}. 
In both cases, when the symmetry subgroups satisfy $\sym_{G, U}(\psi)\subset\sym_{G, V}(\sigma)$, we also present a sufficient condition on the number $N$ of the copies of the input state $\psi$ for $\rho^{\otimes N}\xrightarrow{\left(G, U^{\otimes N}, V^{\otimes \floor{rN}}\right)\text{-cov.}}_\epsilon\sigma^{\otimes \floor{rN}}$ for all $r\in (0, \infty)$ and $\epsilon\in (0, 1)$.

%-------------------------------------------------------------------------------------------------------------------------------------%
\section{Exact i.i.d. conversion}
\label{SMsec:exact_conversion}

In this section, we give a proof of Theorem~\ref{thm:exact_conversion} in the main text.  
First, we prove it in Theorem~\ref{SMthm:exact_conversion}, which is restricted to the case of non-projective unitary representations. 
Next, we show its counterpart in the case of projective unitary representations in Theorem~\ref{SMthm:exact_projective}.

Before describing the detailed statement of Theorem~\ref{SMthm:exact_conversion}, we briefly review important results about single-shot exact covariant conversions in Ref.~\cite{Marvian2013}, which serve as useful tools in the proof of Theorem~\ref{SMthm:exact_conversion}. 
The characteristic function is introduced as a key ingredient for characterizing the convertibility, which is defined for a general state $\rho$, a group $G$, and its projective unitary representation $U$ by 
\begin{align}
    \chi_{\rho, G, U}(g):=\mathrm{tr}(\rho U(g)), 
\end{align}
which corresponds to $\chi_\rho(g)$ in the main text. 
This quantity satisfies the following basic properties.

\begin{lemma} \label{SMlem:characteristic_function}
    (Sec.~5.2 in Ref.~\cite{Marvian2013}.) 
    Let $\rho$, $\rho_1$, and $\rho_2$ be general (not necessarily pure) states on finite-dimensional Hilbert spaces $\calH$, $\calH_1$, and $\calH_2$, respectively, and $U$, $U_1$, and $U_2$ be projective unitary representations of a finite group $G$ on $\calH$, $\calH_1$, and $\calH_2$, respectively. 
    Then, $\chi$ satisfies the following three properties: 
    \begin{enumerate}
        \item $|\chi_{\rho, G, U}(g)|\leq 1$ for all $g\in G$. 
        If $|\chi_{\rho, G, U}(g)|=1$, then $g\in\sym_{G, U}(\rho)$. 
        When $\rho$ is a pure state, for any $g\in G$, $|\chi_{\rho, G, U}(g)|=1$ is equivalent to $g\in\sym_{G, U}(\rho)$. 
        \item If $|\chi_{\rho, G, U}(g)|=1$ for all $g\in G$, then $\rho$ is $(G, U)$-invariant. 
        When $\rho$ is a pure state, $|\chi_{\rho, G, U}(g)|=1$ for all $g\in G$ if and only if $\rho$ is $(G, U)$-invariant. 
        \item $\chi_{\rho_1\otimes \rho_2, G, U_1\otimes U_2}(g)
        =\chi_{\rho_1, G, U_1}(g)\chi_{\rho_2, G, U_2}(g)$ for all $g\in G$. 
    \end{enumerate}
\end{lemma}

These properties directly follow from the definition of the characteristic function. 
Although the latter half of item 2 is not explicitly written in Ref.~\cite{Marvian2013}, it directly follows from item 1.

\bigskip

The characteristic function gives the necessary and sufficient condition for the exact convertibility via covariant operations as follows.

\begin{lemma} \label{SMlem:single_shot}
    (Theorem~6 of Ref.~\cite{Marvian2013}.) 
    Let $U$ and $V$ be non-projective unitary representations of a finite group $G$ on a finite-dimensional Hilbert space $\calH_\mathrm{in}$ and $\calH_\mathrm{out}$, and $\psi$ and $\phi$ be pure states on $\calH_\mathrm{in}$ and $\calH_\mathrm{out}$. 
    Then, $\psi$ can be converted to $\phi$ via some $(G, U, V)$-covariant operation if and only if there exists some positive semi-definite function $f: G\to\mathbb{C}$ satisfying 
    \begin{align}
        \chi_{\psi, G, U}(g)=f(g)\chi_{\phi, G, V}(g) 
    \end{align}
    for all $g\in G$, where $f$ is called positive semi-definite if it satisfies 
    \begin{align}
        \sum_{g, g'\in G} \alpha(g)^* \alpha(g')f(g^{-1}g')\geq 0 \label{SMeq:positive_function_def}
    \end{align}
    for all $\alpha: G\to\mathbb{C}$. 
\end{lemma}

We note that this lemma does not hold for projective unitary representations.

\bigskip

By using the two lemmas above, we can prove the basic properties of $L$ defined by Eq.~\eqref{SMeq:asymmetry_measure}.

\begin{lemma} \label{SMlem:measure_L_properties}
    Let $\psi$, $\psi_1$, and $\psi_2$ be pure states on finite-dimensional Hilbert spaces $\calH$, $\calH_1$, and $\calH_2$, respectively, $U$, $U_1$, and $U_2$ be projective unitary representations of a finite group $G$ on $\calH$, $\calH_1$, and $\calH_2$, respectively. 
    Then, $L$ satisfies the following four properties: 
    \begin{enumerate}
        \item {\it positivity}: For any $g\in G$, $L_{\psi, G, U}(g)\geq 0$, where equality holds if and only if $g\in\sym_{G, U}(\psi)$. 
        \item {\it faithfulness}: $L_{\psi, G, U}(g)=0$ for all $g\in G$ if and only if $\psi$ is $(G, U)$-invariant. 
        \item {\it additivity}: For any $g\in G$, $L_{\psi_1\otimes \psi_2, G, U_1\otimes U_2}(g)=L_{\psi_1, G, U_1}(g)+L_{\psi_2, G, U_2}(g)$. 
        \item {\it monotonicity}: If $\psi_1\xrightarrow{(G, U_1, U_2)\text{-cov.}}\psi_2$, then $L_{\psi_1, G, U_1}(g)\geq L_{\psi_2, G, U_2}(g)$ for all $g\in G$. 
    \end{enumerate}
\end{lemma}

Since $L$ can be expressed as $L_{\rho, G, U}(g)=-\ln(|\chi_{\rho, G, U}(g)|)$ for all $g\in G$ with the characteristic function $\chi$, the first three properties directly follow from the three basic properties of the characteristic function shown in Lemma~\ref{SMlem:characteristic_function}. 
The fourth property follows from Lemma~\ref{SMlem:single_shot} when $U_1$ and $U_2$ are non-projective unitary representations, and we extend it to the case where $U_1$ or $U_2$ is a projective unitary representation, by using Lemma~\ref{SMlem:unitary_rep_construction}.

\begin{proof}
    Since the first three properties directly follow from Lemma~\ref{SMlem:characteristic_function}, we have only to show the monotonicity. 
    We suppose that $\psi_1$ can be converted to $\psi_2$ via some $(G, U_1, U_2)$-covariant operation. 
    First, we restrict ourselves to the case where $U_1$ and $U_2$ are non-projective unitary representations. 
    Then, by Lemma~\ref{SMlem:single_shot}, we can take some positive semi-definite function $f: G\to\mathbb{C}$ such that $\chi_{\psi_1, G, U_1}(h)=f(h)\chi_{\psi_2, G, U_2}(h)$ for all $h\in G$. 
    Since $U_1$ and $U_2$ are non-projective unitary representations, we have $\chi_{\psi_1, G, U_1}(e)=\chi_{\psi_2, G, U_2}(e)=1$, which implies $f(e)=1$. 
    We take arbitrary $g\in G$. 
    By considering the case where $\alpha(h)=0$ for all $h\in G-\{e, g\}$ in the definition of a positive semi-definite function (Eq.~\eqref{SMeq:positive_function_def}), we find that 
    $\begin{pmatrix}
        f(e) & f(g) \\
        f(g^{-1}) & f(e)
    \end{pmatrix}$
    is a positive semi-definite matrix. 
    By noting that a $2\times 2$ complex matrix $M$ is positive semi-definite if and only if $M$ is Hermitian and satisfies $\mathrm{tr}(M)\geq 0$ and $\mathrm{det}(M)\geq 0$, we get $|f(g)|\leq 1$. 
    Therefore, we have $|\chi_{\psi_1, G, U_1}(g)|=|f(g)||\chi_{\psi_2, G, U_2}(g)|\leq|\chi_{\psi_2, G, U_2}(g)|$, which implies that $L_{\psi_1, G, U_1}(g)=-\ln(|\chi_{\psi_1, G, U_1}(g)|)\geq -\ln(|\chi_{\psi_2, G, U_2}(g)|)=L_{\psi_2, G, U_2}(g)$.

    Next, we extend this monotonicity property to the case where $U_1$ or $U_2$ may be a projective unitary representation. 
    We define states $\Psi_1$ and $\Psi_2$ by $\Psi_1:=\psi_1^{\otimes d_1}$ and $\Psi_2:=\psi_2^{\otimes d_2}$, and projective unitary representations $U'_1$ and $U'_2$ by $U'_1(g):=U_1(g)^{\otimes d_1}/\mathrm{det}(U_1(g))$ and $U'_2(g):=U_2(g)^{\otimes d_2}/\mathrm{det}(U_2(g))$ for all $g\in G$, where $d_1$ and $d_2$ are the dimensions of $\calH_1$ and $\calH_2$, respectively. 
    Then, we have $\Psi_1\xrightarrow{(G, U'_1, U_1^{\otimes d_1})\text{-cov.}}\psi_1^{\otimes d_1}$ and $\psi_2^{\otimes d_2}\xrightarrow{(G, U_2^{\otimes d_2}, U'_2)\text{-cov.}}\Psi_2$, which are realized by identity maps. 
    If $\psi_1\xrightarrow{(G, U_1, U_2)\text{-cov.}}\psi_2$, by combining these three conversions, we have $\Psi_1^{\otimes d_2}\xrightarrow{(G, {U'_1}^{\otimes d_2}, {U'_2}^{\otimes d_1})\text{-cov.}}\Psi_2^{\otimes d_1}$. 
    By Lemma~\ref{SMlem:unitary_rep_construction}, $U'_1$ and $U'_2$ are non-projective unitary representations. 
    We can thus use the monotonicity result above for the case of non-projective unitary representations, and obtain for any $g\in G$, 
    \begin{align}
        L_{\Psi_1^{\otimes d_2}, G, {U'_1}^{\otimes d_2}}(g) 
        \geq L_{\Psi_2^{\otimes d_1}, G, {U'_2}^{\otimes d_1}}(g). \label{SMeq:SMlem:measure_L_properties1}
    \end{align}
    By the additivity of $L$, we have 
    \begin{align}
        L_{\Psi_1^{\otimes d_2}, G, {U'_1}^{\otimes d_2}}(g) 
        =d_2L_{\Psi_1, G, U'_1}(g). \label{SMeq:SMlem:measure_L_properties2}
    \end{align}
    By the definition of $\Psi_1$ and $U'_1$, we have 
    \begin{align}
        L_{\Psi_1, G, U'_1}(g) 
        =-\ln\left(\left|\frac{\bra{\psi_1}^{\otimes d_1}U_1^{\otimes d_1}\ket{\psi_1}^{\otimes d_1}}{\mathrm{det}(U_1)}\right|\right) 
        =d_1L_{\psi_1, G, U_1}(g). \label{SMeq:SMlem:measure_L_properties3}
    \end{align}
    By plugging Eq.~\eqref{SMeq:SMlem:measure_L_properties3} into Eq.~\eqref{SMeq:SMlem:measure_L_properties2}, we get $L_{\Psi_1^{\otimes d_2}, G, {U'_1}^{\otimes d_2}}(g)=d_1 d_2 L_{\psi_1, G, U_1}(g)$. 
    Similarly, we get $L_{\Psi_2^{\otimes d_1}, G, {U'_2}^{\otimes d_1}}(g)=d_1 d_2 L_{\psi_2, G, U_2}(g)$. 
    By plugging these relations into Eq.~\eqref{SMeq:SMlem:measure_L_properties1}, we get $L_{\psi_1, G, U_1}(g)\geq L_{\psi_2, G, U_2}(g)$. 
\end{proof}

%Theorem S1
\begin{theorem} \label{SMthm:exact_conversion}
    (Restatement of Theorem~\ref{thm:exact_conversion} in the main text.) 
    Let $\psi$ and $\phi$ be pure states on finite-dimensional Hilbert spaces $\calH_\mathrm{in}$ and $\calH_\mathrm{out}$, and $U$ and $V$ be non-projective unitary representations of a finite group $G$ on $\calH_\mathrm{in}$ and $\calH_\mathrm{out}$. 
    Then, the exact $(G, U, V)$-covariant conversion rate from $\psi$ to $\phi$ is given by 
    \begin{align}
        R_{\mathrm{ex}, G, U, V}(\psi\to\phi) = \min_{g\in G} \frac{L_{\psi, G, U}(g)}{L_{\phi, G, V}(g)}, \label{SMeq:SMthm:exact_conversion1}
    \end{align}
    where we define $c/0:=\infty$ when $0\leq c\leq \infty$ and $\infty/\infty:=\infty$ for the value of $L_{\psi, G, U}(g)/L_{\phi, G, V}(g)$ in extreme cases.
    Moreover, for any $r\in (0, R_{\mathrm{ex}, G, U, V}(\psi\to\phi))$, the conversion rate $r$ is exactly achievable if the number $N$ of copies of $\psi$ satisfies 
    \begin{align}
        \begin{cases}
            N\geq 0 & \textrm{if } \phi \textrm{ is } (G, V) \textrm{-invariant} \\
            N\geq \displaystyle\frac{\ln\left(\displaystyle\frac{|G|}{|\sym_{G, U}(\psi)|}\right)}{\left(1-\displaystyle\frac{r}{R_{\mathrm{ex}, U, V}(\psi\rightarrow\phi)}\right)\Delta_{G, U}(\psi)} & \textrm{otherwise},
        \end{cases} \label{SMeq:SMthm:exact_conversion2}
    \end{align}
    where $\Delta_{G, U}(\psi):=\min_{g\in G-\sym_{G, U}(\psi)} L_{\psi, G, U}(g)$. 
\end{theorem}

We make two remarks about this theorem. 
First, this theorem implies that $R_{\textrm{ex}, G, U, V}(\psi\to\phi)=0$ if $\sym_{G, U}(\psi) \not\subset \sym_{G, V}(\phi)$ or $S_{G, U}^\infty(\psi) \not\supset S_{G, V}^\infty(\phi)$, and that $R_{\textrm{ex}, G, U, V}(\psi\to\phi)=\infty$ if $\sym_{G, U}(\psi)\subset \sym_{G, V}(\phi)$, $S_{G, U}^\infty(\psi)\supset S_{G, V}^\infty(\phi)$, and $G=\sym_{G, U}(\phi) \cup S_{G, U}^\infty(\psi)$, where $S_{G, U}^\infty(\psi):=\{g\in G |L_{\psi, G, U}(g) =\infty \}$. 
Otherwise $R_{\textrm{ex}, G, U, V}(\psi\to\phi)$ takes some non-zero finite value. 
Second, when $R_{\mathrm{ex}, U, V}(\psi\rightarrow\phi)>0$ and $\phi$ is not $(G, V)$-invariant, we have $\sym_{G, U}(\psi) \subset \sym_{G, V}(\phi)\subsetneq G$, which allows the definition of $\Delta_{G, U}(\psi)$. 
By item 1 of Lemma~\ref{SMlem:measure_L_properties}, we can confirm that $\Delta_{G, U}(\psi)$ is the smallest nonzero value of $\{L_{\psi, G, U}(g)\}_{g\in G}$.

\begin{proof}
    First, we show that 
    \begin{align}
        R_{\textrm{ex}, G, U, V}(\psi\to\phi)\leq\min_{g\in G} \frac{L_{\psi, G, U}(g)}{L_{\phi, G, V}(g)} \label{SMeq:SMthm:exact_conversion3}. 
    \end{align}
    Since this inequality trivially holds when $R_{\textrm{ex}, G, U, V}(\psi\to\phi)=0$, we consider the case where $R_{\textrm{ex}, G, U, V}(\psi\to\phi)>0$ in the following. 
    We take arbitrary $r\in(0, R_{\textrm{ex}, G, U, V}(\psi\to\phi))$. 
    Then, there exists some $N_0\in\mathbb{N}$ such that $\psi^{\otimes N}\xrightarrow{(G, U^{\otimes N}, V^{\otimes \floor{rN}})\textrm{-cov.}}\phi^{\otimes \floor{rN}}$ for all $N\geq N_0$. 
    Then, by the additivity and monotonicity of $L$ shown in Lemma~\ref{SMlem:measure_L_properties}, we have $NL_{\psi, G, U}(g) \leq \floor{rN}L_{\phi, G, V}(g)$ for all $g\in G$, which implies that $\floor{rN}/N \leq L_{\psi, G, U}(g)/L_{\phi, G, V}(g)$. 
    Since this holds for all $g\in G$, we have $\floor{rN}/N \leq\min_{g\in G} L_{\psi, G, U}(g)/L_{\phi, G, V}(g)$. 
    By taking the limit of $N\to\infty$, we get $r\leq \min_{g\in G} L_{\psi, G, U}(g)/L_{\phi, G, V}(g)$. 
    Since the choice of $r\in(0, R_{\textrm{ex}, G, U, V}(\psi\to\phi))$ was arbitrary, we get Eq.~\eqref{SMeq:SMthm:exact_conversion3}.

    Next, we show that 
    \begin{align}
        R_{\textrm{ex}, G, U, V}(\psi\to\phi)\geq\min_{g\in G} \frac{L_{\psi, G, U}(g)}{L_{\phi, G, V}(g)}. \label{SMeq:SMthm:exact_conversion4}
    \end{align}
    Since this inequality trivially holds if the right-hand side is $0$, we consider the case where the right-hand side is not $0$ in the following, which implies $\sym_{G, U}(\psi)\subset \sym_{G, V}(\phi)$ and $S_{G, U}^\infty(\psi)\supset S_{G, V}^\infty(\phi)$. 
    For the proof of Eq.~\eqref{SMeq:SMthm:exact_conversion4}, we take arbitrary $r\in(0, \min_{g\in G} L_{\psi, G, U}(g)/L_{\phi, G, V}(g))$ and show that $\psi^{\otimes N}\xrightarrow{(G, U^{\otimes N}, V^{\otimes \floor{rN}})\textrm{-cov.}}\phi^{\otimes\floor{rN}}$ when $N\in\mathbb{N}$ satisfies Eq.~\eqref{SMeq:SMthm:exact_conversion2}. 
    When $\phi$ is $(G, V)$-invariant, this conversion is trivially possible, because we have only to discard the input state $\psi^{\otimes N}$ and bring the output state $\phi^{\otimes\floor{rN}}$. 
    In the following, we consider the case where $\phi$ is not $(G, V)$-invariant. 
    By Theorem~6 of Ref.~\cite{Marvian2013} (Lemma~\ref{SMlem:single_shot}), it is sufficient to find a positive semi-definite function $f: G\to\mathbb{C}$ satisfying $\chi_{\psi^{\otimes N}, G, U^{\otimes N}}(g)=f(g)\chi_{\phi^{\otimes \floor{rN}}, G, V^{\otimes \floor{rN}}}(g)$ for all $g\in G$. 
    We define a function $f: G\to\mathbb{C}$ by 
    \begin{align}
        f(g)=
        \begin{cases}
            \displaystyle\frac{\chi_{\psi, G, U}(g)^N}{\chi_{\phi, G, V}(g)^{\floor{rN}}} & \textrm{ if } \chi_{\phi, G, V}(g)\neq 0 \\
            0 & \textrm{ if } \chi_{\phi, G, V}(g)=0 
        \end{cases} \label{SMeq:SMthm:exact_conversion5}
    \end{align}
    for all $g\in G$. 
    This definition yields $\chi_{\psi^{\otimes N}, G, U^{\otimes N}}(g)=f(g)\chi_{\phi^{\otimes \floor{rN}}, G, V^{\otimes \floor{rN}}}(g)$ for all $g\in G$ such that $\chi_{\phi, G, V}\neq 0$. 
    By noting that we have $S_{G, U}^\infty(\psi)\supset S_{G, V}^\infty(\phi)$, the relation also holds for all $g\in G$ such that $\chi_{\phi, G, V}=0$.

    In the following, we prove that $f$ is positive semi-definite, i.e., 
    \begin{align}
        \sum_{g, g'\in G} \alpha(g)^* \alpha(g') f(g^{-1}g')\geq 0
        \label{SMeq:SMthm:exact_conversion6}
    \end{align}
    for all $\alpha: G\to\mathbb{C}$. 
    We take a set $\Xi$ of left coset representatives of $\sym_{G, U}(\psi)$ in $G$, i.e., $\Xi$ is the set of all representatives of the left coset $G/\sym_{G, U}(\psi)$. 
    Then, we can transform the left-hand side of Eq.~\eqref{SMeq:SMthm:exact_conversion6} as 
    \begin{align}
        \sum_{g, g'\in G} \alpha(g)^* \alpha(g') f(g^{-1}g') 
        =&\sum_{g, g'\in\Xi} \sum_{h, h'\in \sym_{G, U}(\psi)} \alpha(gh)^* \alpha(g'h') f((gh)^{-1}(g'h')) \nonumber\\
        =&\sum_{g, g'\in\Xi} \sum_{h, h'\in \sym_{G, U}(\psi)} \alpha(gh)^* \alpha(g'h') f(h^{-1}(g^{-1}g')h'). \label{SMeq:SMthm:exact_conversion7}
    \end{align}
    We note that 
    \begin{align}
        \chi_{\psi, G, U}(h^{-1} g h')
        =&\bra{\psi}U(h)^\dag U(g)U(h')\ket{\psi} \nonumber\\
        =&\bra{\psi}\chi_{\psi, G, U}(h)^* U(g) \chi_{\psi, G, U}(h')\ket{\psi} \nonumber\\
        =&\chi_{\psi, G, U}(h)^*\chi_{\psi, G, U}(g)\chi_{\psi, G, U}(h') \label{SMeq:SMthm:exact_conversion8}
    \end{align}
    for all $g\in G$ and $h, h'\in \sym_{G, U}(\psi)$. 
    Since we have $\sym_{G, U}(\psi)\subset \sym_{G, V}(\phi)$, we can similarly get 
    \begin{align}
        \chi_{\phi, G, V}(h^{-1} g h')
        =\chi_{\phi, G, V}(h)^*\chi_{\phi, G, V}(g)\chi_{\phi, G, V}(h') \label{SMeq:SMthm:exact_conversion9}
    \end{align}
    for all $g\in G$ and $h, h'\in \sym_{G, U}(\psi)$. 
    We note that Eq.~\eqref{SMeq:SMthm:exact_conversion9} implies that $h^{-1}gh'\in S_{G, V}^\infty(\phi)$ is equivalent to $g\in S_{G, V}^\infty(\phi)$. 
    Then, by the definition of $f$ (Eq.~\eqref{SMeq:SMthm:exact_conversion5}) and Eqs.~\eqref{SMeq:SMthm:exact_conversion8} and \eqref{SMeq:SMthm:exact_conversion9}, we get 
    \begin{align}
        f(h^{-1}gh')=f(h)^*f(g)f(h') 
    \end{align}
    for all $g\in G$ and $h, h'\in \sym_{G, U}(\psi)$. 
    By using this relation, we have 
    \begin{align}
        \sum_{g, g'\in G} \alpha(g)^* \alpha(g') f(g^{-1}g')
        =&\sum_{g, g'\in \Xi} \sum_{h, h'\in \sym_{G, U}(\psi)} \alpha(gh)^* \alpha(g'h') f(h)^*f(g^{-1}g')f(h') \nonumber\\
        =&\sum_{g, g'\in \Xi} \beta(g)^* \beta(g') f(g^{-1}g') \nonumber\\
        =&\sum_{g\in \Xi} |\beta(g)|^2 +\sum_{g, g'\in \Xi: g\neq g'} \beta(g)^* \beta(g') f(g^{-1}g'), \label{SMeq:SMthm:exact_conversion10}
    \end{align}
    where we define $\beta: G\to\mathbb{C}$ by 
    \begin{align}
        \beta(g):=\sum_{h\in \sym_{G, U}(\psi)} \alpha(gh) f(h) 
    \end{align}
    for all $g\in G$. 
    By the triangle inequality, we have 
    \begin{align}
        \sum_{g, g'\in G} \alpha(g)^* \alpha(g') f(g^{-1}g')
        \geq& \sum_{g\in \Xi} |\beta(g)|^2-\left|\sum_{g, g'\in \Xi: g\neq g'} \beta_{N}(g)^* \beta(g') f(g^{-1}g')\right| \nonumber\\
        \geq& \sum_{g\in \Xi} |\beta(g)|^2-\sum_{g, g'\in \Xi: g\neq g'} |\beta(g)| |\beta(g')| |f(g^{-1}g')|. \label{SMeq:SMthm:exact_conversion11}
    \end{align}
    By Eq.~\eqref{SMeq:SMthm:exact_conversion3}, for any $g\in G$, we have 
    \begin{align}
        \frac{L_{\psi, G, U}(g)}{L_{\phi, G, V}(g)}\geq R_{\textrm{ex}, G, U, V}(\psi\to\phi), 
    \end{align}
    which implies that 
    \begin{align}
        |\chi_{\psi, G, U}(g)|\leq |\chi_{\phi, G, V}(g)|^{R_{\textrm{ex}, G, U, V}(\psi\to\phi)}. 
    \end{align}
    We can thus give an upper bound of $|f(g)|$ for $g\in G-\sym_{G, U}(\psi)$ as follows: 
    \begin{align}
        |f(g)|
        \leq |\chi_{\psi, U}(g)|^{N-\frac{\floor{rN}}{R_{\textrm{ex}, G, U, V}(\psi\to\phi)}} 
        \leq |\chi_{\psi, U}(g)|^{\left(1-\frac{r}{R_{\textrm{ex}, G, U, V}(\psi\to\phi)}\right)N} 
        \leq \ee^{-\left(1-\frac{r}{R_{\textrm{ex}, G, U, V}(\psi\to\phi)}\right)\Delta_{G, U}(\psi) N}, 
    \end{align}
    where we used the definition of $\Delta_{G, U}(\psi)$ in the last inequality. 
    If $N$ satisfies Eq.~\eqref{SMeq:SMthm:exact_conversion2}, we can further upper bound $|f(g)|$ as 
    \begin{align}
        |f(g)| 
        \leq \frac{|\sym_{G, U}(\psi)|}{|G|}
        = \frac{1}{|\Xi|}. 
    \end{align}
    Then, we have 
    \begin{align}
        \sum_{g, g'\in\Xi: g\neq g'} |\beta(g)| |\beta(g')| |f(g^{-1}g')| 
        \leq& \frac{1}{|\Xi|} \sum_{g, g'\in\Xi: g\neq g'} |\beta(g)| |\beta(g')| \nonumber\\
        \leq& \frac{1}{|\Xi|} \sum_{g, g'\in\Xi: g\neq g'} \frac{|\beta(g)|^2+|\beta(g')|^2}{2} \nonumber\\
        =&\frac{|\Xi|-1}{|\Xi|} \sum_{g\in \Xi} |\beta(g)|^2, \label{SMeq:SMthm:exact_conversion12}
    \end{align}
    where we used the relation between the arithmetic and geometric means. 
    By plugging Eqs.~\eqref{SMeq:SMthm:exact_conversion12} into Eq.~\eqref{SMeq:SMthm:exact_conversion11}, we get 
    \begin{align}
        \sum_{g, g'\in G} \alpha(g)^* \alpha(g') f(g^{-1}g') 
        \geq \frac{1}{|\Xi|} \sum_{g\in \Xi} |\beta(g)|^2
        \geq 0. 
    \end{align}
    Therefore, we have proven that $\psi^{\otimes N}\xrightarrow{(G, U^{\otimes N}, V^{\otimes \floor{rN}})\textrm{-cov.}}\phi^{\otimes\floor{rN}}$ when $N$ satisfies Eq.~\eqref{SMeq:SMthm:exact_conversion2}, which implies that $r\leq R_{\textrm{ex}, G, U, V}(\psi\to\phi)$. 
    Since the choice for $r\in(0, \min_{g\in G} L_{\psi, G, U}(g)/L_{\phi, G, V}(g))$ was arbitrary, we get Eq.~\eqref{SMeq:SMthm:exact_conversion4}. 
\end{proof}

We show the counterpart of Theorem~\ref{SMthm:exact_conversion} in the case of projective unitary representations, by using Theorem~\ref{SMthm:exact_conversion} and Lemmas~\ref{SMlem:unitary_rep_construction} and \ref{SMlem:multiple_copy}.

%Theorem S2
\begin{theorem} \label{SMthm:exact_projective}
    (Counterpart of Theorem~\ref{thm:exact_conversion} in the main text for projective unitary representations.) 
    Let $\psi$ and $\phi$ be pure states on finite-dimensional Hilbert spaces $\calH_\mathrm{in}$ and $\calH_\mathrm{out}$, and $U$ and $V$ be projective unitary representations of a finite group $G$ on $\calH_\mathrm{in}$ and $\calH_\mathrm{out}$. 
    Then, the exact conversion rate from $\psi$ to $\phi$ via $(G, U, V)$-covariant operations is given by 
    \begin{align}
        R_{\mathrm{ex}, G, U, V}(\psi\rightarrow\phi) = \min_{g\in G} \frac{L_{\psi, G, U}(g)}{L_{\phi, G, V}(g)}, \label{SMeq:SMthm:exact_projective1}
    \end{align}
    where we define $c/0:=\infty$ when $0\leq c\leq \infty$ and $\infty/\infty:=\infty$ for the value of $L_{\psi, G, U}(g)/L_{\phi, G, V}(g)$ in extreme cases.
    Moreover, for any $r\in (0, R_{\mathrm{ex}, G, U, V}(\psi\to\phi))$, the conversion rate $r$ is exactly achievable if the number $N$ of copies of $\psi$ satisfies 
    \begin{align}
        \begin{cases}
            N\geq 0 & \textrm{if } \phi \textrm{ is } (G, V) \textrm{-invariant}\\
            N\geq \displaystyle\frac{\ln\left(\displaystyle\frac{|G|}{|\sym_{G, U}(\psi)|}\right)+\left(d_\mathrm{in}+\displaystyle\frac{d_\mathrm{out}}{R_{\mathrm{ex}, G, U, V}(\psi\rightarrow\phi)}\right)\Delta_{G, U}(\psi)}{\left(1-\displaystyle\frac{r}{R_{\mathrm{ex}, G, U, V}(\psi\rightarrow\phi)}\right)\Delta_{G, U}(\psi)} & \textrm{otherwise}, 
        \end{cases} \label{SMeq:SMthm:exact_projective2}
    \end{align}
    where $d_\mathrm{in}$ and $d_\mathrm{out}$ are the dimensions of $\calH_\mathrm{in}$ and $\calH_\mathrm{out}$, respectively, and $\Delta_{G, U}(\psi):=\min_{g\in G-\sym_{G, U}(\psi)} L_{\psi, G, U}(g)$. 
\end{theorem}

We note that $\Delta_{G, U}(\psi)$ can be defined and satisfies $\Delta_{G, U}(\psi)>0$ when $R_{\mathrm{ex}, G, U, V}(\psi\rightarrow\phi)>0$ and $\phi$ is not $(G, V)$-invariant, as we mentioned just below Theorem~\ref{SMthm:exact_conversion}.

\begin{proof}
    First, we show Eq.~\eqref{SMeq:SMthm:exact_projective1}. 
    We define two pure states $\Psi$ and $\Phi$ on $\mathcal{H}_\mathrm{in}^{\otimes d_\mathrm{in}}$ and $\mathcal{H}_\mathrm{out}^{\otimes d_\mathrm{out}}$ by 
    \begin{align}
        &\Psi:=\psi^{\otimes d_\mathrm{in}}, \\
        &\Phi:=\phi^{\otimes d_\mathrm{out}}. 
    \end{align}
    We also define projective unitary representations $U'$ and $V'$ of $G$ on $\mathcal{H}_\mathrm{in}^{\otimes d_\mathrm{in}}$ and $\mathcal{H}_\mathrm{out}^{\otimes d_\mathrm{out}}$ by 
    \begin{align}
        &U'(g):=\frac{U(g)^{\otimes d_\mathrm{in}}}{\mathrm{det}(U(g))}, \\
        &V'(g):=\frac{V(g)^{\otimes d_\mathrm{out}}}{\mathrm{det}(V(g))} 
    \end{align}
    for all $g\in G$. 
    By Lemma~\ref{SMlem:unitary_rep_construction}, $U'$ and $V'$ are non-projective unitary representations of $G$ on $\mathcal{H}_\mathrm{in}^{\otimes d_\mathrm{in}}$ and $\mathcal{H}_\mathrm{out}^{\otimes d_\mathrm{out}}$. 
    We can therefore use the result of Theorem~\ref{SMthm:exact_conversion} and obtain 
    \begin{align}
        R_{\textrm{ex}, G, U', V'}(\Psi\to\Phi)
        =\min_{g\in G} \frac{L_{\Psi, G, U'}(g)  }{L_{\Phi, G, V'}(g)}. \label{SMeq:SMthm:exact_projective3}
    \end{align}
    As for the left-hand side of Eq.~\eqref{SMeq:SMthm:exact_projective3}, since $(G, U', V')$-covariance is equivalent to $(G, U^{\otimes d_\mathrm{in}}, V^{\otimes d_\mathrm{out}})$-covariance, we have 
    \begin{align}
        R_{\textrm{ex}, G, U', V'}(\Psi\to\Phi) 
        =R_{\textrm{ex}, G, U^{\otimes d_\mathrm{in}}, V^{\otimes d_\mathrm{out}}}(\Psi\to\Phi). \label{SMeq:SMthm:exact_projective4}
    \end{align}
    By Lemma~\ref{SMlem:multiple_copy}, we have 
    \begin{align}
        R_{\textrm{ex}, G, U^{\otimes d_\mathrm{in}}, V^{\otimes d_\mathrm{out}}}(\Psi\to\Phi) 
        =\frac{d_\mathrm{in}}{d_\mathrm{out}} R_{\textrm{ex}, G, U, V}(\psi\to\phi). \label{SMeq:SMthm:exact_projective5}
    \end{align}
    By Eqs.~\eqref{SMeq:SMthm:exact_projective4} and \eqref{SMeq:SMthm:exact_projective5}, we get 
    \begin{align}
        R_{\textrm{ex}, G, U', V'}(\Psi\to\Phi) 
        =\frac{d_\mathrm{in}}{d_\mathrm{out}} R_{\textrm{ex}, G, U, V}(\psi\to\phi). \label{SMeq:SMthm:exact_projective6}
    \end{align}
    As for the right-hand side of Eq.~\eqref{SMeq:SMthm:exact_projective3}, for any $g\in G$, we have 
    \begin{align}
        L_{\Psi, G, U'}(g) 
        =&-\ln\left(\left|\bra{\Psi}U'(g)\ket{\Psi}\right|\right) \nonumber\\
        =&-\ln\left(\frac{\left|\bra{\psi}^{\otimes d_\mathrm{in}}U(g)^{\otimes d_\mathrm{in}}\ket{\psi}^{\otimes d_\mathrm{in}}\right|}{|\mathrm{det}(U(g))|}\right) \nonumber\\
        =&-\ln\left(|\bra{\psi}U(g)\ket{\psi}|^{d_\mathrm{in}}\right) \nonumber\\
        =&d_\mathrm{in}L_{\psi, G, U}(g). \label{SMeq:SMthm:exact_projective7}
    \end{align}
    Similarly, we have 
    \begin{align}
        L_{\phi^{\otimes d_\mathrm{out}}, G, V'}(g)=d_\mathrm{out}L_{\phi, G, V}(g) \label{SMeq:SMthm:exact_projective8}
    \end{align}
    for all $g\in G$. 
    By Eqs.~\eqref{SMeq:SMthm:exact_projective7} and \eqref{SMeq:SMthm:exact_projective8}, we get 
    \begin{align}
        \min_{g\in G} \frac{L_{\psi^{\otimes d_\mathrm{in}}, G, U'}(g)  }{L_{\phi^{\otimes d_\mathrm{out}}, G, V'}(g)} 
        =\frac{d_\mathrm{in}}{d_\mathrm{out}} \min_{g\in G} \frac{L_{\psi, G, U}(g)}{L_{\phi, G, V}(g)}. \label{SMeq:SMthm:exact_projective9}
    \end{align}
    By plugging Eqs.~\eqref{SMeq:SMthm:exact_projective6} and \eqref{SMeq:SMthm:exact_projective9} into Eq.~\eqref{SMeq:SMthm:exact_projective3}, we get Eq.~\eqref{SMeq:SMthm:exact_projective1}.

    Next, we show that Eq.~\eqref{SMeq:SMthm:exact_projective2} gives a sufficient condition on the number of copies of $\psi$ for achieving the exact conversion rate $r$. 
    Concretely, we take arbitrary $r\in (0, R_{\mathrm{ex}, G, U, V}(\psi\rightarrow\phi))$ and $N\in\mathbb{N}$ satisfying Eq.~\eqref{SMeq:SMthm:exact_projective2}, and show $\psi^{\otimes N}$ can be converted to $\phi^{\floor{rN}}$ via some $(G, U^{\otimes N}, V^{\otimes \floor{rN}})$-covariant map. 
    When $\phi$ is $(G, V)$-invariant, the conversion is trivially realized just by discarding the input state and bringing the output state. 
    In the following, we consider the case where $\phi$ is not $(G, V)$-invariant. 
    Equation~\eqref{SMeq:SMthm:exact_projective2} implies that $N>d_\mathrm{in}$ and 
    \begin{align}
        \frac{r\displaystyle\frac{N}{d_\mathrm{out}}+1}{\displaystyle\frac{N}{d_\mathrm{in}}-1} 
        \leq \left[1-\frac{\ln\left(\displaystyle\frac{|G|}{|\sym_{G, U}(\psi)|}\right)}{\left(\displaystyle\frac{N}{d_\mathrm{in}}-1\right)d_\mathrm{in}\Delta_{G, U}(\psi)}\right]\frac{d_\mathrm{in}}{d_\mathrm{out}}R_{\mathrm{ex}, G, U, V}(\psi\to\phi). \label{SMeq:SMthm:exact_projective10}
    \end{align}
    By Eq.~\eqref{SMeq:SMthm:exact_projective7}, we have 
    \begin{align}
        \Delta_{G, U'}(\Psi)=d_\mathrm{in}\Delta_{G, U}(\psi). \label{SMeq:SMthm:exact_projective11}
    \end{align}
    By plugging Eqs.~\eqref{SMeq:SMthm:exact_projective6} and \eqref{SMeq:SMthm:exact_projective11} into Eq.~\eqref{SMeq:SMthm:exact_projective10}, we get 
    \begin{align}
        \frac{r\displaystyle\frac{N}{d_\mathrm{out}}+1}{\displaystyle\frac{N}{d_\mathrm{in}}-1} 
        \leq \left[1-\frac{\ln\left(\displaystyle\frac{|G|}{|\sym_{G, U}(\psi)|}\right)}{\left(\displaystyle\frac{N}{d_\mathrm{in}}-1\right)\Delta_{G, U'(\Psi)}}\right]R_{\textrm{ex}, G, U', V'}(\Psi\to \Phi). \label{SMeq:SMthm:exact_projective12}
    \end{align}
    We can thus take $r'\in (0, \infty)$ such that 
    \begin{align}
        &r'\leq \left[1-\frac{\displaystyle\ln\left(\frac{|G|}{|\sym_{G, U}(\psi)|}\right)}{\left(\displaystyle\frac{N}{d_\mathrm{in}}-1\right)\Delta_{G, U'}(\Psi)}\right]R_{\textrm{ex}, G, U', V'}(\Psi\to\Phi), \\
        &r'\geq \frac{r\displaystyle\frac{N}{d_\mathrm{out}}+1}{\displaystyle\frac{N}{d_\mathrm{in}}-1}, 
    \end{align}
    which imply that 
    \begin{align}
        &\left\lfloor\frac{N}{d_\mathrm{in}}\right\rfloor
        \geq\frac{N}{d_\mathrm{in}}-1 \geq\frac{\displaystyle\ln\left(\frac{|G|}{|\sym_{G, U}(\psi)|}\right)}{\left(1-\displaystyle\frac{r'}{R_{\mathrm{ex}, G, U', V'}(\Psi\to\Phi)}\right)\Delta_{G, U'}(\Psi)}, \label{SMeq:SMthm:exact_projective13}\\
        &d_\mathrm{out}\left\lfloor r'\left\lfloor \frac{N}{d_\mathrm{in}}\right\rfloor\right\rfloor
        \geq d_\mathrm{out}\left[r'\left(\frac{N}{d_\mathrm{in}}-1\right)-1\right]\geq rN, \label{SMeq:SMthm:exact_projective14}
    \end{align}
    respectively. 
    Therefore, $\phi^{\otimes N}$ can be converted to $\psi^{\otimes \floor{rN}}$ via some $(G, U^{\otimes N}, V^{\otimes \floor{rN}})$-covariant operation consisting of the following four steps: 
    \begin{align}
        \psi^{\otimes N}
        &\xrightarrow{\left(G, U^{\otimes N}, {U'}^{\otimes \left\lfloor\frac{N}{d_\mathrm{in}}\right\rfloor}\right)\text{-cov.}} \Psi^{\otimes \left\lfloor\frac{N}{d_\mathrm{in}}\right\rfloor} \nonumber\\
        &\xrightarrow{\left(G, {U'}^{\otimes \left\lfloor\frac{N}{d_\mathrm{in}}\right\rfloor}, {V'}^{\otimes \left\lfloor r'\left\lfloor \frac{N}{d_\mathrm{in}}\right\rfloor\right\rfloor}\right)\text{-cov.}} \Phi^{\otimes \left\lfloor r'\left\lfloor \frac{N}{d_\mathrm{in}}\right\rfloor\right\rfloor} \nonumber\\
        &\xrightarrow{\left(G, {V'}^{\otimes \left\lfloor r'\left\lfloor \frac{N}{d_\mathrm{in}}\right\rfloor\right\rfloor}, V^{\otimes d_\mathrm{out}\left\lfloor r'\left\lfloor \frac{N}{d_\mathrm{in}}\right\rfloor \right\rfloor}\right)\text{-cov.}} \phi^{\otimes d_\mathrm{out}\left\lfloor r'\left\lfloor \frac{N}{d_\mathrm{in}}\right\rfloor \right\rfloor} \nonumber\\
        &\xrightarrow{\left(G, V^{\otimes d_\mathrm{out}\left\lfloor r'\left\lfloor \frac{N}{d_\mathrm{in}}\right\rfloor \right\rfloor}, V^{\otimes \floor{rN}}\right)\text{-cov.}} \phi^{\otimes \floor{rN}}, 
    \end{align}
    where the first conversion is always possible, and the possibility of the second one is guaranteed by Theorem~\ref{SMthm:exact_conversion} and Eq.~\eqref{SMeq:SMthm:exact_projective13}, the third one is realized by the identity map, and the last one is realized by partially discarding the input state, which is guaranteed by Eq.~\eqref{SMeq:SMthm:exact_projective14}. 
\end{proof}

%-----------------------------------------------------------------------------------%
\section{Approximate i.i.d. conversion} \label{SMsec:approx_conversion}

In this section, we give a proof of Theorem~\ref{thm:approx_conversion} in the main text.  
First, we prove it in Theorem~\ref{SMthm:approx_conversion}, which is restricted to the case of non-projective unitary representations. 
Next, we show its counterpart in the case of projective unitary representations in Theorem~\ref{SMthm:approx_projective}. 
We note that the output states may not be pure, in contrast to Theorem~\ref{thm:exact_conversion}.

%Theorem S3
\begin{theorem} \label{SMthm:approx_conversion}
    (Restatement of Theorem~\ref{thm:approx_conversion} in the main text.)
    Let $\psi$ be a pure state on a finite-dimensional Hilbert space $\calH_\mathrm{in}$ and $\sigma$ be a general (not necessarily pure) state on a finite-dimensional Hilbert space $\calH_\mathrm{out}$, and $U$ and $V$ be (non-projective) unitary representations of a finite group $G$ on $\calH_\mathrm{in}$ and $\calH_\mathrm{out}$. 
    Then, the approximate $(G, U, V)$-covariant conversion rate from $\psi$ to $\sigma$ is given by 
    \begin{align}
        R_{\mathrm{ap}, G, U, V}(\psi\to\sigma)
        =\begin{cases}
            \infty & \textrm{if }\sym_{G, U}(\psi)\subset\sym_{G, V}(\sigma)\\
            0 & \textrm{otherwise}. \label{SMeq:SMthm:approx_conversion1}
        \end{cases}
    \end{align}
    Moreover, when $\sym_{G, U}(\psi)\subset\sym_{G, V}(\sigma)$, for any $r\in (0, \infty)$ and $\epsilon\in (0, 1)$, conversion rate $r$ is $\epsilon$-approximately achievable if the initial number $N$ of copies of state $\psi$ satisfies 
    \begin{align}
        \begin{cases}
            N\geq 0 & \textrm{if } \sigma \textrm{ is } (G, V) \textrm{-invariant} \\
            N\geq\displaystyle\frac{-\ln\left(2\displaystyle\frac{|\sym_{G, U}(\psi)|}{|G|}\left[\frac{|\sym_{G, V}(\sigma)|}{|G|}\epsilon(1-\epsilon)\right]^{\frac{1}{2}}\right)}{\Delta_{G, U}(\psi)} & \textrm{otherwise,} 
        \end{cases}
        \label{SMeq:SMthm:approx_conversion2}
    \end{align}
    where $\Delta_{G, U}(\psi):=\min_{g\in G-\sym_{G, U}(\psi)} L_{\psi, G, U}(g)$. 
\end{theorem}

We note that when $\sym_{G, U}(\psi)\subset\sym_{G, V}(\sigma)$ and $\sigma$ is not $(G, V)$-invariant, we have $\sym_{G, U}(\psi)\subset\sym_{G, V}(\sigma)\subsetneq G$, which allows us to define $\Delta_{G, U}(\psi)$. 
By item 1 of Lemma~\ref{SMlem:measure_L_properties}, we can confirm that $\Delta_{G, U}(\psi)>0$.

\bigskip

For the proof of this theorem, we define ``maximally asymmetric states''. 
For a subgroup $H\subset G$ and a projective unitary representation $W$ of $G$, we say that a state $\xi$ is a maximally $(G, W)$-asymmetric state with symmetry subgroup $H$ if $\xi$ is a pure state with a characteristic function satisfying 
\begin{align}
    |\chi_{\xi, G, W}(g)| 
    =\begin{cases}
        1 & \textrm{if }g\in H,\\
        0 & \textrm{otherwise}.
    \end{cases} \label{SMeq:max_asymm_condition}
\end{align}
We note that the symmetry subgroup of this state is $H$ by item 5 in Sec.~5.2 in Ref.~\cite{Marvian2013} (item 1 in Lemma~\ref{SMlem:characteristic_function}). 
The reason why such a state $\xi$ is important is that $\xi$ can be converted to any (not necessarily pure) state $\sigma$ with symmetry subgroup $\sym_{G, V}(\sigma)\supset H$ via some $(G, W, V)$-covariant map, which we prove in Lemma~\ref{SMlem:conversion_from_max_asymm}. 
We note that although maximally asymmetric states do not necessarily exist for a given set of a projective unitary representation and a Hilbert space, we can show that maximally asymmetric states always exist for the left regular representation $F$ of $G$ and its representation space $\calK$, which has an orthonormal basis $\{\ket{\eta_g}\}_{g\in G}$ satisfying $F(g)\ket{\eta_h}=\ket{\eta_{gh}}$ for all $g, h\in G$. 
Concretely, if we define a pure state $\ket{\gamma}$ on $\calK$ by 
\begin{align}
    \ket{\gamma}:=\frac{1}{\sqrt{|H|}}\sum_{h\in H} \ket{\eta_h}, \label{SMeq:max_asymm_left_regular_special}
\end{align}
then $\gamma$ is a maximally $(G, F)$-asymmetric state with symmetry subgroup $H$. 
We present the proof of a more general statement in Lemma~\ref{SMlem:max_asymm_construction}.

The key in the proof is to use the maximally asymmetric state as an intermediate state in the conversion of copies of $\psi$ to copies of $\sigma$. 
First, we show that $N$ copies of $\ket{\psi}$ can be approximately transformed to the maximally asymmetric state $\gamma$ with symmetry subgroup $\sym_{G, V}(\sigma)$ within an error exponentially small in $N$. 
Next, we take a covariant map that transforms $\gamma$ into $\sigma^{\otimes \floor{rN}}$ for all $r\in (0, \infty)$. 
By combining these two maps, we can construct a covariant map that transforms $\psi^{\otimes N}$ into $\sigma^{\otimes \floor{rN}}$ within an arbitrarily small error.

\begin{proof}[proof of Theorem~\ref{SMthm:approx_conversion}]
    First, we prove that $R_{\mathrm{ap}, G, U, V}(\psi\to\sigma)=\infty$ if $\sym_{G, U}(\psi)\subset\sym_{G, V}(\sigma)$. 
    We take arbitrary $r>0$, $\epsilon\in (0, 1)$, and $N\in\mathbb{N}$ satisfying Eq.~\eqref{SMeq:SMthm:approx_conversion2}. 
    It is sufficient to show that there exists a $(G, U^{\otimes N}, V^{\otimes \floor{rN}})$-covariant map $\calE$ satisfying $T(\calE(\psi^{\otimes N}), \sigma^{\otimes \floor{rN}})\leq \epsilon$. 
    When $\sigma$ is $(G, V)$-invariant, we can take the map that discards every input state $\psi^{\otimes N}$ and adds $\sigma^{\otimes \floor{rN}}$ as an example of a $(G, U^{\otimes N}, V^{\otimes \floor{rN}})$-covariant map $\calE$ satisfying $\calE(\psi^{\otimes N})=\sigma^{\otimes \floor{rN}}$. 
    In the following, we consider the case where $\sigma$ is not $(G, V)$-invariant, and we are going to construct a $(G, U^{\otimes N}, V^{\otimes \floor{rN}})$-covariant map $\calE$ satisfying $T(\calE(\psi^{\otimes N}), \sigma^{\otimes \floor{rN}})\leq \epsilon$ in two steps. 
    In the first step, we construct a map $\calE_1$ that transforms $N$ copies of $\psi$ into a state $\zeta$ that is close to a maximally $(G, F)$-asymmetric state with symmetry subgroup $\sym_{G, V}(\sigma)$, where $F$ is the left regular representation of $G$ on $\calK$ mentioned above. 
    Specifically, we define a pure state $\zeta$ by 
    \begin{align}
        \ket{\zeta}:=\sqrt{1-\epsilon^2}\ket{\gamma}+\epsilon\ket{\overline{\gamma}} 
    \end{align}
    with 
    \begin{align}
        &\ket{\gamma}:=\frac{1}{\sqrt{|\sym_{G, V}(\sigma)|}}\sum_{g\in \sym_{G, V}(\sigma)}\ket{\eta_g}, \label{SMeq:SMthm:approx_conversion3}\\
        &\ket{\overline{\gamma}}:=\frac{1}{\sqrt{|G|-|\sym_{G, V}(\sigma)|}}\sum_{g\in G-\sym_{G, V}(\sigma)} \ket{\eta_g}, 
    \end{align}
    where $\{\ket{\eta_g}\}_{g\in G}$ is an orthonormal basis of $\calK$ satisfying $F(g)\ket{\eta_h}=\ket{\eta_{gh}}$ for all $g, h\in G$. 
    Then, for any $g\in \sym_{G, V}(\sigma)$, we have $L_{\zeta, G, F}(g)=0$, and for any $g\in G-\sym_{G, V}(\sigma)$, we have 
    \begin{align}
        |\mathrm{tr}(\zeta F(g))| 
        =2\sqrt{\frac{|\sym_{G, V}(\sigma)|}{|G|-|\sym_{G, V}(\sigma)|}}\epsilon\sqrt{1-\epsilon^2}+\frac{|G|-2|\sym_{G, V}(\sigma)|}{|G|-|\sym_{G, V}(\sigma)|}\epsilon^2 
        \geq 2\left[\frac{|\sym_{G, V}(\sigma)|}{|G|}(1-\epsilon^2)\right]^{\frac{1}{2}}\epsilon, 
    \end{align}
    which implies that 
    \begin{align}
        L_{\zeta, G, F}(g)
        =-\ln(|\mathrm{tr}(\zeta F(g))|)
        \leq -\ln\left(2\left[\frac{|\sym_{G, V}(\sigma)|}{|G|}(1-\epsilon^2)\right]^{\frac{1}{2}}\epsilon\right). 
    \end{align}
    We also note that $L_{\psi, G, U}(g)=0$ for all $g\in \sym_{G, U}(\psi)$ and $L_{\psi, G, U}(g)\geq \Delta_{G, U}(\psi)$ for all $g\in G-\sym_{G, U}(\psi)$. 
    By Theorem~\ref{SMthm:exact_conversion}, we have 
    \begin{align}
        R_{\mathrm{ex}, G, U, F}(\psi\to\zeta) 
        =\min_{g\in G}\frac{L_{\psi, G, U}(g)}{L_{\zeta, G, F}(g)} 
        \geq\frac{\Delta_{G, U}(\psi)}{-\ln\left(2\left[\displaystyle\frac{|\sym_{G, V}(\sigma)|}{|G|}(1-\epsilon^2)\right]^{\frac{1}{2}}\epsilon\right)}. \label{SMeq:Rap_MAS_UpperBound}
    \end{align}
    We set $s:=1/N$ with $N$ that satisfies Eq.~\eqref{SMeq:SMthm:approx_conversion2}. 
    Then, we have 
    \begin{align}
        \frac{1}{R_{\mathrm{ex}, G, U, F}(\psi\to\zeta)}
        \leq \frac{-\ln\left(2\left[\displaystyle\frac{|\sym_{G, V}(\sigma)|}{|G|}(1-\epsilon^2)\right]^{\frac{1}{2}}\epsilon\right)}{\Delta_{G, U}(\psi)} 
        \leq \frac{1}{s}-\frac{1}{\Delta_{G, U}(\psi)} \ln\left(\frac{|G|}{|\sym_{G, U}(\psi)|}\right), 
    \end{align}
    which implies that 
    \begin{align}
        \frac{\ln\left(\displaystyle\frac{|G|}{|\sym_{G, U}(\psi)|}\right)}{\left(1-\displaystyle\frac{s}{R_{\mathrm{ex}, G, U, F}(\psi\to\zeta)}\right)\Delta_{G, U}(\psi)} 
        \leq \frac{1}{s} 
        =N. 
    \end{align}
    Noting that $s\le R_{\mathrm{ex}, G, U, F}(\psi\to\zeta)$ from Eqs.~\eqref{SMeq:SMthm:approx_conversion2} and \eqref{SMeq:Rap_MAS_UpperBound}, by Theorem~\ref{SMthm:exact_conversion}, rate $s$ is is exactly achievable in the conversion from $\psi$ to $\zeta$ if the initial number $N$ of copies of $\psi$ satisfies Eq.~\eqref{SMeq:SMthm:approx_conversion2}, i.e., we can take some $(G, U^{\otimes N}, F)$-covariant map $\calE_1$ such that $\calE_1(\psi^{\otimes N})=\zeta$. 
    In the second step, we take a $(G, F, V^{\otimes \floor{rN}})$-covariant map $\calE_2$ such that $\calE_2(\gamma)=\sigma^{\otimes \floor{rN}}$. 
    By combining these two maps, we define a $(G, U^{\otimes N}, V^{\otimes \floor{rN}})$-covariant map $\calE:=\calE_2\circ\calE_1$. 
    Then, by the information-processing inequality, we get 
    \begin{align}
        T(\calE(\psi^{\otimes N}), \sigma^{\otimes \floor{rN}})
        =T(\calE_2(\calE_1(\psi^{\otimes N})), \calE_2(\gamma))
        \leq T(\calE_1(\psi^{\otimes N}), \gamma)
        =T(\zeta, \gamma) 
        =\sqrt{1-|\braket{\zeta|\gamma}|^2}
        =\epsilon, \label{SMeq:information_processing}
    \end{align}

    Next, we prove that $R_{\textrm{ap}, G, U, V}(\psi\to\sigma)=0$ if $\sym_{G, U}(\psi)\not\subset\sym_{G, V}(\sigma)$. 
    It is sufficient to show that there exists some $\epsilon>0$ such that for any $N\in\mathbb{N}$, we cannot convert $N$ copies of $\psi$ into even a single copy of $\sigma$ within error $\epsilon$ by any $(G, U^{\otimes N}, V)$-covariant maps. 
    Since we have $\sym_{G, U}(\psi)\not\subset\sym_{G, V}(\sigma)$, we can take some $g\in\sym_{G, U}(\psi)-\sym_{G, V}(\sigma)$. 
    By items 4 and 5 in Sec.~5.2 in Ref.~\cite{Marvian2013} (item 1 in Lemma~\ref{SMlem:characteristic_function}), since $g\not\in\sym_{G, V}(\sigma)$, we have $|\chi_{\sigma, G, V}(g)|<1$. 
    We set $\epsilon:=(1-|\chi_{\sigma, G, V}(g)|)/4>0$.
    We take arbitrary $N\in\mathbb{N}$ and $(G, U^{\otimes N}, V)$-covariant map $\calE$. 
    Then, we have $\sym_{G, V}(\calE(\psi^{\otimes N}))\supset\sym_{G, U^{\otimes N}}(\psi^{\otimes N})=\sym_{G, U}(\psi)$. 
    We thus have $g\in\sym_{G, V}(\calE(\psi^{\otimes N}))$, implying that $|\chi_{\calE(\psi^{\otimes N}), G, V}(g)|=1$ by item 5 in Sec.~5.2 in Ref.~\cite{Marvian2013} (item 1 in Lemma~\ref{SMlem:characteristic_function}). 
    We can give a lower bound of the trace distance between $\calE(\psi^{\otimes N})$ and $\sigma$ as follows: 
    \begin{align}
        T(\calE(\psi^{\otimes N}), \sigma)
        =&\frac{1}{2}\|\calE(\psi^{\otimes N})-\sigma\|_1 \nonumber\\
        \geq&\frac{1}{2}\|\calE(\psi^{\otimes N})-\sigma\|_1\|V(g)\|_\infty \nonumber\\
        \geq&\frac{1}{2}\|(\calE(\psi^{\otimes N})-\sigma)V(g)\|_1 \nonumber\\
        \geq&\frac{1}{2}|\mathrm{tr}((\calE(\psi^{\otimes N})-\sigma)V(g))| \nonumber\\
        =&\frac{1}{2}|\chi_{\calE(\psi^{\otimes N}), V}(g)-\chi_{\sigma, V}(g)| \nonumber\\
        \geq&\frac{1}{2}(|\chi_{\calE(\psi^{\otimes N}), V}(g)|-|\chi_{\sigma, V}(g)|) \nonumber\\
        =&2\epsilon \nonumber\\
        >&\epsilon, \label{SMeq:SMthm:approx_conversion4}
    \end{align}
    where we used H\"{o}lder's inequality for the Schatten norm in the third line and the triangle inequality in the fourth and sixth lines.  
\end{proof}

We show the counterpart of Theorem~\ref{SMthm:approx_conversion} in the case of projective unitary representations, by using Theorem~\ref{SMthm:approx_conversion} and Lemmas~\ref{SMlem:unitary_rep_construction} and \ref{SMlem:multiple_copy}.

%Theorem S4
\begin{theorem} \label{SMthm:approx_projective}
    (Counterpart of Theorem~\ref{thm:approx_conversion} in the main text for projective unitary representations.) 
    Let $\psi$ be a pure state on a finite-dimensional Hilbert space $\calH_\mathrm{in}$ and $\sigma$ be a general (not necessarily pure) state on a finite-dimensional Hilbert space $\mathcal{H}_\mathrm{out}$, and $U$ and $V$ be projective unitary representations of a finite group $G$ on $\calH_\mathrm{in}$ and $\calH_\mathrm{out}$, respectively. 
    Then, the approximate $(G, U, V)$-covariant conversion rate from $\psi$ to $\sigma$ is given by 
    \begin{align}
        R_{\mathrm{ap}, G, U, V}(\psi\to\sigma)
        =\begin{cases}
            \infty & \textrm{if }\sym_{G, U}(\psi)\subset\sym_{G, V}(\sigma)\\
            0 & \textrm{otherwise}.
        \end{cases} \label{SMeq:SMthm:approx_projective1}
    \end{align}
    Moreover, when $\sym_{G, U}(\psi)\subset\sym_{G, V}(\sigma)$, for any $r\in(0, \infty)$ and $\epsilon\in (0, 1)$, conversion rate $r$ is $\epsilon$-approximately achievable if the initial number $N$ of copies of state $\psi$ satisfies 
    \begin{align}
        \begin{cases}
            N\geq 0 & \textrm{if } \sigma \textrm{ is } (G, V) \textrm{-invariant} \\
            N\geq\displaystyle\frac{-\ln\left(2\displaystyle\frac{|\sym_{G, U}(\psi)|}{|G|}\left[\frac{|\sym_{G, V}(\sigma)|}{|G|}(1-\epsilon^2)\right]^{\frac{1}{2}}\epsilon\right)}{\Delta_{G, U}(\psi)}+d_\mathrm{in} & \textrm{otherwise},
        \end{cases} \label{SMeq:SMthm:approx_projective2}
    \end{align}
    where $d_\mathrm{in}$ is the dimension of $\calH_\mathrm{in}$ and $\Delta_{G, U}(\psi):=\min_{g\in G-\sym_{G, U}(\psi)} L_{\psi, G, U}(g)$. 
\end{theorem}

We note that $\Delta_{G, U}(\psi)$ can be defined and satisfies $\Delta_{G, U}(\psi)>0$ when $\sym_{G, U}(\psi)\subset\sym_{G, V}(\sigma)$ and $\sigma$ is not $(G, V)$-invariant, as we mentioned just below Theorem~\ref{SMthm:approx_conversion}.

\begin{proof}
    First, we show that the approximate conversion rate is given by Eq.~\eqref{SMeq:SMthm:approx_projective1}. 
    In the same way as in the proof of Theorem~\ref{SMthm:exact_projective}, we define $\Psi:=\psi^{\otimes d_\mathrm{in}}$ and $\Gamma:=\sigma^{\otimes d_\mathrm{out}}$, and define $U'$ and $V'$ by $U'(g):=U(g)^{\otimes d_\mathrm{in}}/\mathrm{det}(U(g))$ and $V'(g):=V(g)^{\otimes d_\mathrm{out}}/\mathrm{det}(V(g))$ for all $g\in G$. 
    Then, by Lemma~\ref{SMlem:unitary_rep_construction}, $U'$ and $V'$ are (non-projective) unitary representations. 
    By Theorem~\ref{SMthm:approx_conversion}, we have 
    \begin{align}
        R_{\mathrm{ap}, G, U', V'}(\Psi\to\Gamma) 
        =\begin{cases}
            \infty & \textrm{if }\sym_{G, U'}(\Psi)\subset\sym_{G, V'}(\Gamma),\\
            0 & \textrm{otherwise}.
        \end{cases} \label{SMeq:SMthm:approx_projective3}
    \end{align}
    We note that we have $\sym_{G, U'}(\Psi)=\sym_{G, U^{\otimes d_\mathrm{in}}}(\psi^{\otimes d_\mathrm{in}})=\sym_{G, U}(\psi)$, and we similarly have $\sym_{G, V'}(\Gamma)=\sym_{G, V}(\sigma)$. 
    Thus Eq.~\eqref{SMeq:SMthm:approx_projective3} is equivalent to 
    \begin{align}
        R_{\mathrm{ap}, G, U', V'}(\Psi\to\Gamma) 
        =\begin{cases}
            \infty & \textrm{if }\sym_{G, U}(\psi)\subset\sym_{G, V}(\sigma),\\
            0 & \textrm{otherwise}.
        \end{cases} \label{SMeq:SMthm:approx_projective4}
    \end{align}
    By Lemma~\ref{SMlem:multiple_copy}, we have 
    \begin{align}
        R_{\mathrm{ap}, G, U, V}(\psi\to\sigma) 
        =\frac{d_\mathrm{in}}{d_\mathrm{out}} R_{\mathrm{ap}, G, U^{\otimes d_\mathrm{in}}, V^{\otimes d_\mathrm{out}}}(\Psi\to\Gamma) 
        =\frac{d_\mathrm{in}}{d_\mathrm{out}} R_{\mathrm{ap}, G, U', V'}(\Psi\to\Gamma), \label{SMeq:SMthm:approx_projective5}
    \end{align}
    where we used the fact that the conditions for $(G, U^{\otimes d_\mathrm{in}}, V^{\otimes d_\mathrm{out}})$-covariance and $(G, U', V')$-covariance are the same. 
    By plugging Eq.~\eqref{SMeq:SMthm:approx_projective4} into Eq.~\eqref{SMeq:SMthm:approx_projective5}, we get Eq.~\eqref{SMeq:SMthm:approx_projective1}.

    Next, when $\sym_{G, U}(\psi)\subset\sym_{G, V}(\sigma)$, we take arbitrary $r\in (0, \infty)$, $N\in\mathbb{N}$, and $\epsilon\in (0, 1)$ satisfying Eq.~\eqref{SMeq:SMthm:approx_projective2}, and show that conversion rate $r$ is $\epsilon$-approximately achievable if we initially have $N$ copies of state $\psi$. 
    We take arbitrary $N\in\mathbb{N}$ satisfying Eq.~\eqref{SMeq:SMthm:approx_projective2} and suppose that we have $N$ copies of $\psi$. 
    It is sufficient to show that there exists some $(G, U^{\otimes N}, V^{\otimes \floor{rN}})$-covariant map $\calE$ satisfying $T(\calE(\psi^{\otimes N}), \sigma^{\otimes \floor{rN}})\leq\epsilon$. 
    When $\sigma$ is $(G, V)$-invariant, as we mentioned in the proof of Theorem~\ref{SMthm:approx_conversion}, we can easily take a $(G, U^{\otimes N}, V^{\otimes \floor{rN}})$-covariant map $\calE$ satisfying $\calE(\psi^{\otimes N})=\sigma^{\otimes \floor{rN}}$. 
    In the following, we consider the case where $\sigma$ is not $(G, V)$-invariant, and we are going to construct $(G, U^{\otimes N}, V^{\otimes \floor{rN}})$-covariant map $\calE$ satisfying $T(\calE(\psi^{\otimes N}), \sigma^{\otimes \floor{rN}})\leq\epsilon$ in three steps. 
    As the first step, we take a $(G, U^{\otimes N}, {U'}^{\otimes \floor{N/d_\mathrm{in}}})$-covariant map $\calE_1$ that transforms $\psi^{\otimes N}$ to $\Psi^{\otimes \floor{N/d_\mathrm{in}}}$, which is a map just discarding $N-d_\mathrm{in}\floor{N/d_\mathrm{in}}$ copies of $\psi$. 
    As the second step, we take a $(G, {U'}^{\otimes \floor{N/d_\mathrm{in}}}, F)$-covariant map that transforms $\Psi^{\otimes \floor{N/d_\mathrm{in}}}$ into $\gamma$ within error $\epsilon$, where $F$ is the left regular representation of $G$ and $\gamma$ is a maximally $(G, F)$-asymmetric state with symmetry subgroup $\sym_{G, V}(\sigma)$, defined by Eq.~\eqref{SMeq:SMthm:approx_conversion3}. 
    We note that Eq.~\eqref{SMeq:SMthm:approx_projective2} implies that 
    \begin{align}
        \left\lfloor\frac{N}{d_\mathrm{in}}\right\rfloor
        \geq &\frac{N}{d_\mathrm{in}}-1 \nonumber\\
        \geq & \frac{-\ln\left(2\displaystyle\frac{|\sym_{G, U}(\psi)|}{|G|}\left[\frac{|\sym_{G, V}(\sigma)|}{|G|}(1-\epsilon^2)\right]^{\frac{1}{2}}\epsilon\right)}{d_\mathrm{in}\Delta_{G, U}(\psi)} \nonumber\\
        =& \frac{-\ln\left(2\displaystyle\frac{|\sym_{G, U'}(\Psi)|}{|G|}\left[\frac{|\sym_{G, W}(\gamma)|}{|G|}(1-\epsilon^2)\right]^{\frac{1}{2}}\epsilon\right)}{\Delta_{G, U'}(\Psi)}, \label{SMeq:SMthm:approx_projective6}
    \end{align}
    where we used $\Delta_{G, U'}(\Psi)=d_\mathrm{in}\Delta_{G, U}(\psi)$ in the last line. 
    By Theorem~\ref{SMthm:approx_conversion} and Eq.~\eqref{SMeq:SMthm:approx_projective6}, there exists a $(G, {U'}^{\otimes \floor{N/d_\mathrm{in}}}, F^{\otimes s\floor{N/d_\mathrm{in}}})$-covariant map that transforms $\Psi^{\otimes \floor{N/d_\mathrm{in}}}$ into $\gamma^{\otimes s\floor{N/d_\mathrm{in}}}$ within error $\epsilon$ for all $s\in (0, \infty)$. 
    By setting $s=\floor{N/d_\mathrm{in}}^{-1}$, we take a $(G, U^{\otimes N}, F)$-covariant map $\calE_2$ satisfying $T(\calE_2(\psi^{\otimes N}), \gamma)\leq\epsilon$. 
    By Lemma~\ref{SMlem:conversion_from_max_asymm}, we can take a $(G, W, V^{\otimes \floor{rN}})$-covariant map $\calE_3$ such that $\calE_3(\gamma)=\sigma^{\otimes \floor{rN}}$. 
    By combining the three maps above, we define a $(G, U^{\otimes N}, V^{\otimes \floor{rN}})$-covariant map $\calE:=\calE_3 \circ \calE_2 \circ \calE_1$. 
    Then, by the information-processing inequality, we get 
    \begin{align}
        T(\calE(\psi^{\otimes N}), \sigma^{\otimes \floor{rN}})
        =T(\calE_3(\calE_2(\Psi^{\otimes \floor{N/d_\mathrm{in}}})), \calE_3(\gamma)) 
        \leq T(\calE_2(\Psi^{\otimes \floor{N/d_\mathrm{in}}}), \gamma) 
        \leq \epsilon. 
    \end{align}
\end{proof}

\section{Reduction to non-projective unitary representations} \label{SMsec:reduction_to_nonprojective}

In this section, we show the lemmas that are useful for generalizing the results about the conversion rates obtained for non-projective unitary representations to the case of projective unitary representations.

First, we show that for any projective unitary representation $U$ on a $d$-dimensional space, there exists a non-projective unitary representation $U'$ on the $d$-fold tensor product of the representation space such that $U'$ corresponds with $U^{\otimes d}$ up to the global phase.

\begin{lemma} \label{SMlem:unitary_rep_construction}
    Let $U$ be a projective unitary representation of a group $G$ on a Hilbert space $\mathcal{H}$ with dimension $d(<\infty)$, and a map $U'$ from $G$ to $\calL(\calH^{\otimes d})$ be defined by 
    \begin{align}
        U'(g):=\frac{U(g)^{\otimes d}}{\mathrm{det}(U(g))}\ \forall g\in G. 
    \end{align}
    Then, $U'$ is a non-projective unitary representation of $G$ on $\mathcal{H}^{\otimes d}$. 
\end{lemma}

\begin{proof}
    We denote the cocycle of $U$ by $\omega$, i.e., $\omega$ is a map from $G\times G$ to $\mathbb{C}$ satisfying $U(g)U(h)=\omega(g, h)U(gh)$ for all $g, h\in G$. 
    Then, by the definition of $U'$, we have 
    \begin{align}
        U'(g)U'(h)
        =\frac{(U(g)U(h))^{\otimes d}}{\mathrm{det}(U(g)U(h))}
        =\frac{(\omega(g, h)U(gh))^{\otimes d}}{\mathrm{det}(\omega(g, h)U(gh))}
        =\frac{\omega(g, h)^d U(gh)^{\otimes d}}{\omega(g, h)^d\mathrm{det}(U(gh))}
        =U'(gh)
    \end{align}
    for all $g, h\in G$. 
\end{proof}

Next, we show a lemma about the convertibility between multiple copies of states.

\begin{lemma} \label{SMlem:multiple_copy}
    Let $\rho$ and $\sigma$ be general (not necessarily pure) states on finite-dimensional Hilbert spaces $\mathcal{H}_\mathrm{in}$ and $\mathcal{H}_\mathrm{out}$, $U$ and $V$ be projective unitary representations of a group $G$ on $\mathcal{H}_\mathrm{in}$ and $\mathcal{H}_\mathrm{out}$, and $p, q, p', q'\in\mathbb{N}$. 
    Then, 
    \begin{align}
        \frac{q}{p}R_{\textrm{ex}, G, U^{\otimes p}, V^{\otimes q}} (\rho^{\otimes p}\to\sigma^{\otimes q})
        =\frac{q'}{p'}R_{\textrm{ex}, G, U^{\otimes p'}, V^{\otimes q'}} (\rho^{\otimes p'}\to\sigma^{\otimes q'}), \\
        \frac{q}{p}R_{\textrm{ap}, G, U^{\otimes p}, V^{\otimes q}} (\rho^{\otimes p}\to\sigma^{\otimes q})
        =\frac{q'}{p'}R_{\textrm{ap}, G, U^{\otimes p'}, V^{\otimes q'}} (\rho^{\otimes p'}\to\sigma^{\otimes q'}). 
    \end{align}
\end{lemma}

\begin{proof} 
    We only give a proof for the approximate case, because the proof for the exact case can be obtained by setting the allowed error $\epsilon$ to be zero. 
    Noting that the statement of this lemma is symmetric under the exchange of $(p, q)$ and $(p', q')$, it is sufficient to show that 
    \begin{align}
        \frac{q}{p}R_{\textrm{ap}, G, U^{\otimes p}, V^{\otimes q}} (\rho^{\otimes p}\to\sigma^{\otimes q})
        \leq\frac{q'}{p'}R_{\textrm{ap}, G, U^{\otimes p'}, V^{\otimes q'}} (\rho^{\otimes p'}\to\sigma^{\otimes q'}). \label{SMeq:SMlem:multiple_copy1}
    \end{align}
    Since this inequality trivially holds when $R_{\textrm{ap}, G, U^{\otimes p}, V^{\otimes q}} (\rho^{\otimes p}\to\sigma^{\otimes q})=0$, we assume that $R_{\textrm{ap}, G, U^{\otimes p}, V^{\otimes q}} (\rho^{\otimes p}\to\sigma^{\otimes q})>0$ in the following. 
    We take arbitrary $r\in (0, R_{\textrm{ap}, G, U^{\otimes p}, V^{\otimes q}} (\rho^{\otimes p}\to\sigma^{\otimes q}))$ and $r'\in(r, R_{\textrm{ap}, G, U^{\otimes p}, V^{\otimes q}} (\rho^{\otimes p}\to\sigma^{\otimes q}))$. 
    Then, for any $\epsilon>0$, we can take some $N_0\in\mathbb{N}$ such that 
    \begin{align}
        (\rho^{\otimes p})^{\otimes N}\xrightarrow{(G, U^{\otimes pN}, V^{\otimes q\floor{r'N}})\textrm{-cov.}}_\epsilon (\sigma^{\otimes q})^{\otimes \floor{r'N}} \label{SMeq:SMlem:multiple_copy2}
    \end{align}
    for all $N\geq N_0$. 
    We take $M_0\in\mathbb{N}$ such that $M_0\geq \max \{(p/p')(N_0+1) , p(1+r')/[p'(r'-r)] \}$, which implies that for any $M\ge M_0$, the following inequalities hold: 
    \begin{align}
        &\left\lfloor\frac{p'}{p}M\right\rfloor
        \geq \frac{p'}{p}M-1
        \geq N_0, \\
        &q\left\lfloor r'\left\lfloor \frac{p'}{p}M\right\rfloor \right\rfloor
        \geq q\left[r'\left(\frac{p'}{p}M-1\right)-1\right] 
        \geq q'\cdot\frac{qp'}{pq'}rM 
        \geq q'\left\lfloor \frac{qp'}{pq'}rM\right\rfloor. 
    \end{align}
    Thus for any $M\geq M_0$, we can transform $(\rho^{\otimes p'})^{\otimes M}$ into $(\sigma^{\otimes q'})^{\otimes \left\lfloor\frac{qp'}{pq'}rM\right\rfloor}$ with the error $\epsilon$ by the $(G, U^{\otimes p'M}, V^{\otimes \left\lfloor\frac{qp'}{pq'}rM\right\rfloor})$-covariant process consisting of the following three processes: 
    \begin{align}
        (\rho^{\otimes p'})^{\otimes M}
        &\xrightarrow{\left(G, U^{\otimes p'M}, U^{\otimes p\left\lfloor\frac{p'}{p}M\right\rfloor}\right)\textrm{-cov.}} 
        (\rho^{\otimes p})^{\otimes \left\lfloor\frac{p'}{p}M\right\rfloor} \nonumber\\
        &\xrightarrow{\left(G, U^{\otimes p\left\lfloor\frac{p'}{p}M\right\rfloor}, V^{\otimes q\left\lfloor r'\left\lfloor\frac{p'}{p}M\right\rfloor\right\rfloor}\right)\textrm{-cov.}}_\epsilon
        (\sigma^{\otimes q})^{\otimes \left\lfloor r'\left\lfloor\frac{p'}{p}M\right\rfloor\right\rfloor} \nonumber\\
        &\xrightarrow{\left(G, V^{\otimes q\left\lfloor r'\left\lfloor\frac{p'}{p}M\right\rfloor\right\rfloor}, V^{\otimes q'\left\lfloor\frac{qp'}{pq'}rM\right\rfloor}\right)\textrm{-cov.}}
        (\sigma^{\otimes q'})^{\otimes \left\lfloor\frac{qp'}{pq'}rM\right\rfloor}, 
    \end{align}
    where we discard redundant subsystems in the first and the third processes, and we use the transformation shown in Eq.~\eqref{SMeq:SMlem:multiple_copy2} in the second process. 
    Since this holds for all $\epsilon>0$, we get $\frac{qp'}{pq'}r\leq R_{\textrm{ap}, G, U^{\otimes p'}, U^{\otimes q'}} (\rho^{\otimes p'}\to\sigma^{\otimes q'})$. 
    By noting that the choice for $r\in (0, R_{\textrm{ap}, G, U^{\otimes p}, U^{\otimes q}} (\rho^{\otimes p}\to\sigma^{\otimes q}))$ was arbitrary, we get Eq.~\eqref{SMeq:SMlem:multiple_copy1}, which holds even when $R_{\textrm{ap}, G, U^{\otimes p}, V^{\otimes q}} (\rho^{\otimes p}\to\sigma^{\otimes q})=\infty$. 
\end{proof}

\section{Properties of maximally asymmetric states} \label{SMsec:max_asymm_map_construction}

In this section, we present two important properties of maximally asymmetric states defined by Eq.~\eqref{SMeq:max_asymm_condition}.

First, we show that there exists a covariant map that transforms maximally asymmetric states to arbitrary states with the same or larger symmetry subgroup by explicitly constructing such a map.

\begin{lemma} \label{SMlem:conversion_from_max_asymm}
    Let $W$ and $V$ be projective unitary representations of a finite group $G$ on finite-dimensional Hilbert spaces $\calQ$ and $\calH$, $\xi$ be a maximally $(G, W)$-asymmetric pure state on $\calQ$ with symmetry subgroup $H\subset G$, and $\sigma$ be a general (not necessarily pure) state on $\calH$ satisfying $\sym_{G, V}(\sigma)\supset H$. 
    Then, $\xi$ can be converted to $\sigma$ via some $(G, W, V)$-covariant operation. 
\end{lemma}

We note that this lemma holds even when $W$ or $V$ may be a projective unitary representation, which is important for the proof of Theorem~\ref{SMthm:approx_projective}.

\begin{proof}
    We define a map $\calE: \calL(\calQ)\to\calL(\calH)$ by 
    \begin{align}
        \calE(X):=\sum_{a\in A} \mathrm{tr}\left(W(a)\xi W(a)^\dag X\right)\left(V(a)\sigma V(a)^\dag-\frac{I_\calH}{d} \right)+\mathrm{tr}(X)\frac{I_\calH}{d}\ \forall X\in\calL(\calQ), \label{SMeq:SMlem:conversion_from_max_asymm1}
    \end{align}
    where $A$ is the set of representatives of the left cosets of $H$ in $G$, and $d$ is the dimension of $\calH$. 
    In the following, we show that $\mathcal{E}$ is a $(G, W, V)$-covariant CPTP map and satisfies $\mathcal{E}(\xi)=\sigma$.

    First, we show that $\mathcal{E}$ is well defined, i.e., the definition is independent of the choice of representatives $A$ of the left cosets $G/H$. 
    We take an arbitrary set $A'$ of representatives of $G/H$. 
    Then, $A'$ can be written as $\{ah_a\}_{a\in A}$ with some $h_a\in H$. 
    We thus have 
    \begin{align}
        &\sum_{a'\in A'} \mathrm{tr}\left(W(a')\xi W(a')^\dag X\right)\left(V(a')\sigma V(a')^\dag-\frac{I_\calH}{d}\right)+\mathrm{tr}(X)\frac{I_\calH}{d} \nonumber\\
        =&\sum_{a\in A} \mathrm{tr}\left(W(ah_a)\xi W(ah_a)^\dag X\right)\left(V(ah_a)\sigma V(ah_a)^\dag-\frac{I_\calH}{d}\right)+\mathrm{tr}(X)\frac{I_\calH}{d} \nonumber\\
        =&\sum_{a\in A} \mathrm{tr}\left(W(a)W(h_a)\xi W(h_a)^\dag W(a)^\dag X\right)\left(V(a)V(h_a)\sigma V(h_a)^\dag V(a)^\dag-\frac{I_\calH}{d}\right)+\mathrm{tr}(X)\frac{I_\calH}{d} \nonumber\\
        =&\sum_{a\in A} \mathrm{tr}\left(W(a)\xi W(a)^\dag X\right)\left(V(a)\sigma V(a)^\dag-\frac{I_\calH}{d}\right)+\mathrm{tr}(X)\frac{I_\calH}{d}, 
    \end{align}
    where we used $H=\sym_{G, W}(\xi)\subset\sym_{G, V}(\sigma)$ in the last line.

    Next, we show that $\mathcal{E}$ is $(G, W, V)$-covariant CPTP map. 
    We note that $\calE$ can be written as 
    \begin{align}
        \calE(X)=\sum_{a\in A} \mathrm{tr}\left(W(a)\xi W(a)^\dag X\right)V(a)\sigma V(a)^\dag +\mathrm{tr}\left(\left(I_\calQ -\sum_{a\in A} W(a)\xi W(a)^\dag\right)X\right)\frac{I_\calH}{d}, 
    \end{align}
    and that $\{W(a)\xi W(a)^\dag\}_{a\in A}$ is the set of projections onto orthogonal spaces, since $\xi$ is a maximally $(G, W)$-asymmetric state. 
    We can thus confirm that $\calE$ is a measure-and-prepare channel, which means that it is a CPTP map. 
    For any $g\in G$, we have 
    \begin{align}
        &V(g)\calE(W(g)^\dag X W(g)) V(g)^\dag \nonumber\\
        =&V(g)\left[\sum_{a\in A} \mathrm{tr}\left(W(a)\xi W(a)^\dag W(g)^\dag XW(g)\right)\left(V(a)\sigma V(a)^\dag -\frac{I_\calH}{d}\right)+\mathrm{tr}(X)\frac{I_\calH}{d}\right]V(g)^\dag \nonumber\\
        =&\sum_{a\in A} \mathrm{tr}\left(W(g)W(a)\xi W(a)^\dag W(g)^\dag X\right)\left(V(g)V(a)\sigma V(a)^\dag V(g)^\dag-\frac{I_\calH}{d}\right)+\mathrm{tr}(X)\frac{I_\calH}{d} \nonumber\\
        =&\sum_{a\in A} \mathrm{tr}\left(W(ga)\xi W(ga)^\dag X\right)\left(V(ga)\sigma V(ga)^\dag-\frac{I_\calH}{d}\right)+\mathrm{tr}(X)\frac{I_\calH}{d} \nonumber\\
        =&\calE(X), 
    \end{align}
    where we used the fact that $\{ga\}_{a\in A}$ is a set of representatives of the left cosets in the last line. 
    Therefore, $\calE$ is a $(G, W, V)$-covariant map.

    Finally, we confirm that $\calE$ satisfies $\calE(\xi)=\sigma$. 
    By the definition of $\calE$, we have 
    \begin{align}
        \mathcal{E}(\xi) 
        =&\sum_{a\in A} \mathrm{tr}\left(W(a)\xi W(a)^\dag \xi\right)\left(V(a)\sigma V(a)^\dag-\frac{I_\calH}{d}\right)+\mathrm{tr}(\xi)\frac{I_\calH}{d} \nonumber\\
        =&\sum_{a\in A} \widetilde{\delta}_{a, e} \left(V(a)\sigma V(a)^\dag-\frac{I_\calH}{d}\right)+\frac{I_\calH}{d} \nonumber\\
        =&\left(V(a_0)\sigma V(a_0)^\dag-\frac{I_\calH}{d}\right)+\frac{I_\calH}{d} \nonumber\\
        =&\sigma, 
    \end{align}
    where $a_0$ is the single element in $A\cap H$, $\widetilde{\delta}_{g_1, g_2}:=1$ if $g_2^{-1}g_1\in H$ and $\widetilde{\delta}_{g_1, g_2}:=0$ otherwise, and we used 
    \begin{align}
        \mathrm{tr}\left(W(a)\xi W(a)^\dag \xi\right) 
        =\bra{\xi}W(a)\ket{\xi}\bra{\xi}W(a)\ket{\xi} 
        =|\bra{\xi}W(a)\ket{\xi}|^2 
        =|\mathrm{tr}(W(a)\xi)|^2 
        =\widetilde{\delta}_{a, e} 
    \end{align}
    in the last line. 
\end{proof}

We note that by using this lemma, we can show that maximally asymmetric states can serve as catalysts, though we do not directly use it in the proofs of the main theorems.

\begin{corollary} \label{SMcor:catalytic_map}
    Let $U$, $V$, and $W$ be projective unitary representations of a finite group $G$ on finite-dimensional Hilbert spaces $\calH_\mathrm{in}$, $\calH_\mathrm{out}$, and $\calQ$, $\xi$ be a maximally $(G, W)$-asymmetric pure state with symmetry subgroup $H\subset G$, $\rho$ and $\sigma$ be general (not necessarily pure) states satisfying $\sym_{G, V}(\sigma)\supset H$. 
    Then, $\rho\otimes\xi$ can be converted to $\sigma\otimes \xi$ via some $(G, U\otimes W, V\otimes W)$-covariant operation. 
\end{corollary}

\begin{proof}
    Since the map tracing out $\calH_\mathrm{in}$ on $\calH_\mathrm{in}\otimes \calQ$ is $(G, U\otimes W, W)$-covariant, we have $\rho\otimes \xi\xrightarrow{(G, U\otimes W, W)\textrm{-cov.}} \xi$. 
    Since $\sigma$ and $\xi$ satisfies $\sym_{G, V}(\sigma)\supset H$ and $\sym_{G, W}(\xi)=H$, we have $\sym_{G, V\otimes W}(\sigma\otimes \xi)\supset H$. 
    Theorem~\ref{SMlem:conversion_from_max_asymm} thus implies that $\xi\xrightarrow{(G, W, V\otimes W)\textrm{-cov.}} \sigma\otimes \xi$. 
    By combining these two maps, we get $\rho\otimes \xi\xrightarrow{(G, U\otimes W, V\otimes W)\textrm{-cov.}} \sigma\otimes \xi$. 
\end{proof}

We can explicitly present a map $\calE$ realizing this conversion as follows. 
The first step of the conversion is realized by the map $\calD: \calL(\calH_\mathrm{in}\otimes \calQ)\to\calL(\calH_\mathrm{in})$ as the partial trace over $\calH_\mathrm{in}$. 
As for the second step, by substituting $\sigma$ and $V$ with $\sigma\otimes \xi$ and $V\otimes W$, respectively in the definition of $\calE$ in Eq.~\eqref{SMeq:SMlem:conversion_from_max_asymm1}, we define 
\begin{align}
    \calE'(X)=\sum_{a\in A} \mathrm{tr}\left(U(a)\xi U(a)^\dag X\right)\left[V(a)(\sigma\otimes \xi) V(a)^\dag-\frac{I_{\calH_\mathrm{out}\otimes\calQ}}{d_\mathrm{out}q}\right]+\mathrm{tr}(X)\frac{I_{\calH_\mathrm{out}\otimes\calQ}}{d_\mathrm{out}q}, 
\end{align}
where $d_\mathrm{out}$ and $q$ are the dimensions of $\calH_\mathrm{out}$ and $\calQ$, respectively. 
By combining these two maps, we define a map $\calE$ by 
\begin{align}
    \calE(X)
    =&\calE'\circ\calD(X) \nonumber\\
    =&\sum_{a\in A} \mathrm{tr}\left(U(a)\xi U(a)^\dag \calD(X)\right)\left[V(a)(\sigma\otimes \xi) V(a)^\dag-\frac{I_{\calH_\mathrm{out}\otimes\calQ}}{d_\mathrm{out}q}\right]+\mathrm{tr} (\calD(X))\frac{I_{\calH_\mathrm{out}\otimes\calQ}}{d_\mathrm{out}q} \nonumber\\
    =&\sum_{a\in A} \mathrm{tr}\left(\left(I_{\calH_\mathrm{in}}\otimes U(a)\xi U(a)^\dag\right) X\right)\left[V(a)(\sigma\otimes \xi) V(a)^\dag-\frac{I_{\calH_\mathrm{out}\otimes\calQ}}{d_\mathrm{out}q}\right]+\mathrm{tr}(X)\frac{I_{\calH_\mathrm{out}\otimes\calQ}}{d_\mathrm{out}q}. 
\end{align}
Then, $\calE$ is $(G, U\otimes W, V\otimes W)$-covariant and satisfies $\calE(\rho\otimes \xi)=\sigma\otimes \xi$.

\bigskip

Next, we show that there exists a maximally asymmetric state in the representation space of the left regular representation by the explicit construction of the state.

\begin{lemma} \label{SMlem:max_asymm_construction}
    Let $F$ be the left regular representation of a finite group $G$ on a Hilbert space $\calK$, $\{\ket{\eta_g}\}_{g\in G}$ be an orthonormal basis of $\calK$ satisfying $F(g)\ket{\eta_h}=\ket{\eta_{gh}}$ for all $g, h\in G$, $H$ be a subgroup of $G$, $k\in G$ be in the normalizer of $H$ in $G$, i.e., $kHk^{-1}=H$, and $\gamma_k$ be a pure state on $\calK$ defined by 
    \begin{align}
        \ket{\gamma_k}:=\frac{1}{\sqrt{|H|}}\sum_{h\in H} \ket{\eta_{kh}}, \label{SMeq:max_asymm_left_regular}
    \end{align}
    Then, $\gamma_k$ is a maximally $(G, F)$-asymmetric pure state with symmetry subgroup $H$. 
\end{lemma}

We note that the condition $kHk^{-1}=H$ is satisfied for all $k\in H$ in general, and it holds for all $g\in G$ when $H$ is a normalizer of $G$. 
Equation~\eqref{SMeq:max_asymm_left_regular} includes Eq.~\eqref{SMeq:max_asymm_left_regular_special} as a special case of $k=e$.

\begin{proof}
    By the definition of $\gamma_e$, for any $g\in G$, we have 
    \begin{align}
        \chi_{\gamma_e, G, F}(g) 
        =\frac{1}{|H|}\sum_{h, h'\in H} \braket{\eta_{h}|\eta_{gh'}}
        =\begin{cases}
            1 & \textrm{if }g\in H \\
            0 & \textrm{if }g\notin H.
        \end{cases}
    \end{align}
    By using this result, for any $k\in G$ satisfying $kHk^{-1}=H$, we get 
    \begin{align}
        \chi_{\gamma_k, G, F}(g) 
        =\braket{\gamma_k| F(g)| \gamma_k}
        =\braket{\gamma_e| F(k)^\dag F(g)F(k)| \gamma_e}
        =\braket{\gamma_e| F(k^{-1}gk)| \gamma_e}
        =\chi_{\gamma_e, G, F}(k^{-1}gk)
        =\begin{cases}
            1 & \textrm{if }g\in H \\
            0 & \textrm{if }g\notin H, 
        \end{cases}
    \end{align}
    where we used the fact that $g\in H$ is equivalent to $k^{-1}gk\in H$. 
\end{proof}

We can construct a covariant catalytic map that is different from the one in Corollary~\ref{SMcor:catalytic_map} when the catalyst is the maximally asymmetric state defined by Eq.~\eqref{SMeq:max_asymm_left_regular}.

\begin{proposition} 
    Let $U$ and $V$ be projective unitary representations of a finite group $G$ on finite-dimensional Hilbert spaces $\calH_\mathrm{in}$ and $\calH_\mathrm{out}$, $F$ be the left regular representation of $G$ on a Hilbert space $\calK$, $\{\ket{\eta_g}\}_{g\in G}$ be an orthonormal basis of $\calK$ satisfying $F(g)\ket{\eta_h}=\ket{\eta_{gh}}$ for all $g, h\in G$, $k\in G$ be in the normalizer of $H$ in $G$, i.e., $kHk^{-1}=H$, $\gamma_k$ be a maximally $(G, F)$-asymmetric pure state with symmetry subgroup $H\subset G$ on $\calK$ defined by Eq.~\eqref{SMeq:max_asymm_left_regular}, $\rho$ and $\sigma$ be general (not necessarily pure) states satisfying $\sym_{G, V}(\sigma)\supset H$, and $\calE_k$ be defined by 
    \begin{align}
        \calE_k(X):=\sum_{a\in A} \mathrm{tr}\left(\left(I_{\calH_\mathrm{in}}\otimes \sum_{h\in H} \eta_{ah}\right)X\right)V(ak^{-1})\sigma V(ak^{-1})^\dag\otimes \gamma_a, 
    \end{align}
    where $A$ is the set of representatives of the left cosets of $H$ in $G$. 
    Then, $\mathcal{E}_k$ is $(G, U\otimes F, V\otimes F)$-covariant and satisfies $\mathcal{E}_k(\rho\otimes \gamma_k)=\sigma\otimes \gamma_k$. 
\end{proposition}

We note that when $H=\{e\}$, the map $\calE_k$ reduces to the map presented in Sec.~8.2 of Ref.~\cite{Marvian2013}. 
The proof is almost the same as that in Lemma~\ref{SMlem:conversion_from_max_asymm}.

\begin{proof}
First, we show that $\mathcal{E}_k$ is well defined, i.e., the definition is independent of the choice of representatives $A$ of the left cosets. 
We take an arbitrary set $A'$ of representatives of the left cosets. 
Then, $A'$ can be written as $\{ah_a\}_{a\in A}$ with some $h_a\in H$. 
We thus have 
\begin{align}
    &\sum_{a'\in A'} \mathrm{tr}\left(\left(I_{\calH_\mathrm{in}}\otimes\sum_{h\in H} \eta_{a'h}\right)X\right)V(a'k^{-1})\sigma V(a'k^{-1})^\dag\otimes \gamma_{a'} \nonumber\\
    =&\sum_{a\in A} \mathrm{tr}\left(\left(I_{\calH_\mathrm{in}}\otimes\sum_{h\in H} \eta_{ah_a h}\right)X\right)V(ak^{-1})V(kh_a k^{-1})\sigma V(kh_a k^{-1})^\dag V(ak^{-1})^\dag\otimes \gamma_{ah_a} \nonumber\\
    =&\sum_{a\in A} \mathrm{tr}\left(\left(I_{\calH_\mathrm{in}}\otimes\sum_{h\in H} \eta_{ah}\right)X\right)V(ak^{-1})\sigma V(ak^{-1})^\dag\otimes \gamma_{a}, 
\end{align}
where we used $k h_a k^{-1}\in H$ and $H\subset\sym_{G, V}(\sigma)$ in the last line.

Next, we show that $\mathcal{E}_k$ is a $(G, U\otimes F, V\otimes F)$-covariant CPTP map. 
Since $\{\sum_{h\in H} \eta_{ah}\}_{a\in A}$ is the set of projections onto orthogonal spaces and satisfies $\sum_{a\in A} (\sum_{h\in H} \eta_{ah})=I_\calK$, $\calE_k$ is a measure-and-prepare channel, which is a CPTP map. 
For any $g\in G$, we have 
\begin{align}
    &(V(g)\otimes F(g))\mathcal{E}_k((U(g)\otimes F(g))^\dag X (U(g)\otimes F(g))) (V(g)\otimes F(g))^\dag \nonumber\\
    =&\sum_{a\in A} \mathrm{tr}\left(\left[I_{\calH_\mathrm{in}}\otimes F(g)\left(\sum_{h\in H} \eta_{ah}\right) F(g)^\dag\right] X\right) V(g)\left(V(ak^{-1})\sigma V(ak^{-1})^\dag\right) V(g)^\dag\otimes F(g)\gamma_a F(g)^\dag \nonumber\\
    =&\sum_{a\in A} \mathrm{tr}\left(\left(I_{\calH_\mathrm{in}}\otimes\sum_{h\in H} \eta_{ga\cdot h}\right)X\right)V(ga\cdot k^{-1})\sigma V(ga\cdot k^{-1})^\dag\otimes \gamma_{ga} \nonumber\\
    =&\mathcal{E}_k(X), 
\end{align}
where we used the fact that $\{ga\}_{a\in A}$ is a set of representatives of the left cosets $G/H$ in the last line.

Finally, we show that $\calE_k$ satisfies $\calE_k(\rho\otimes \gamma_k)=\sigma\otimes \gamma_k$. 
Since the definition of $\mathcal{E}_k$ is independent of the choice of the representatives $A$, we assume that $k\in A$. 
Then, we have 
\begin{align}
    \mathcal{E}_k(\rho\otimes \gamma_k)
    =\sum_{a\in A} \widetilde{\delta}_{a, k} V(ak^{-1})\sigma V(ak^{-1})^\dag \otimes \gamma_a 
    =\sigma\otimes \gamma_k, 
\end{align}
where $\widetilde{\delta}_{g_1, g_2}:=1$ if $g_2^{-1}g_1\in H$ and $\widetilde{\delta}_{g_1, g_2}:=0$ otherwise. 
\end{proof}

\end{document}